\documentclass[a4paper,11pt]{article}
\usepackage{jheppub}
\usepackage{bm}
\usepackage{booktabs}
\usepackage{tabularx}
\usepackage{placeins}
\usepackage{mathrsfs}
\usepackage{braket}
\usepackage{dsfont}
\usepackage{xcolor}
\newcommand{\tr}{\mathrm{tr}}
\usepackage[noabbrev]{cleveref}
\crefname{equation}{Eq.}{Eqs.}
\Crefname{equation}{Eq.}{Eqs.}
\crefrangeformat{equation}{(#3#1#4)--(#5#2#6)}
\usepackage{slashed}

\usepackage{empheq}
\usepackage{ytableau}
\usepackage{tikz}
\usetikzlibrary{arrows.meta,decorations.markings}
\DeclareMathOperator{\Tr}{Tr}
\newcommand{\TrF}{\operatorname{Tr}_F}
\newcommand{\trsigma}{\operatorname{tr}_\sigma}

\providecommand{\Tr}{\operatorname{Tr}}

\usepackage{amsmath,amssymb,longtable,array}
\usepackage[T1]{fontenc}
\usepackage[utf8]{inputenc}

\newcommand{\dchi}{d_\chi}
\newcommand{\ii}{\mathrm{i}}
\newcommand{\trsig}{\operatorname{tr}_{\sigma}}
\newcommand{\LRD}{\overleftrightarrow{D}}

\newcolumntype{Y}{>{\raggedright\arraybackslash}X}
\newcolumntype{L}[1]{>{\raggedright\arraybackslash}p{#1}}

\allowdisplaybreaks[4]
\title{Static-Recoil Factorization in Heavy-Baryon Chiral EFTs}

\author[a]{Ziyu Dong}
\author[b,c]{Teng Ma}
\author[b,c]{Chengjie Yang}
\author[d, e]{Zizheng Zhou}

\affiliation[a]{The Abdus Salam ICTP, Strada Costiera 11, 34135 Trieste, Italy}
\affiliation[b]{International Centre for Theoretical Physics Asia-Pacific (ICTP-AP),\\ University of Chinese Academy of Sciences (UCAS), 100190 Beijing, China}
\affiliation[c]{Taiji Laboratory for Gravitational Wave Universe, University of Chinese Academy of Sciences (UCAS), Beijing, 100190, China}
\affiliation[d]{CAS Key Laboratory of Theoretical Physics, Institute of Theoretical Physics, \\Chinese Academy of Sciences, Beijing 100190, China}
\affiliation[e]{School of Physical Sciences, University of Chinese Academy of Sciences, Beijing 100049, China}

\emailAdd{zdong@ictp.it}
\emailAdd{mateng@ucas.ac.cn}
\emailAdd{yangchengjie@ucas.ac.cn}
\emailAdd{zhouzizheng@itp.ac.cn}

\abstract{
Extending heavy-baryon \(\chi\)PT to higher-spin resonances introduces unphysical lower-spin admixtures, leading to costly path-integral projections.
To resolve this, we develop an on-shell implementation of the heavy-baryon expansion based on recoil-channel partial-wave eigenoperators.
These eigenoperators separate the static mass dependence from recoil structures and assign chiral order directly in amplitude space.
The resulting static--recoil factorization systematically generates non-redundant local operators for heavy-baryon sectors involving higher-spin resonances.
Flavor structures and identical-particle constraints are imposed through a new linear-algebraic reduction that uses von Neumann alternating projections to extract the corresponding common physical subspace.
Beyond this specific application, the framework can be broadly applied to general nonrelativistic effective field theories.
}
\begin{document}

\maketitle

\section{Introduction}
Chiral perturbation theory (\(\chi\)PT) provides the universal low-energy effective description of quantum chromodynamics (QCD) in the regime where hadrons, rather than quarks and gluons, are the relevant degrees of freedom \cite{Goldstone:1962es, Glashow:1967rx, Weinberg:1968de, Coleman:1969sm, Callan:1969sn, Weinberg:1978kz, Gasser:1983yg, Gasser:1984gg,Bernard:2006gx}.
Its organizing principle is the spontaneous breaking of chiral symmetry in the light-quark sector, which gives rise to pseudo-Goldstone bosons and constrains their interactions in a derivative and quark-mass expansion.
In the purely mesonic sector, this leads to a clean power counting in external momenta and light quark masses \cite{Scherer:2005ri}, making \(\chi\)PT one of the most successful applications of effective field theory (EFT) in hadronic physics \cite{Pich:1995bw}.
Once baryons are kept as explicit degrees of freedom, however, the chiral expansion becomes qualitatively more subtle.

The essential complication is that the baryon mass does not vanish in the chiral limit, but remains of order the chiral-symmetry-breaking scale.
A naive relativistic formulation therefore obscures the relation between loop order, operator dimension, and chiral order.
This motivates heavy-baryon \(\chi\)PT (HB\(\chi\)PT), or heavy-baryon projection (HBP)~\cite{Gasser:1987rb,Jenkins:1990jv}, in which the baryon momentum is decomposed as
\[
P^\mu = M v^\mu + k^\mu \,.
\]
The static contribution carried by the baryon mass is separated from the residual momentum $k^\mu$, and the effective theory is reorganized as an expansion in the residual scale, such as $k/M$ or $k/\Lambda_\chi$.
This separation decouples the heavy mass from the low-energy dynamics and restores a transparent chiral power counting, making HB\(\chi\)PT the standard framework for baryonic \(\chi\)PT \cite{Jenkins:1990jv,Bernard:1995dp,Bernard:2006gx}.

The need for this organization extends beyond the spin-\(1/2\) sector.
In baryon chiral dynamics, massive spin-\(3/2\) resonances often have to be retained explicitly rather than absorbed into local low-energy constants \cite{Napsuciale:1996gv,Belkov:1996djr,Hemmert:1997ye}.
The canonical example is the $\Delta(1232)$, whose excitation energy above the nucleon is comparable to other low scales in chiral dynamics, and whose strong coupling to the $\pi N$ channel makes its effects numerically important in many observables.
Thus the explicit inclusion of spin-\(3/2\) baryons is not merely optional, but phenomenologically necessary in many applications.
More generally, a systematic treatment of baryon resonances and their interactions with pseudo-Goldstone bosons and external sources requires a framework that accommodates higher-spin states while preserving controlled chiral power counting.

For higher-spin baryons, the main difficulty lies in the HBP itself.
In the conventional Rarita--Schwinger formulation \cite{Rarita:1941mf}, a relativistic spin-\(3/2\) field contains not only the physical spin-\(3/2\) modes but also auxiliary spin-\(1/2\) sectors.
The HBP must therefore combine the usual velocity projection, which separates large and small components, with additional spin projections that remove these lower-spin admixtures.
The difficulty is not simply the presence of additional fields, but the implementation of these projections within the richer Lorentz structure of higher-spin variables \cite{Napsuciale:1996gv,Belkov:1996djr,Hemmert:1997ye,Pascalutsa:1998pw}.

Modern on-shell and representation-theoretic methods have streamlined EFT operator-basis construction by replacing the reduction of an overcomplete off-shell Lagrangian with local on-shell structures modulo equations of motion (EoM), integration by parts (IBP), and other off-shell redundancies \cite{Arkani-Hamed:2017jhn,Dittmaier:1998nn,Shadmi:2018xan,Durieux:2019eor,Henning:2019enq,DeAngelis:2022qco,Dong:2021vxo,Dong:2022mcv,Dong:2024dce}, with the resulting bases connecting UV theories to IR EFTs through one-loop matching and running \cite{Henning:2016lyp,Dong:2025wvf}, and allowing unitarity, analyticity, and causality to relate the associated low-energy coefficients to consistent UV dynamics \cite{Adams:2006sv,Bertucci:2024qzt,Dong:2024omo,Dong:2025dpy}.
Such reductions are already familiar in meson-baryon chiral Lagrangians, even in the spin-\(1/2\) baryon sector \cite{Ecker:1995rk,Fettes:2000gb,Frink:2006hx}.
In baryonic \(\chi\)PT, a direct on-shell construction faces a mismatch of gradings, since relativistic amplitudes are graded by canonical mass dimension, whereas the heavy-baryon expansion is graded by the residual soft scale.
The chiral power counting must therefore usually be supplied by a field-space HBP analysis before the operator structures are enumerated.

For spin-\(1/2\) baryons this procedure is effective, because the heavy projection operator is simple and the small component can be eliminated straightforwardly.
For spin-\(3/2\) and higher-spin baryons, however, the relativistic tensor-spinor contains unphysical lower-spin sectors, so the projection must both select the heavy component and remove these admixtures \cite{Rarita:1941mf,Pascalutsa:1998pw,Pascalutsa:2000kd,Pascalutsa:2006up}.
At the field level, this requires inverting the kinetic operator inside the subspace selected by the combined velocity and spin projectors, and then expanding the resulting light-component operators according to chiral order.
The size of this projected space grows rapidly: it is already \(27\)-dimensional for spin-\(3/2\), while the corresponding spin-\(5/2\) tensor space contains \(552\) structures before the projected inverse is evaluated.
The symbolic inversion of this projected matrix, followed by its \(1/M\) expansion, becomes practically unmanageable.
This is the field-theoretic bottleneck that the on-shell construction is designed to avoid.

The purpose of this work is to implement the heavy-baryon expansion directly in on-shell amplitude space.
Instead of using an off-shell Lagrangian analysis to assign chiral dimensions, we construct a static--recoil decomposition for relativistic on-shell building blocks.
By expanding massive spinor-helicity variables in the residual momentum \(k/M\), we separate heavy-pair structures into static factors and recoil insertions: the former carry the universal \(\mathcal O(M)\) mass dependence and are absorbed into low-energy constants, while the latter carry explicit powers of the residual scale \(\mathcal O(p)\) and determine the chiral order.

In this on-shell basis, the chiral dimension is obtained by subtracting the static \(M\)-weight from the canonical dimension of the Lorentz-covariant basis.
This \(M\)-weight is fixed by the total angular momentum in the \(\bar B B\to N\) light particles channel and by the factorizable spinor contractions within the heavy pair.
The Young-tableau construction naturally organizes the \(2\to N\) structures into eigenstates of this total angular momentum, obtained by coupling the heavy-pair spins to their orbital angular momentum.
The remaining heavy-pair contractions form the static--recoil factors, which correspond to irreducible representations of the common \(\mathrm{SO}(3)\) little group preserving \(Mv^\mu\) in the heavy-baryon limit.
Thus the static \(M\)-scaling, and hence the HB\(\chi\)PT chiral order, is determined directly from on-shell representation data.

The rest of this paper is organized as follows.
Sec.~\ref{sec:chpt-intro} reviews the chiral effective theory framework, including the building blocks and the chiral counting rules used in the operator enumeration.
Sec.~\ref{sec:heavy-projection} revisits the conventional heavy-baryon projection and identifies the field-space bottleneck for higher-spin resonances.
Sec.~\ref{sec:onshell-heavy-dealing} develops the on-shell static--recoil factorization and the corresponding chiral grading.
Sec.~\ref{sec:example} illustrates the construction in representative operator sectors and explains the treatment of flavor and identical-particle constraints.
Sec.~\ref{sec:pheno} discusses phenomenological implications and possible applications.
Sec.~\ref{sec:conclusion} summarizes the results and outlines future directions.
App.~\ref{app:inverse_C} details the projected spin-\(3/2\) algebra and the field-space bottleneck.
App.~\ref{app:flavor_projectors} presents the flavor projectors and alternating-projection reduction.
App.~\ref{app:conventions} collects conventions and the heavy-pair amplitude--operator correspondence.
App.~\ref{app:comparison_chi_fchi} compares selected sectors with existing Dirac/heavy-baryon representatives.
App.~\ref{app:format-operator-bases} lists representative operator bases.

\section{Chiral Symmetry, Sources, and Resonance Building Blocks}
\label{sec:chpt-intro}

\(\chi\)PT is the low-energy effective description of QCD associated with the spontaneous breaking pattern \(\mathrm{SU}(3)_L\times\mathrm{SU}(3)_R\to\mathrm{SU}(3)_V\).
The corresponding pseudo-Goldstone bosons, \(\pi\), \(K\), and \(\eta\), are described by a nonlinear realization and their interactions are organized in a derivative and quark-mass expansion.
In this work we only need the covariant building blocks used in the subsequent operator enumeration, together with their transformation properties and chiral orders.

The operator sectors considered below are not restricted to the pure pseudo-Goldstone theory, but may also contain explicit vector and axial-vector resonances such as the \(\rho\) and the \(a_1\), a spectrum also familiar from composite-Higgs models \cite{Bellazzini:2014yua,Marzocca:2012zn,Dong:2020eqy}.
Several formulations have been developed to incorporate such spin-1 degrees of freedom, including the Coleman--Callan--Wess--Zumino (CCWZ) matter-field approach~\cite{Callan:1969sn,Coleman:1969sm}, Massive Yang--Mills formulations~\cite{Schwinger:1967tc,Wess:1967jq,Gasiorowicz:1969kn,Kaymakcalan:1983qq,Meissner:1987ge}, the antisymmetric tensor formalism~\cite{Gasser:1983yg,Ecker:1988te}, and the hidden local symmetry (HLS) framework together with its generalized form (GHLS)~\cite{Bando:1987ym,Bando:1985rf,Bando:1987br}.
In this work we use the HLS/GHLS language as a convenient way to organize the vector- and axial-vector-meson building blocks while keeping their chiral transformation properties manifest.
Only transformation laws and chiral counting of HLS/GHLS enter the local operator enumeration~\cite{Song:2024fae}.
We begin with the standard CCWZ nonlinear realization and then introduce the GHLS building blocks needed for the spin-1 resonance sector.

We denote elements of the global chiral group by \(g_L\in\mathrm{SU}(3)_L\) and \(g_R\in\mathrm{SU}(3)_R\).
The pseudo-Goldstone field is described by an \(\mathrm{SU}(3)\)-valued chiral matrix \(U(x)\), which transforms as \(U\to g_L U g_R^\dagger\).
Equivalently, one may introduce two \(\mathrm{SU}(3)\)-valued chiral-frame variables \(\xi_L\) and \(\xi_R\) through \(U=\xi_L^\dagger\xi_R\).
This factorization contains a local redundancy associated with the unbroken subgroup, which leaves \(U\) unchanged,
\begin{equation}
	\xi_L\to h(x)\xi_L,\qquad
	\xi_R\to h(x)\xi_R,\qquad
	h(x)\in \mathrm{SU}(3)_V,
\end{equation}
where \(h(x)\equiv h(u,g_L,g_R)\) is the compensator.
In the unitary representation \(\xi_L^\dagger=\xi_R=u\),
\begin{equation}
	\label{eq:unitary-u-def}
	U(x)=u^2(x), \qquad u(x)=e^{i\Phi(x)/2F}, \qquad u\to h(x)\,u\,g_R^\dagger =g_L\,u\,h^\dagger(x).
\end{equation}
The external sources and spurion are
\begin{equation}
	\label{eq:external-source-defs}
	v_\mu=\frac{1}{2}(r_\mu+\ell_\mu),
	\qquad
	a_\mu=\frac{1}{2}(r_\mu-\ell_\mu),
	\qquad
	\chi=2B_0(s+ip).
\end{equation}
Here \(v_\mu,a_\mu\) are nondynamical external vector and axial-vector sources coupled to the QCD currents \(\bar q\gamma_\mu q\) and \(\bar q\gamma_\mu\gamma_5q\), while \(s,p\) are scalar and pseudoscalar sources.
In the physical limit, the scalar source is set to the quark-mass matrix and the pseudoscalar source is switched off, so that \(\chi\to2B_0\,\mathrm{diag}(m_u,m_d,m_s)\).
The external sources transform as chiral gauge connections,
\begin{equation}
	\label{eq:external-source-transform-short}
	\ell_\mu
	\to
	g_L(\ell_\mu+i\partial_\mu)g_L^\dagger,
	\qquad
	r_\mu
	\to
	g_R(r_\mu+i\partial_\mu)g_R^\dagger,
	\qquad
	\chi\to g_L\chi g_R^\dagger .
\end{equation}
The corresponding field strengths are the curvatures of these connections, e.g.~\(\ell_{\mu\nu}\equiv -i[\ell_\mu+i\partial_\mu,\ell_\nu+i\partial_\nu]\), and similarly for \(r_{\mu\nu}\).
Then the basic CCWZ covariant building blocks are
\begin{equation}
	\label{eq:CCWZ-building-blocks-short}
	\begin{aligned}
		\Gamma_\mu
		 & =
		\frac12\big[
			u^\dagger(\partial_\mu-i\ell_\mu)u
			+u(\partial_\mu-ir_\mu)u^\dagger
			\big],
		\\[3pt]
		u_\mu
		 & =
		i\big[
			u^\dagger(\partial_\mu-i\ell_\mu)u
			-u(\partial_\mu-ir_\mu)u^\dagger
			\big],
		\\[3pt]
		\chi_\pm
		 & =
		u^\dagger\chi \, u^\dagger
		\pm u\, \chi^\dagger u,
		\\[3pt]
		f_\pm^{\mu\nu}
		 & =
		u^\dagger\ell^{\mu\nu}u
		\pm u \, r^{\mu\nu}u^\dagger .
	\end{aligned}
\end{equation}
The objects \(u_\mu\), \(\chi_\pm\), \(f_\pm^{\mu\nu}\), and the covariant derivative \((D_\mu X)=\partial_\mu X+[\Gamma_\mu,X]\) transform homogeneously as \(X\to hXh^\dagger\).

We next introduce the GHLS building blocks for the explicit spin-1 resonance sector.
The hidden local group is
\begin{equation}
	\label{eq:GHLS-hidden-group-short}
	H_{\rm local}^{\rm GHLS} = \mathrm{SU}(3)_{H,L}\times \mathrm{SU}(3)_{H,R}, \qquad G_L(x)\in \mathrm{SU}(3)_{H,L}, \qquad G_R(x)\in \mathrm{SU}(3)_{H,R}.
\end{equation}
The corresponding hidden gauge fields \(L_\mu\) and \(R_\mu\) transform as
\begin{equation}
	\label{eq:LR-hidden-transform}
	L_\mu \to G_L (L_\mu + i\partial_\mu) G_L^\dagger,\qquad R_\mu \to G_R (R_\mu + i\partial_\mu) G_R^\dagger.
\end{equation}
The uppercase hidden fields \(L_\mu,R_\mu\) should not be confused with the lowercase external sources \(\ell_\mu,r_\mu\).

In GHLS the chiral field is factorized as \(U=\xi_L^\dagger\xi_M\xi_R\), with hidden local \(G_{L/R}\) and global chiral \(g_{L/R}\) transformations
\begin{equation}
	\label{eq:GHLS-xi-transform-short}
	\xi_L\to G_L\xi_L g_L^\dagger,
	\qquad
	\xi_R\to G_R\xi_R g_R^\dagger,
	\qquad
	\xi_M\to G_L\xi_M G_R^\dagger .
\end{equation}
Here \(\xi_M\) is a link field between the two hidden local frames associated with \(G_L\) and \(G_R\).
It carries only their relative local orientation, which can be removed by a hidden gauge choice.
Before gauge fixing, covariant derivatives in GHLS are
\begin{equation}
	\label{eq:GHLS-cov-derivs-short}
	\begin{aligned}
		D_\mu\xi_L & = \partial_\mu\xi_L-iL_\mu\xi_L+i\xi_L\ell_\mu, \\
		D_\mu\xi_R & = \partial_\mu\xi_R-iR_\mu\xi_R+i\xi_R r_\mu,   \\
		D_\mu\xi_M & = \partial_\mu\xi_M-iL_\mu\xi_M+i\xi_MR_\mu ,
	\end{aligned}
\end{equation}
which transform in the same hidden-local and global-chiral frames as the corresponding \(\xi\) fields in Eq.~\eqref{eq:GHLS-xi-transform-short}.
Their associated Maurer--Cartan one-forms are
\begin{equation}
	\label{eq:ahatLRM}
	\hat\alpha_{L\mu}
	=
	\frac{1}{i}(D_\mu\xi_L)\,\xi_L^\dagger,
	\qquad
	\hat\alpha_{R\mu}
	=
	\frac{1}{i}(D_\mu\xi_R)\,\xi_R^\dagger,
	\qquad
	\hat\alpha_{M\mu}
	=
	\frac{1}{2i}(D_\mu\xi_M)\,\xi_M^\dagger .
\end{equation}
These one-forms transform adjointly only under the hidden local frames,
\begin{equation}
	\hat\alpha_{L\mu}\to
	G_L\hat\alpha_{L\mu}
	G_L^\dagger, \qquad \hat\alpha_{R\mu}\to G_R\hat\alpha_{R\mu}G_R^\dagger, \qquad \hat\alpha_{M\mu} \to G_L\hat\alpha_{M\mu}G_L^\dagger .
\end{equation}

We now work in the GHLS unitary gauge \(\xi_M=\mathbf{1}\).
From Eq.~\eqref{eq:GHLS-xi-transform-short}, the remaining hidden transformations satisfy
\begin{equation}
	\label{eq:xiM-unitary-gauge}
	G_L(x)=G_R(x)\equiv h(x).
\end{equation}
This gauge choice identifies the two hidden local frames, so all building blocks below transform under the residual diagonal compensator \(h(x)\in\mathrm{SU}(3)_V\).
The covariant vector and axial combinations are
\begin{equation}
	\label{eq:ahatperp-par}
	\hat\alpha_{\parallel\mu}^{\rm GHLS}
	=
	\frac12
	\left(
	\hat\alpha_{R\mu}
	+
	\hat\alpha_{L\mu}
	\right),
	\qquad
	\hat\alpha_{\perp\mu}^{\rm GHLS}
	=
	\frac12
	\left(
	\hat\alpha_{R\mu}
	-
	\hat\alpha_{L\mu}
	\right).
\end{equation}
The link one-form in Eq.~\eqref{eq:GHLS-cov-derivs-short} becomes
\begin{equation}
	\label{eq:alphaM-A}
	\hat\alpha_{M\mu}
	=
	\frac12(R_\mu-L_\mu)
	\equiv A_\mu ,
	\qquad
	A_\mu\to hA_\mu h^\dagger .
\end{equation}
Thus, in the unitary gauge, \(\hat\alpha_{M\mu}\) is identified with the axial-vector meson field, which transforms homogeneously as an adjoint matter field and represents the axial-vector-meson multiplet, such as the \(a_1\)-type resonances.
The vector combination
\begin{equation}
	\frac{1}{2}(R_\mu+L_\mu)\equiv V_\mu,
	\qquad
	V_\mu\to h(V_\mu +i\partial_\mu) h^\dagger
\end{equation}
is the hidden gauge connection of the \(\mathrm{SU}(3)_V\) and represents the vector-meson multiplet, such as the \(\rho\)-type resonances.
Thus the uppercase fields \(V_\mu\) and \(A_\mu\) describe the dynamical spin-1 meson sector.
The vector and axial-vector field strengths are defined as linear combinations of the Yang--Mills curvatures of \(L_\mu\) and \(R_\mu\), e.g.~\(L_{\mu\nu}\equiv -i[L_\mu+i\partial_\mu,L_\nu+i\partial_\nu]\).
Explicitly,
\begin{equation}
	\label{eq:VA-fieldstrength-defs-short}
	\begin{aligned}
		V_{\mu\nu} & = \frac12(R_{\mu\nu}+L_{\mu\nu}), \qquad V_{\mu\nu}\to hV_{\mu\nu}h^\dagger, \\ A_{\mu\nu} &= \frac12(R_{\mu\nu}-L_{\mu\nu}),\qquad A_{\mu\nu}\to hA_{\mu\nu}h^\dagger.
	\end{aligned}
\end{equation}
Hence \(V_{\mu\nu}\) and \(A_{\mu\nu}\) are covariant tensors under the residual \(\mathrm{SU}(3)_V\).

Using Eqs.~\eqref{eq:ahatperp-par} and \eqref{eq:alphaM-A}, the relation between the GHLS and ordinary HLS one-forms is
\begin{equation}
	\label{eq:GHLS-HLS-CCWZ-reduction-short}
	\hat\alpha_{\perp\mu}^{\rm GHLS}
	+
	A_\mu
	=
	\hat\alpha_{\perp\mu}^{\rm HLS}.
\end{equation}
With the standard HLS normalization \cite{Bando:1987br,Harada:2003jx}, the HLS axial one-form is proportional to the CCWZ Goldstone one-form,
\begin{equation}
	\hat\alpha_{\perp\mu}^{\rm HLS}=-\frac{1}{2}u_\mu .
\end{equation}
Thus, removing the explicit axial-vector resonance, \(A_\mu=0\), reduces the GHLS building blocks to the ordinary HLS set.
If the vector resonance is further integrated out at low energies, its leading EoM imposes \(\hat\alpha_{\parallel\mu}^{\rm HLS}=0\), leaving only the CCWZ Goldstone one-form \(u_\mu\).
At the level of covariant building blocks,
\begin{equation}
	\label{eq:GHLS-HLS-CCWZ-chain-short}
	\text{GHLS}:
	\{u_\mu,\hat\alpha_{\perp\mu}^{\rm GHLS},
	\hat\alpha_{\parallel\mu}^{\rm GHLS}\}
	\;\xrightarrow{A_\mu=0}\;
	\text{HLS}:
	\{u_\mu,\hat\alpha_{\parallel\mu}^{\rm HLS}\}
	\;\xrightarrow{\hat\alpha_{\parallel\mu}^{\rm HLS}=0}\;
	\text{CCWZ}:
	\{u_\mu\}.
\end{equation}

The building blocks used in the subsequent operator construction are summarized in Table~\ref{tab:building_blocks}.
\begin{table}[htb]
	\centering
	\small
	\setlength{\tabcolsep}{4pt}
	\renewcommand{\arraystretch}{1.12}
	\begin{tabularx}{\linewidth}{@{} l l X l @{}}
		\toprule
		Symbol & Role                                                                                      & Transformation & Order \\
		\midrule
		\multicolumn{4}{@{}l}{\textit{Goldstone and GHLS one-form structures}}                                                      \\
		$u_\mu$
		       & CCWZ 1-form
		       & $u_\mu\to h\,u_\mu\,h^\dagger$
		       & $\mathcal O(p)$                                                                                                    \\

		$\hat\alpha_{\parallel\mu}$
		       & HLS 1-form
		       & $\hat\alpha_{\parallel\mu}\to h\,\hat\alpha_{\parallel\mu}\,h^\dagger$
		       & $\mathcal O(p)$                                                                                                    \\

		$\hat\alpha_{\perp\mu}$
		       & GHLS 1-form
		       & $\hat\alpha_{\perp\mu}\to h\,\hat\alpha_{\perp\mu}\,h^\dagger$
		       & $\mathcal O(p)$                                                                                                    \\

		\midrule
		\multicolumn{4}{@{}l}{\textit{Spin-1 resonance field strengths}}                                                            \\
		$V_{\mu\nu}$
		       & vector field strength
		       & $V_{\mu\nu}\to h\,V_{\mu\nu}\,h^\dagger$
		       & $\mathcal O(p^2)$                                                                                                  \\

		$A_{\mu\nu}$
		       & axial-vector field strength
		       & $A_{\mu\nu}\to h\,A_{\mu\nu}\,h^\dagger$
		       & $\mathcal O(p^2)$                                                                                                  \\

		\midrule
		\multicolumn{4}{@{}l}{\textit{Sources and spurions}}                                                                        \\
		$\chi_\pm$
		       & scalar/pseudoscalar spurion
		       & $\chi_\pm\to h\,\chi_\pm\,h^\dagger$
		       & $\mathcal O(p^2)$                                                                                                  \\

		$f_{\pm}^{\mu\nu}$
		       & external field strength
		       & $f_{\pm}^{\mu\nu}\to h\,f_{\pm}^{\mu\nu}\,h^\dagger$
		       & $\mathcal O(p^2)$                                                                                                  \\

		\midrule
		\multicolumn{4}{@{}l}{\textit{Covariant Derivative}}                                                                        \\
		$D_\mu X$
		       & On Meson
		       & $D_\mu X\to h\,(D_\mu X)\,h^\dagger$
		       & $\mathcal O(p)$                                                                                                    \\

		$\bar B\overleftrightarrow{D_\mu} B$
		       & On Baryon
		       & $\bar B\overleftrightarrow{D_\mu} B \to  h(\bar B \overleftrightarrow{D_\mu} B)h^\dagger$
		       & $\mathcal O(p^0)$                                                                                                  \\

		\midrule
		\multicolumn{4}{@{}l}{\textit{Matter fields}}                                                                               \\
		$B$
		       & baryon octet
		       & $B\to h\,B\,h^\dagger$
		       & $\mathcal O(p^0)$                                                                                                  \\

		$T^\mu_{ijk}$
		       & baryon decuplet
		       & $T^\mu_{ijk}\to h_i{}^l h_j{}^m h_k{}^n T^\mu_{lmn}$
		       & $\mathcal O(p^0)$                                                                                                  \\
		\bottomrule
	\end{tabularx}
	\caption{Building blocks used in the subsequent operator construction.
	The one-forms \(u_\mu\), \(\hat\alpha_{\perp\mu}\), and \(\hat\alpha_{\parallel\mu}\) organize the GHLS, HLS, and CCWZ descriptions.
	The tensors \(V_{\mu\nu}\) and \(A_{\mu\nu}\) are the covariant field strengths of the vector and axial-vector resonance sectors.}
	\label{tab:building_blocks}
\end{table}

The baryon entry in Table~\ref{tab:building_blocks} is the relative derivative of the baryon bilinear and carries hard \(\mathcal O(M)\) momentum.

\FloatBarrier

Chiral order in the operator basis is carried by soft momenta and light-sector fields.
One-forms and covariant derivatives acting on light-sector building blocks count as \(\mathcal O(p)\), while spurions and source/resonance field strengths count as \(\mathcal O(p^2)\).
Baryon fields count as \(\mathcal O(p^0)\).
The relative derivative \(\bar B\overleftrightarrow D_\mu B\) carries hard \(\mathcal O(M)\) momentum; once this mass factor is absorbed into the corresponding Wilson coefficient, the derivative contributes no additional power of \(p\).
No independent residual derivative acting on a baryon field is introduced in the operator bases below.

In the nonlinear realization, baryon matter fields transform under the unbroken \(\mathrm{SU}(3)_V\).
The spin-\(1/2\) baryon octet is represented by a traceless \(3\times3\) matrix \(B\), with
\begin{equation}
	B\to hBh^\dagger,\qquad D_\mu B=\partial_\mu B+[\Gamma_\mu,B] .
\end{equation}
For the decuplet, we use a totally symmetric flavor tensor \(T_{\{ijk\}}\), with each index transforming in the fundamental representation of \(\mathrm{SU}(3)_V\),
\begin{equation}
	T_{ijk}\to h_i{}^l h_j{}^m h_k{}^n T_{lmn},\qquad (D_\mu T)_{ijk} = \partial_\mu T_{ijk} +(\Gamma_\mu)_i{}^l T_{ljk} +(\Gamma_\mu)_j{}^l T_{ilk} +(\Gamma_\mu)_k{}^l T_{ijl}.
\end{equation}
The physical-state component assignment is
\begin{equation}
	\label{eq:decuplet-tensor}
	\begin{aligned}
		 & T_{111} = \Delta^{++},\quad
		T_{112} = \tfrac{1}{\sqrt{3}}\Delta^{+},\quad
		T_{122} = \tfrac{1}{\sqrt{3}}\Delta^{0},\quad
		T_{222} = \Delta^{-},\quad
		T_{113} = \tfrac{1}{\sqrt{3}}\Sigma^{*+},         \\
		 & T_{123} = \tfrac{1}{\sqrt{6}}\Sigma^{*0},\quad
		T_{223} = \tfrac{1}{\sqrt{3}}\Sigma^{*-},\quad
		T_{133} = \tfrac{1}{\sqrt{3}}\Xi^{*0},\quad
		T_{233} = \tfrac{1}{\sqrt{3}}\Xi^{*-},\quad
		T_{333} = \Omega^{-}.
	\end{aligned}
\end{equation}
\section{Field-Space Heavy-Baryon Projection}
\label{sec:heavy-projection}

In the baryon sector, \(\chi\)PT faces the familiar power-counting problem that the baryon mass \(M\) remains nonzero in the chiral limit and is numerically of the same order as the chiral scale \(\Lambda_\chi\sim 4\pi F_\pi\).
As a result, loop integrals in a relativistic formulation contain analytic pieces whose coefficients involve the hard scale \(M\), obscuring the direct relation between loop order, operator dimension, and chiral order.
One standard remedy is the heavy-baryon projection (HBP), in which the large rest-mass dependence is removed from the propagator and the theory is reorganized as an expansion in residual momenta and light-meson insertions.
Other relativistic formulations, such as infrared regularization and the extended-on-mass-shell scheme, are also available, but for the purposes of the present work the heavy-baryon organization is the most direct starting point because the operator bases constructed later are explicitly graded by heavy-baryon chiral order~\cite{Krause:1990xc,Jenkins:1990jv,Bernard:1995dp,Ecker:1995rk,Hemmert:1997ye,Fettes:2000gb,Frink:2006hx,Bijnens:2018lez}.

This section has two limited goals.
First, we review the field-space HBP in the language of \(\chi\)PT.
Second, we isolate the source of the higher-spin bottleneck that motivates the on-shell construction of the next section.

\subsection{Field-Space Setup of the Heavy-Baryon Expansion}
\label{sec:Framework_HBP}

HBP starts by decomposing a relativistic baryon momentum into a fixed four-velocity and a residual momentum,
\begin{equation}
	\label{eq:hb-momentum-decomp}
	P^\mu = M v^\mu + k^\mu, \qquad v^2=1, \qquad k^\mu=\mathcal O(p).
\end{equation}
After removing the universal rest-mass phase, the relativistic field is split into its large and small components,
\begin{equation}
	\label{eq:light-heavy-def}
	\begin{aligned}
		\Psi(x) & = e^{-i M v\cdot x}\,\bigl(N(x)+h(x)\bigr), \\[3pt]
		N(x)    & \equiv e^{+i M v\cdot x} P_+ \Psi(x),\qquad
		h(x)\equiv e^{+i M v\cdot x} P_- \Psi(x),
	\end{aligned}
\end{equation}
where
\begin{equation}
	\label{eq:velocity-projectors}
	P_\pm=\frac{1\pm \slashed v}{2}, \qquad P_\pm^2=P_\pm, \qquad P_+P_-=0 .
\end{equation}
The field \(N\) is the large, or light, component that contains the low-energy baryon degrees of freedom, whereas \(h\) is the small, or heavy, component with an excitation energy of order \(2M\).
The latter is eliminated in the heavy-baryon effective theory.

A crucial point is that the chiral counting applies only to residual derivatives acting on the phase-redefined fields.
Writing \(D_\perp^\mu \equiv D^\mu-v^\mu(v\!
\cdot\!
D)\), the standard counting is
\begin{equation}
	\label{eq:hb-counting}
	D_\perp^\mu \sim \mathcal O(p), \qquad N\sim \mathcal O(p^0).
\end{equation}
By contrast, a full relativistic derivative acting on \(\Psi\) still contains the static contribution \(Mv^\mu\) and therefore does not carry a definite chiral order before the heavy-baryon reorganization.

The transverse derivative \(D_\perp\) belongs exclusively to the intermediate field-space elimination of the small component.
The on-shell operator bases in Secs.~\ref{sec:onshell-heavy-dealing} and~\ref{sec:example} instead use the baryon relative-derivative convention of Table~\ref{tab:building_blocks}.

The elimination of the small component \(h\) may be described either by solving its equation of motion order by order in \(1/M\), or equivalently by performing the Gaussian Grassmann integral over \(h\) in the generating functional.
For a generic quadratic relativistic term,
\begin{equation}
	\label{eq:relativistic-operator}
	\mathcal L = \bar\Psi\,\Lambda\,\Psi ,
\end{equation}
let \(\Lambda\) denote the phase-rotated operator.
The decomposition of Eq.~\eqref{eq:light-heavy-def} gives
\begin{equation}
	\label{eq:block-L}
	\mathcal{L}
	= \bar N \,\mathcal{A}\, N
	+ \bar h \,\mathcal{B}\, N
	+ \bar N \,\widetilde{\mathcal{B}}\, h
	- \bar h \,\mathcal{C}\, h ,
\end{equation}
where
\begin{equation}
	\label{eq:ABC-def}
	\mathcal{A}\equiv P_+\Lambda P_+,\qquad
	\mathcal{B}\equiv P_-\Lambda P_+,\qquad
	\mathcal{C}\equiv -P_-\Lambda P_-,
	\qquad
	\widetilde{\mathcal{B}}\equiv \gamma^0\mathcal{B}^\dagger\gamma^0 .
\end{equation}
The corresponding generating functional is
\begin{equation}
	\label{eq:hb-generating-functional}
	Z = \int \mathcal D N\,\mathcal D\bar N\,\mathcal D h\,\mathcal D\bar h\; \exp\!
	\left\{i\!\int\! d^4x\,\mathcal L[\bar N,N,\bar h,h]\right\}.
\end{equation}
Completing the square by the change of variables
\begin{equation}
	\label{eq:shift-h}
	h=h'+\mathcal C^{-1}\mathcal B\,N,
	\qquad
	\bar h=\bar h'+\bar N\,\widetilde{\mathcal B}\,\mathcal C^{-1},
\end{equation}
one obtains
\begin{equation}
	\label{eq:diag-L}
	\mathcal L
	=
	\bar N\Bigl(\mathcal A+\widetilde{\mathcal B}\,\mathcal C^{-1}\,\mathcal B\Bigr)N
	-\bar h'\,\mathcal C\,h' .
\end{equation}
The functional integral over \(h'\) is Gaussian and yields a fermionic determinant,
\begin{equation}
	\label{eq:gaussian-heavy-det}
	\int \mathcal D h'\,\mathcal D\bar h'\;
	\exp\!\left\{ -i\!\int\! d^4x\,\bar h'\mathcal C h' \right\}
	\propto \det \mathcal C .
\end{equation}
Separating off this baryon-independent determinant, one obtains
\begin{equation}
	\label{eq:Leff-final}
	Z \propto \det \mathcal C \int \mathcal D N\,\mathcal D\bar N\; \exp\!
	\left\{ i\!\int\! d^4x\,
	\bar N\Bigl(\mathcal A+\widetilde{\mathcal B}\,\mathcal C^{-1}\,\mathcal B\Bigr)N
	\right\}.
\end{equation}
The combination \(\mathcal A+\widetilde{\mathcal B}\,\mathcal C^{-1}\mathcal B\) generates the heavy-baryon effective Lagrangian order by order in \(1/M\).

This conventional field-space HBP is standard and remains the appropriate route for deriving explicit heavy-baryon Lagrangians and their \(1/M\) corrections.
However, it introduces a clear bottleneck for basis construction.
Operators generated after integrating out the heavy component still have to be reduced by EoM and IBP relations.
Moreover, for high-spin resonances in Sec.~\ref{sec:HBP_highspin}, the intermediate field-space decomposition contains unphysical lower-spin sectors.
These representation-dependent structures must be controlled by spin projectors and algebraic identities before a minimal EFT basis can be obtained.

\subsection{Spin-\(1/2\) Projection as a Solvable Case}
\label{sec:HBP_spin12}

For a relativistic Dirac field, the heavy-baryon projection is straightforward.
Using
\begin{equation}
	\label{eq:dirac-projection-identities}
	P_\pm \slashed D P_\pm = \pm (v\!
	\cdot\! D)\,P_\pm,
	\qquad
	P_\pm \slashed D P_\mp = \slashed D_\perp P_\mp,
\end{equation}
the phase-rotated kinetic operator associated with \((i\slashed D-M)\) gives
\begin{equation}
	\label{eq:spin12-blocks}
	\mathcal A = i\,v\!\cdot\!
	D, \qquad \mathcal B = i\,\slashed D_\perp, \qquad \mathcal C = 2M+i\,v\!
	\cdot\!
	D.
\end{equation}
The crucial simplification is that \(\mathcal C\) is scalar-like in spinor space.
Its inverse is therefore obtained immediately as
\begin{equation}
	\label{eq:spin12-Cinv}
	\mathcal C^{-1}
	= \frac{1}{2M}\sum_{n=0}^\infty
	\left(\frac{-i\,v\!\cdot\!
		D}{2M}\right)^n .
\end{equation}
Substituting Eq.~\eqref{eq:spin12-Cinv} into Eq.~\eqref{eq:Leff-final} yields
\begin{equation}
	\label{eq:spin12-effective-lagrangian}
	\mathcal L_{\rm eff}^{(1/2)}
	=
	\bar N
	\left(
	i\,v\!\cdot\!
	D - \slashed D_\perp\, \frac{1}{2M+i\,v\!
		\cdot\!
		D}\, \slashed D_\perp \right)N .
\end{equation}
Expanding the denominator reproduces the familiar heavy-baryon \(1/M\) series.

A standard relativistic interaction reduces equally simply.
For example, the axial bilinear obeys
\begin{equation}
	\label{eq:spin12-axial-map}
	\bar\Psi\,\gamma^\mu\gamma_5\,u_\mu\,\Psi
	\xrightarrow{\mathrm{integrating\ out}\ h}
	2\,\bar N\,S^\mu u_\mu\,N
	+ \mathcal O(1/M),\quad  S^\mu \equiv \frac{i}{2}\gamma_5\sigma^{\mu\nu}v_\nu .
\end{equation}
Accordingly, the leading \(\pi N\) heavy-baryon Lagrangian takes the standard form
\begin{equation}
	\label{eq:spin12-leading-HB}
	\mathcal L_{\pi N}^{(1)}
	=
	\bar N\bigl(i\,v\!\cdot\! D + g_A\,S^\mu u_\mu\bigr)N
	+ \mathcal O(1/M),
\end{equation}
resulting in the leading nucleon propagator \(D_N(k)=i/(v\!\cdot\! k+i\epsilon)\), with \(k\) denoting the residual momentum.
The spin-\(1/2\) sector is therefore simple: once the velocity projection is made, the small component can be integrated out in closed form, and the residual \(1/M\) expansion is straightforward.

\subsection{Higher-Spin Projection and Tensor-Spinor Complexity}
\label{sec:HBP_highspin}

The situation changes qualitatively for higher-spin baryons.
A spin-\(J\) fermion is represented by a tensor-spinor carrying one Dirac index and \(J-1/2\) Lorentz indices.
The heavy-baryon projection must therefore separate not only the small velocity component but also the lower-spin sectors contained in the relativistic tensor-spinor representation.
This is already nontrivial for the spin-\(3/2\) Rarita--Schwinger (RS) field \(\Psi^\mu\)~\cite{Rarita:1941mf,Hemmert:1997ye}.
A convenient free quadratic term may be written as
\begin{equation}
	\label{eq:RS}
	\mathcal L_{3/2}
	=
	\bar\Psi^\mu\,\Lambda_{\mu\nu}\,\Psi^\nu ,
	\qquad
	\Lambda_{\mu\nu}^{\rm free}
	=
	-\Bigl(i\slashed\partial-M\Bigr)g_{\mu\nu}
	+\frac14\,\gamma_\mu\gamma_\rho
	\Bigl(i\slashed\partial-M\Bigr)
	\gamma^\rho\gamma_\nu .
\end{equation}
Its Lorentz content decomposes as
\begin{equation}
	\label{eq:RS-representation}
	\bigl[(\tfrac12,0)\oplus(0,\tfrac12)\bigr]\otimes(\tfrac12,\tfrac12)
	=
	(1,\tfrac12)\oplus(\tfrac12,1)\oplus(\tfrac12,0)\oplus(0,\tfrac12).
\end{equation}
Thus the vector-spinor contains, besides the desired spin-\(3/2\) sector, unphysical spin-\(1/2\) sectors.
A naive velocity projection therefore does not by itself isolate the physical spin-\(3/2\) component.

In the fixed-velocity RS formalism one introduces a complete set of projectors onto the spin-\(3/2\) sector and the complementary spin-\(1/2\) sectors.
For the purpose of separating the retained spin-\(3/2\) mode from the eliminated remainder, it is sufficient to collect the latter into a single lower-spin projector and define~\cite{Hemmert:1997ye}
\begin{equation}
	\label{eq:spin32-projectors}
	P_{\mu\nu}^{3/2} = g_{\mu\nu} -\frac13\gamma_\mu\gamma_\nu -\frac13\bigl(\slashed v\gamma_\mu v_\nu+v_\mu\gamma_\nu\slashed v\bigr), \qquad P_{\mu\nu}^{1/2} \equiv g_{\mu\nu}-P_{\mu\nu}^{3/2} .
\end{equation}
Correspondingly, the retained and eliminated subspaces are
\begin{equation}
	\label{eq:spin32-light-heavy-projectors}
	P_{\mu\nu}^{\rm L}\equiv P_+\,P_{\mu\nu}^{3/2}, \qquad P_{\mu\nu}^{\rm H}\equiv g_{\mu\nu}-P_{\mu\nu}^{\rm L}.
\end{equation}
The phase-redefined fields are then
\begin{equation}
	\label{eq:spin32-fields}
	T_\mu(x) = e^{iMv\cdot x}\,P_{\mu\nu}^{\rm L}\,\Psi^\nu(x), \qquad G_\mu(x) = e^{iMv\cdot x}\,P_{\mu\nu}^{\rm H}\,\Psi^\nu(x).
\end{equation}
Here \(T_\mu\) is the large spin-\(3/2\) component retained in the heavy-baryon theory, while \(G_\mu\) contains the small spin-\(3/2\) component together with the lower-spin sectors to be integrated out.

Inserting Eq.~\eqref{eq:spin32-fields} into Eq.~\eqref{eq:RS} gives
\begin{equation}
	\label{eq:spin_3/2_eft_L}
	\mathcal{L}_\Delta^{\mathrm{eff}}
	=
	\bar T^\mu\,\Big(
	\mathcal A_{\mu\nu}
	+ \widetilde{\mathcal B}_{\mu\rho}\,(\mathcal C^{-1})^{\rho\sigma}\,
	\mathcal B_{\sigma\nu}
	\Big)\,T^\nu,
\end{equation}
with
\begin{equation}
	\label{eq:spin32-ABC}
	\mathcal A_{\mu\nu}
	\equiv
	P^{\rm L}_{\mu\rho}\Lambda^{\rho\sigma}P^{\rm L}_{\sigma\nu},
	\qquad
	\mathcal B_{\mu\nu}
	\equiv
	P^{\rm H}_{\mu\rho}\Lambda^{\rho\sigma}P^{\rm L}_{\sigma\nu},
	\qquad
	\mathcal C_{\mu\nu}
	\equiv
	-P^{\rm H}_{\mu\rho}\Lambda^{\rho\sigma}P^{\rm H}_{\sigma\nu},
\end{equation}
and \(\widetilde{\mathcal B}\) the corresponding Dirac adjoint block.
Conceptually, Eq.~\eqref{eq:spin_3/2_eft_L} parallels the spin-\(1/2\) case, except that \(\mathcal C^{-1}\) is not a scalar-like denominator but an inverse on the projected tensor-spinor space.
At leading order, the mass block \(\mathcal C_\Delta\) has a simple diagonal structure, so its \(O(1/M)\) inverse can be written explicitly in the usual heavy-\(\Delta\) formalism~\cite{Hemmert:1997ye}.
This simplification is special to the leading mass block.
For an effective Lagrangian through chiral order \(\chi\), higher terms up to \(O(1/M^{\chi+1})\) also contribute, and \(\Lambda\) becomes the full Lorentz-covariant operator with undetermined Wilson coefficients, rather than only the kinematic block in Eq.~\eqref{eq:spin12-blocks}.

Beyond leading order, the diagonal structure is lost.
A direct field-space calculation must expand \(\Lambda\) in a \(40\)-dimensional two-index tensor-spinor basis and invert \(\mathcal C\) on the \(27\)-dimensional projected heavy subspace, as detailed in App.~\ref{app:inverse_C}.
For numerical matrices this would scale as \(O(27^3\chi^2)\), but here the entries are symbolic.
If \(N\) covariant operators contribute through order \(\chi\), the ordered noncommuting insertions give a schematic workload \(O(27^3\chi^2N^\chi)\), with \(N\) itself rapidly growing in ordinary operator counting~\cite{Henning:2015alf,Lehman:2015coa,Marinissen:2020jmb}.
The combination in Eq.~\eqref{eq:spin_3/2_eft_L} then produces redundant polynomial combinations of Wilson coefficients, still requiring EoM, IBP, and spin-projector reductions.
Thus the spin-\(3/2\) inverse is useful at leading order, but not a scalable symbolic route to a general higher-order effective Lagrangian.
For higher spin the problem worsens: the spin-\(5/2\) tensor-spinor basis already contains \(552\) structures before projection.
This is the field-space bottleneck avoided by the on-shell construction below.
\section{On-Shell Static--Recoil Factorization}
\label{sec:onshell-heavy-dealing}

In the previous section, we employed the HBP path-integral formalism to implement a systematic power counting for \(\chi\)PT with baryons.
The core step is a velocity-projection decomposition of the relativistic baryon field, followed by removing the non-propagating component via a field redefinition.
At the level of effective action, this requires inverting the quadratic fluctuation operator in the projected subspace, schematically $\bar\Psi\,\Lambda\,\Psi$.
For higher-spin resonances $\Lambda$ becomes a dense operator acting on a high-dimensional tensor-spinor space, and the inversion rapidly turns into an algebraic bottleneck; the spin-$3/2$ case already illustrates the practical obstruction.

This section replaces that field-space inversion by an on-shell construction.
Local EFT interactions correspond to polynomial contact structures in the on-shell S-matrix, so the heavy-baryon organization can be implemented directly in amplitude space and translated back to local operators only at the end.
The goal is to obtain the heavy-baryon operator basis without explicitly inverting the projected higher-spin Lagrangian.

The only extra ingredient needed for chiral EFT is the separation of the hard baryon mass from the soft recoil expansion.
The usual on-shell organization by canonical dimension counts spinors and derivatives relativistically, whereas chiral power counting treats residual momenta as $\mathcal O(p) \ll\Lambda_\chi$ while the baryon mass satisfies $M\sim \Lambda_\chi$.
Thus the amplitude basis must be filtered after all universal powers of $M$ have been factorized.

We first recall the amplitude--operator correspondence in the canonical-dimension on-shell construction and identify where it fails to display chiral order in the baryon sector.
We then expand massive spinors around a common heavy velocity and derive the two linear combinations that isolate the static $\mathcal O(M)$ contribution and the recoil $\mathcal O(p)$ contribution.
Finally, these building blocks are used to construct the heavy-baryon chiral basis directly in on-shell variables, including arbitrary soft insertions and higher-spin external states.

\subsection{Canonical On-Shell Bases and Local Operators}
\label{subsec:amp_kinematics}

An EFT is defined by local contact operators encoding short-distance physics,
\begin{equation}
	\label{eq:eft_lagrangian}
	\mathcal{L}_{\text{low}}
	=
	\mathcal{L}_{0}
	+
	\sum_{d>4}\sum_{a}\frac{C^{(d)}_{a}}{\Lambda^{\,d-4}}\,
	\mathcal{O}^{(d)}_{a},
\end{equation}
where $\mathcal{O}^{(d)}_{a}$ are independent operators of canonical mass dimension $d$.

We use standard massive spinor-helicity notation~\cite{Arkani-Hamed:2017jhn,Dixon:2013uaa,Dittmaier:1998nn}.
A massless momentum factorizes as $p_{\alpha\dot\alpha}=\lambda_\alpha\tilde\lambda_{\dot\alpha}$, while a massive momentum $p^2=m^2$ is represented by spinors $\lambda^I_\alpha$, $\tilde\lambda^I_{\dot\alpha}$ carrying an $SU(2)_{\rm LG}$ little-group index $I=1,2$ and obeying the on-shell EoM
\begin{equation}
	p_{\alpha\dot\alpha}\tilde\lambda^{I\dot\alpha}=m\lambda^I_\alpha,
	\qquad
	p^{\dot\alpha\alpha}\lambda^I_\alpha=m\tilde\lambda^{I\dot\alpha}.
\end{equation}
Lorentz-invariant contact structures are then polynomials in spinor contractions and momenta.

For a given external field content and canonical dimension $d$, the on-shell algorithm enumerates all local polynomials compatible with Lorentz covariance, little-group covariance, momentum conservation, and the permutation symmetries of identical legs.
Relations invisible to on-shell matrix elements, such as EoM, IBP, and Fierz redundancies, are then quotiented out, yielding a basis of independent contact amplitudes in the on-shell space\cite{Henning:2019enq,Dong:2022mcv,Dong:2021vxo,Dong:2022jru,Balkin:2021dko,Goldberg:2024eot,Christensen:2024bdt,DeAngelis:2022qco}.
This basis is translated back into local operators through the amplitude--operator dictionary in Table~\ref{tab:dm_buildingblocks}, which assigns the corresponding spinor building blocks to each field and derivative insertion.

\begin{table}[tbp]
	\centering
	\renewcommand{\arraystretch}{2}
	\begin{tabular}{c|c|c}
		\hline
		Hadron                                                                                                              & Fields                           & Spinors                                          \\
		\hline
		Spin-$1/2$                                                                                                          & $ \psi^L \Leftrightarrow \psi^R$ & $\lambda^{I}\Leftrightarrow \tilde{\lambda}^{I}$ \\
		\hline
		Vector                                                                                                              &
		$X^L_{\mu\nu}\sigma^{\mu\nu} \Leftrightarrow m X_{\mu}\sigma^\mu \Leftrightarrow X^R_{\mu\nu}\bar{\sigma}^{\mu\nu}$ &
		$\lambda^{\{I}\lambda^{J\}} \Leftrightarrow \lambda^{\{I}\tilde{\lambda}^{J\}} \Leftrightarrow \tilde{\lambda}^{\{I}\tilde{\lambda}^{J\}}$                                                                \\
		\hline
		Spin-$3/2$                                                                                                          &
		$\begin{aligned} & \psi^L_{\mu \nu}\sigma^{\mu\nu} \Leftrightarrow m \psi^L_{\mu}\sigma^\mu \Leftrightarrow \\
                & m \psi^R_{\mu}\bar{\sigma}^\mu \Leftrightarrow \psi^R_{\mu\nu}\bar{\sigma}^{\mu\nu}\end{aligned}$      &
		$\begin{aligned} & \lambda^{\{I}\lambda^{J}\lambda^{K\}} \Leftrightarrow \lambda^{\{I}\lambda^{J}\tilde{\lambda}^{K\}} \Leftrightarrow                 \\
                & \lambda^{\{I}\tilde{\lambda}^{J}\tilde{\lambda}^{K\}} \Leftrightarrow \tilde{\lambda}^{\{I}\tilde{\lambda}^{J}\tilde{\lambda}^{K\}}\end{aligned}$                                                    \\
		\hline
	\end{tabular}
	\caption{Representative field/spinor building blocks for the canonical dimension on-shell operator--amplitude dictionary used below.
	Here \(\Leftrightarrow\) denotes EoM-equivalent representatives, with mass factors shown explicitly; braces \(\{\cdots\}\) symmetrize \(SU(2)_{\rm LG}\) indices, and \(\psi^{L,R}_\mu\equiv (1\mp\gamma^5)\psi_\mu/2\).}
	\label{tab:dm_buildingblocks}
\end{table}

The heavy-pair factors extracted in the static--recoil filtration are translated using the conventions in App.~\ref{app:heavy_pair_translation}.

This pipeline is especially natural for canonical-dimension expansions and has been widely used in SMEFT~\cite{Li:2020gnx} and Higgs EFT operator constructions.
In chiral EFT the light bosonic entries can be used directly in the meson and tensor sectors.
The baryon sector requires an additional reorganization, because the canonical and chiral gradings no longer coincide: a relativistic massive-spinor structure is organized by mass dimension, while in the heavy-baryon regime the same structure can contain both static \(\mathcal O(M)\) pieces and recoil \(\mathcal O(p)\) pieces.
Thus the on-shell basis must be filtered so that universal hard powers of \(M\) are factorized at the amplitude level, leaving only recoil structures with definite chiral order.
The remainder of this section first derives this separation from the heavy expansion of massive spinors, and then uses the resulting static--recoil factorization to build the heavy-baryon on-shell amplitude basis.

\subsection{Static--Recoil Decomposition of Massive Spinors}
\label{sec:heavy_spinor_projection}

We now express the heavy-baryon static--recoil separation in massive spinor-helicity variables.
For a heavy pair \(\bar B B\), the universal static dependence is carried by \(Mv^\mu\), while the residual momentum \(k^\mu\) describes recoil.
The goal of this subsection is to isolate the static \(\mathcal O(M)\) dependence at the level of massive spinors, so that the remaining \(k^\mu\)-dependent structures can be assigned their chiral order \(\mathcal O(p)\).

Consider a massive particle with momentum \(P^\mu=Mv^\mu+k^\mu\).
The on-shell condition \(P^2=M^2\) implies \(v \cdot k=-k^2/(2M)=\mathcal{O}(k^2/M)\).
The massive spinors obey
\begin{equation}
	(M v \cdot \sigma + k \cdot \sigma)_{\alpha\dot{\alpha}} \tilde{\lambda}^{I\dot{\alpha}} = M \lambda^I_\alpha, \qquad
	(M v \cdot \bar{\sigma} + k \cdot \bar{\sigma})^{\dot{\alpha}\alpha} \lambda^I_\alpha = M \tilde{\lambda}^{I\dot{\alpha}}.
\end{equation}
In the static limit \(k^\mu\to0\), one has \(\lambda^I=\sqrt{M}\,\xi^I\) and \(\tilde{\lambda}^I=\sqrt{M}\,\tilde{\xi}^I\), with \( (v\cdot \bar{\sigma})^{\dot\alpha\alpha}\xi^I_\alpha = \tilde{\xi}^{I\dot\alpha}.
\) Here \(\xi^I\) and \(\tilde{\xi}^I\) are reference spinors defined in the center-of-mass frame.

For nonzero residual momentum, the physical spinors are obtained by boosting from \(Mv^\mu\) to \(Mv^\mu+k^\mu\).
In the \((\frac12,0)\) representation, the boost matrix expanded through first order in \(k/M\) is
\begin{equation}
	\Lambda_{Mv \to P}
	=
	\frac{M+(P\cdot\sigma)(v\cdot\bar{\sigma})}
	{\sqrt{2M(M+v\cdot P)}}
	=
	\frac{2M+(k\cdot\sigma)(v\cdot\bar{\sigma})}
	{\sqrt{4M^2+2M v\cdot k}}
	=
	\mathbf{1}
	+
	\frac{1}{2M}(k\cdot\sigma)(v\cdot\bar{\sigma})
	+
	\mathcal O(k^2/M^2).
\end{equation}

We now specialize to a heavy baryon--antibaryon pair in the all-incoming convention.
We take \(P_1^\mu=Mv^\mu+k^\mu/2\) and \(-P_2^\mu=Mv^\mu-k^\mu/2\), so that the physical momentum transfer is \(k^\mu\).
Applying the boost gives
\begin{equation}
	\label{eq:rigorous_expansion}
	\begin{aligned}
		\lambda^I_{1,\alpha} & = \sqrt{M} \left[ \xi^I_\alpha + \frac{1}{4M} (k \cdot \sigma \, v \cdot \bar{\sigma})_\alpha{}^\beta \xi^I_\beta + \mathcal{O}(k^2/M^{2})\right ],  \\
		\lambda^J_{2,\alpha} & = \sqrt{M} \left[ \xi^J_\alpha - \frac{1}{4M} (k \cdot \sigma \, v \cdot \bar{\sigma})_\alpha{}^\beta \xi^J_\beta + \mathcal{O}(k^2/M^{2}) \right ],
	\end{aligned}
\end{equation}
and similarly for \(\tilde{\lambda}^I\), with \(k\cdot\bar{\sigma}\, v \cdot \sigma\) acting on \(\tilde{\xi}^I\).
Since local contact amplitudes are polynomials in the spinors and momenta, Eq.~\eqref{eq:rigorous_expansion} induces a systematic expansion in powers of \(k/M\).

For the two-leg bilinear \([1^I2^J]\equiv\epsilon^{\dot\alpha\dot\beta} \tilde\lambda^I_{1,\dot\alpha}\tilde\lambda^J_{2,\dot\beta}\), this gives
\begin{equation}
	\begin{aligned}
		[1^I 2^J]
		 & = M [\tilde{\xi}^I \tilde{\xi}^J]
		+ \frac{1}{4}\Big([\tilde{\xi}^I (k \cdot \bar{\sigma}\, v \cdot \sigma) \tilde{\xi}^J]
		- [\tilde{\xi}^I (v \cdot \bar{\sigma} \, k \cdot \sigma)\tilde{\xi}^J]\Big)
		+ \mathcal{O}(k^2/M)                 \\
		 & = M [\tilde{\xi}^I \tilde{\xi}^J]
		+ [\tilde{\xi}^I|(k\cdot S)|\tilde{\xi}^J]
		+ \mathcal{O}(k^2/M),
	\end{aligned}
\end{equation}
where \(S^\mu = \frac{i}{2}\gamma^5 \sigma^{\mu\nu} v_\nu\) is the heavy-baryon spin operator.
The expansion of \(\langle 1^I2^J\rangle\) is analogous.

To assign chiral power counting cleanly, we form linear combinations that either isolate the static \(\mathcal O(M)\) limit or cancel it and expose the \(\mathcal O(k)\) recoil term.
For the baryon pair,
\begin{equation}
	\begin{aligned}
		[1^I 2^J] -\langle 1^I 2^J \rangle
		 & = M\Big([\tilde{\xi}^I \tilde{\xi}^J] - \langle \xi^I \xi^J \rangle\Big)
		+ \mathcal O(k^2/M),                                                                 \\
		[1^I 2^J] + \langle 1^I 2^J \rangle
		 & = [\tilde{\xi}^I|(k\cdot S)|\tilde{\xi}^J] + \langle\xi^I|(k\cdot S)|\xi^J\rangle
		+ \mathcal O(k^2/M).
	\end{aligned}
\end{equation}
Through the amplitude--operator correspondence, these structures map onto EFT operators, translating the \(k\)-expansion directly into chiral counting:
\begin{equation}\label{eq:xy_scaling_polarization}
	\begin{aligned}
		 & [1^I 2^J] - \langle 1^I 2^J \rangle\leftrightarrow  M\bar N^I N^J \sim \mathcal O(M)\,\epsilon^{IJ},                       \\
		 & [1^I 2^J] + \langle 1^I 2^J  \rangle \leftrightarrow  \partial_\mu(\bar N^I S^\mu N^J) \sim \mathcal O(p)\,\sigma^{IJ}_3 .
	\end{aligned}
\end{equation}
Here \(\sigma_3^{IJ}\) denotes the Pauli matrix along the chosen \(\hat z\)-axis.
Since \(k/M\ll1\), the baryon and antibaryon spin projections can be referred to the same quantization axis.
Choosing this axis along the recoil direction, the two heavy states transform under the same \(SU(2)_{\rm LG}\).
The first combination in Eq.~\eqref{eq:xy_scaling_polarization} gives the universal static contribution, while the second gives the recoil-sensitive \(\mathcal O(p)\) structure.

The same static--recoil factorization applies to scalar products of momenta.
The relative derivative on a baryon bilinear carries a hard \(\mathcal O(M)\) momentum factor.
After this factor is absorbed into the Wilson coefficient, it contributes no additional soft chiral power, whereas derivatives on light fields retain their usual \(\mathcal O(p)\) counting.
For example, in the \(\bar N N \phi\phi\) sector,
\begin{equation}
	\renewcommand{\arraystretch}{1.35}
	\begin{array}{r@{\;}c@{\;}l@{\qquad\Longleftrightarrow\qquad}l@{\;}c@{\;}l}
		2\,P_1\!\cdot P_2
		                             & =    &
		-M^2[\tilde{\xi}^{I}\tilde{\xi}^{J}]
		\langle \xi_I \xi_J\rangle
		                             &
		M^2 v^2\,\bar N N\,\phi^2
		                             & \sim &
		\mathcal O(M^2),
		\\
		(P_1-P_2)\!\cdot p_3
		                             & =    &
		M[\tilde{\xi}^{I}3]\langle \xi_I 3\rangle
		                             &
		\bar N \overleftrightarrow{\partial^\mu}
		N\, \partial_{\mu}\phi\,\phi & \sim & \mathcal O(Mp), \\ 2\,p_3\!
		\cdot p_4
		                             & =    &
		[34]\langle 34\rangle
		                             &
		\bar N N\,\partial^{\mu}\phi\,\partial_{\mu}\phi
		                             & \sim &
		\mathcal O(p^2).
	\end{array}
	\label{eq:heavy_scalar_product_scaling}
\end{equation}

In summary, the on-shell static--recoil factorization isolates the universal \(\mathcal O(M)\) factor of the static baryon configuration, while the remaining kinematic structures count as \(\mathcal O(p)\) recoil insertions.
This rule is the basic ingredient for constructing the heavy-baryon chiral basis in the next subsection.

\subsection{Chiral Grading by Static--Recoil Factorization}
\label{HBP-amp}

We now use the static--recoil factorization derived above to turn the on-shell basis at fixed canonical dimension into a chiral basis.
Canonical dimension correctly grades local relativistic spinor polynomials.
In the heavy-baryon regime the same grading combines two physical scales.
The baryon mass satisfies \(M\sim\Lambda_\chi\), while residual momenta and light-particle momenta are soft, \(\mathcal O(p)\).
A canonical-dimension monomial therefore does not by itself tell us how many powers of the chiral expansion parameter it carries.
To obtain the heavy-baryon basis, one must factor the hard \(M\)-dependence in a way that is invariant under EoM relations, Schouten identities, and momentum conservation.

The correct invariant datum is the total angular momentum in the heavy-pair channel.
We consider an \((N+2)\)-point local contact amplitude with a heavy pair on legs \(1,2\) and \(N\) light particles.
For the heavy momenta,
\begin{equation}
	p_+^\mu \equiv P_1^\mu+P_2^\mu \sim \mathcal O(p),
	\qquad
	p_-^\mu \equiv P_1^\mu-P_2^\mu \sim \mathcal O(M),
	\label{eq:momentum_decompose}
\end{equation}
so \(p_+\) is the momentum carried by the pair into the light sector, whereas \(p_-\) is a hard relative vector.
In the static limit \(p_-\) is transverse to \(p_+\) and behaves as the orbital-polarization vector of the heavy pair.
Heavy spinors on legs \(1,2\), together with powers of \(p_-\), should therefore be organized into irreducible \(SO(3)\) angular-momentum channels rather than assigned an \(M\)-weight from a building-block table term by term, which is not invariant under the amplitude relations.

\begin{figure}[t]
	\centering
	\begin{tikzpicture}[
		x=1cm,y=1cm,
		ext/.style={
				line width=0.8pt,
				postaction={
						decorate,
						decoration={
								markings,
								mark=at position 0.55 with {
										\arrow{Stealth[length=2.2mm,width=1.6mm]}
									}
							}
					}
			},
		midarrow/.style={line width=1.15pt,-{Stealth[length=3.2mm,width=2.2mm]}},
		blob/.style={draw=black,line width=0.8pt,fill=gray!25},
		lab/.style={font=\small,inner sep=1pt}
		]

		\filldraw[blob] (-3.6,0) circle (0.95);

		\draw[ext] (-4.05, 1.55) -- (-3.95, 0.87);
		\node[lab] at (-4.28, 1.70) {$\bar B_2$};

		\draw[ext] (-4.08,-1.55) -- (-3.92,-0.87);
		\node[lab] at (-4.28,-1.72) {$B_1$};

		\draw[ext] (-2.20, 1.18) -- (-2.87, 0.6);
		\node[lab] at (-2.02, 1.30) {$3$};

		\draw[ext] (-2.18,-1.18) -- (-2.87,-0.6);
		\node[lab] at (-1.95,-1.30) {$N+2$};

		\fill (-2.45, 0.34) circle (0.8pt);
		\fill (-2.35, 0.00) circle (0.8pt);
		\fill (-2.45,-0.34) circle (0.8pt);

		\draw[midarrow] (-1.15,0) -- (0.10,0);

		\filldraw[blob]
		(1.46,0.364)
		arc[start angle=60, delta angle=240, radius=0.42]

		.. controls (1.52,-0.34) and (1.56,-0.22) .. (1.70,-0.18)
		-- (3.62,-0.18)

		.. controls (3.68,-0.18) and (3.71,-0.32) .. (3.771,-0.450)

		arc[start angle=210, delta angle=300, radius=0.90]

		.. controls (3.71,0.32) and (3.68,0.18) .. (3.62,0.18)
		-- (1.70,0.18)

		.. controls (1.56,0.22) and (1.52,0.34) .. (1.46,0.364)
		-- cycle;

		\node[lab] at (2.65,0.58) {$\mathcal{J}(p_{+};J_{12},\omega)$};

		\draw[ext] (0.78, 1.45) -- (1.05, 0.36);
		\node[lab] at (0.62,1.60) {$\bar B_2$};

		\draw[ext] (0.76,-1.43) -- (1.05,-0.36);
		\node[lab] at (0.60,-1.58) {$B_1$};

		\draw[ext] (5.95, 1.13) -- (5.28, 0.55);
		\node[lab] at (6.12,1.26) {$3$};

		\draw[ext] (5.98,-1.13) -- (5.28,-0.55);
		\node[lab] at (5.73,-1.27) {$N+2$};

		\fill (5.67, 0.34) circle (0.8pt);
		\fill (5.75, 0.00) circle (0.8pt);
		\fill (5.67,-0.34) circle (0.8pt);

	\end{tikzpicture}
	\caption{Schematic static--recoil factorization of a local \(\bar B B\) operator with \(N\) light mesons.
	The light-sector contact structure is resolved through an auxiliary off-shell current \(\mathcal J(p_+;J_{12},\omega)\), labelled by the heavy-pair momentum \(p_+\), the \(12\)-channel angular momentum \(J_{12}\), and the Lorentz chiral weight \(\omega\).}
	\label{fig:eft-current-factorization}
\end{figure}
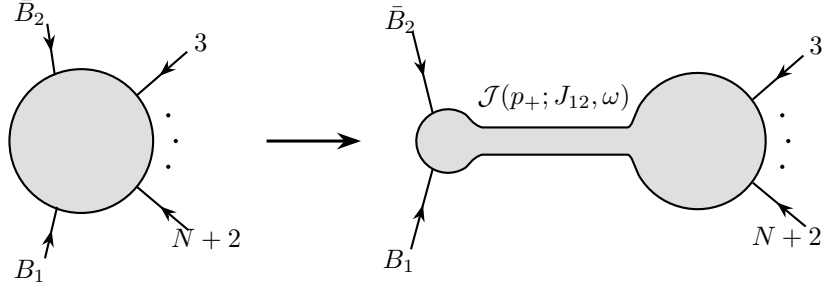

This motivates the factorized bookkeeping shown in Fig.~\ref{fig:eft-current-factorization}.
We organize the contact amplitude as a sum over intermediate currents,
\begin{equation}
	\mathcal A_{B_1\bar B_2\,3\dots (N+2)}
	=
	\sum_{J_{12},\,\omega}\mathcal A_3(12\to\mathcal J(J_{12},\omega))
	\,
	\odot
	\,
	\mathcal R(\mathcal J\to 3\dots (N+2)) .
	\label{eq:current_factorization}
\end{equation}
The current \(\mathcal J(J_{12},\omega)\) is labelled by the \(12\)-channel angular momentum \(J_{12}\) and by a Lorentz chiral weight \(\omega\), equivalently the anti-holomorphic minus holomorphic spinor weight,
\begin{equation}
	\omega\equiv n_{\tilde\lambda}-n_{\lambda},
	\label{eq:current_chiral_weight}
\end{equation}
so the two sides match both the total \(SO(3)\) angular momentum and the \(SU(2)_L\times SU(2)_R\) spinor content of the current.
For fixed \((J_{12},\omega)\), the symbol \(\odot\) denotes the unique singlet contraction obtained after stripping the same current polarization from the left and right tensors.
The auxiliary off-shell current \(\mathcal J\) is only used to diagonalize the \(12\)-channel angular momentum.
It is not a physical propagating state, although in the left three-point amplitude it is treated as a massive on-shell particle.

The off-shell-current coupling \(\mathcal R\) is constructed inside the same semistandard Young-tableau embedding as the full \((N+2)\)-point contact space, rather than as an ordinary on-shell \((N+1)\)-point amplitude.
The pair \(1,2\) is constrained to a fixed \((J_{12},\omega)\) current while the right block retains \(p_+^2\) and the mixed heavy--light Mandelstam data of the original \((N+2)\)-point local interaction.
This preserves the full local kinematic information without duplicating it on the two sides of Eq.~\eqref{eq:current_factorization}.

After the current polarization is stripped from the right block, the remaining right tensor contains only light spinors, light momenta, and the soft momentum \(p_+\).
Its power counting is therefore already the chiral one.
All hard \(M\)-dependence to be factorized resides in the left three-point amplitude \(\mathcal A_3\).
The construction of the heavy-baryon chiral basis is thus reduced to a finite three-point problem for a heavy pair and an auxiliary current of spin \(J_{12}\) and chiral weight \(\omega\).

The finiteness and counting of these three-point structures are the same kinematic facts that appear in massive spinor-helicity amplitudes and in spinning conformal three-point tensor structures~\cite{Arkani-Hamed:2017jhn,Costa:2011mg}.
For heavy spins \(s_1,s_2\) and current spin \(J\), the number of independent three-point tensor structures is most transparently written after decomposing the common heavy-pair little group:
\begin{equation}
	s_1\otimes s_2
	=
	\bigoplus_{\ell=|s_1-s_2|}^{s_1+s_2}\ell .
	\label{eq:heavy_pair_little_group_factorization}
\end{equation}
Here \(\ell\) labels an irreducible representation of the common heavy-pair little group.
Physically, it is the angular momentum obtained by coupling the two intrinsic spins \(s_1\) and \(s_2\), not the total \(12\)-channel angular momentum.
The total number is
\begin{equation}
	N_3(s_1,s_2,J) = \sum_{\ell=|s_1-s_2|}^{s_1+s_2}\left(2\min(\ell,J)+1\right).
	\label{eq:three_point_count}
\end{equation}
Equivalently, the \(\ell\) sector contributes \(2\min(\ell,J)+1\) tensor structures.
The same total count agrees with the standard three-spin form familiar from \(\mathrm{CFT}_3\) tensor-structure counting,
\begin{equation}
	N_3(s'_1,s'_2,s'_3)
	=(2s'_1+1)(2s'_2+1)-q(q+1),
	\qquad
	q=\max(0,s'_1+s'_2-s'_3),
	\label{eq:three_point_sorted_count}
\end{equation}
where \(s'_1\le s'_2\le s'_3\) are the sorted values of \(s_1,s_2,J\).
Once \((J,\ell)\) and the chiral weight \(\omega\) are fixed, the left three-point tensor structure is fixed.
In the saturated regime \(J\ge s_1+s_2\), all heavy-spin components are fully resolved; the remaining growth with \(J\) is purely orbital and carries the expected hard factor \(M^{J}\) in spinor units.
For \(J<s_1+s_2\), part of the heavy spin must be contracted into scalar structures, and the apparent \(M\)-weight of individual brackets is not stable under linear recombination.

The spin-\(\frac12\) pair gives the minimal example.
From Eq.~\eqref{eq:three_point_count},
\begin{equation}
	N_3\!
	\left(\frac12,\frac12,J\right)
	=
	\begin{cases}
		1+1, & J=0,    \\
		1+3, & J\ge 1.
	\end{cases}
	\label{eq:spin_half_three_point_count}
\end{equation}
The two \(J=0\) structures are most usefully chosen as
\begin{equation}
	x \equiv [\mathbf{1}\mathbf{2}]
	-\langle\mathbf{1}\mathbf{2}\rangle
	\sim \mathcal O(M),
	\qquad
	y \equiv [\mathbf{1}\mathbf{2}]
	+\langle\mathbf{1}\mathbf{2}\rangle
	\sim \mathcal O(p).
	\label{eq:xy_def}
\end{equation}
Thus \(x\) is the static singlet while \(y\) is the recoil-sensitive triplet projected to a scalar by the soft momentum.
This is the simplest place where a naive bracket-by-bracket \(M\)-count fails: both \([\mathbf{1}\mathbf{2}]\) and \(\langle\mathbf{1}\mathbf{2}\rangle\) look static, but their sum cancels the static limit and begins at recoil order.
For \(J\ge1\), the triplet is fully resolved into the spin-\(J\) current, so the four tensor structures have definite angular-momentum interpretation.

This fixes the chiral grading used in the basis generation.
The amplitude dimension \(d_{\rm amp}\) counts the spinor-polynomial degree and therefore combines hard and soft powers.
The physical chiral dimension is obtained by subtracting the hard degree and including the field-dimension compensation,
\begin{equation}
	d_\chi=d_{\rm amp}-d_M+\Delta_{\rm field},
	\qquad
	d_M=J_{12}+n_x .
	\label{eq:hard_degree_rule}
\end{equation}
Here \(\Delta_{\rm field}\) compensates the difference between the dimension of a field operator and that carried by its on-shell spinor representative.
Thus \(u_\mu\) requires no compensation, whereas \(f_\pm^{\mu\nu}\) contributes one additional unit.
The term \(J_{12}\) is the universal hard scaling of a spin-\(J_{12}\) left three-point current: its polarization carries \(2J_{12}\) heavy spinor indices, each of order \(M^{1/2}\).
The integer \(n_x\) counts explicit static singlet factors \(x\), normalized as \(x/M\); \(y\) is already recoil order and is not subtracted.
Such extra \(x,y\) prefactors can appear only in unsaturated sectors \(J_{12}<s_1+s_2\).
For a spin-\(\frac12\) heavy pair this means only the \(J_{12}=0\) current; once \(J_{12}\ge1\), \(d_M=J_{12}\).

For the phenomenologically important \(\Delta\)--\(\bar\Delta\) pair, \(s_1=s_2=\frac32\), the common heavy-pair little group decomposes as
\begin{equation}
	\frac32\otimes\frac32=0\oplus1\oplus2\oplus3 .
	\label{eq:spin_threehalf_little_group_factorization}
\end{equation}
Equation~\eqref{eq:three_point_count} gives
\begin{equation}
	N_3\!
	\left(\frac32,\frac32,J\right)
	=
	\begin{cases}
		1+1+1+1, & J=0,   \\
		1+3+3+3, & J=1,   \\
		1+3+5+5, & J=2,   \\
		1+3+5+7, & J\ge3.
	\end{cases}
	\label{eq:spin_threehalf_three_point_count}
\end{equation}
Each entry is the contribution of \(\ell=0,1,2,3\).
When \(\ell>J\), the excess intrinsic spin must be compensated by orbital angular momentum, so only \(2J+1\) components are resolved.
When \(J\ge\ell\), the whole \(2\ell+1\)-dimensional irrep is unfolded.
Beyond \(J=3\), increasing \(J\) only adds orbital powers of \(p_-\), so the chiral grading is fixed by the same left three-point classification.

For spin-\(\frac32\), this decomposition is formulated entirely in terms of physical external states: the symmetric rank-three massive little-group representation excludes lower-spin components from the outset.
The \(J_{12}=0,1,2\) left-current structures translate into the \(x/y\)-graded tensors \(\mathcal P_{xx}\), \(\mathcal P_{xy}\), and \(\mathcal P_{yy}\) derived in App.~\ref{app:conventions}, while the \(J_{12}\ge3\) channels use the standard spin-\(\frac32\) amplitude--operator dictionary.
The resulting explicit \(\bar T T\) chiral operator bases are collected in App.~\ref{app:format-operator-bases}.

The right panel of Fig.~\ref{fig:eft-current-factorization} represents this angular-momentum bookkeeping.
The left blob is the original local \((N+2)\)-point contact interaction.
The right blob displays the same local data after isolating the two-particle current \(\mathcal J(p_+;J_{12},\omega)\) built from the heavy pair.
The right subamplitude keeps the light-sector local basis and the \(p_+\)-dependent Mandelstam variables, while the left three-point substructure contains all static and recoil recombinations associated with the heavy pair.

The current factorization defines a complete and independent basis in its own right.
For each \((J_{12},\ell,\omega)\), the left three-point currents form an independent basis.
The label \(\ell\) belongs only to this left heavy-spin coupling; the right block \(\mathcal R\) is indexed solely by the shared current data \((J_{12},\omega)\).
At fixed \((J_{12},\omega)\), the same right-block basis is sewn to every allowed \(\ell\) sector.
The right block is constructed as a complete and independent local basis within the original \((N+2)\)-point SSYT embedding \cite{Dong:2021vxo,Dong:2022mcv,Dong:2022jru}.
Sewing these two bases through the unique current contraction and summing over distinct channel quantum numbers therefore yields the complete contact basis without cross-channel redundancies.
Its span is equivalent to the original relativistic SSYT contact space, but the factorized basis is a new construction: it exposes the hard mass degree and recoil order, produces the heavy-baryon chiral grading directly, and extends uniformly to higher-spin heavy states and more general nonrelativistic kinematics.

The practical advantage is that no chiral order is assigned by eye to a relativistic monomial.
Such an assignment is not invariant under momentum conservation, EoM relations, or Schouten identities; a structure that appears to start at \(\mathcal O(M^k)\) can have a linear combination whose static part cancels, reducing it to \(\mathcal O(M^{k-1}p)\).
The angular-momentum construction lowers the hard \(M\)-power as far as allowed by the fixed \(J_{12}\), \(\ell\), and \(\omega\) quantum numbers.
After those unavoidable hard factors are absorbed into Wilson coefficients, the remaining powers give the chiral expansion in \(p/\Lambda_\chi\).

In summary, the on-shell construction performs two tasks.
First, it provides a Lorentz-covariant local basis at fixed canonical dimension, now refined by the angular momentum \(J_{12}\) of a chosen two-to-\(N\) channel.
Second, for heavy-baryon kinematics it shows that all hard static--recoil information is captured by the finite three-point problem \(\bar B B\mathcal J(J_{12},\omega)\).
After the corresponding \(M\)-dependent factors are absorbed into Wilson coefficients, the remaining right block gives the physical chiral-order series.
This is the reason the same method applies uniformly to spin-\(\frac12\) baryons and to higher-spin resonances such as the \(\Delta\), as illustrated in the examples below.
\section{Flavor Sectors and Factorized Sewing}
\label{sec:example}
\subsection{Flavor Projection and Exchange Symmetry}
\label{sec:flavor_symmetry}

The static--recoil filtration described above fixes the independent Lorentz structures at a given chiral order.
The remaining constraints are internal: the operator must be an \(SU(3)_V\) singlet, decuplet or antidecuplet legs must be projected onto the desired irreducible flavor representation, and identical external particles must obey the appropriate exchange symmetry.
We impose these conditions only after the kinematic on-shell reduction.
This keeps the recoil grading separate from the flavor algebra and turns the remaining problem into a finite-dimensional linear-algebra reduction.

In the trace formalism, an octet building block is represented by a traceless \(3\times3\) matrix \(X=X^aT^a\), where \(T^a=\lambda^a/2\) with \(\lambda^a\) the Gell-Mann matrices.
Mesons and spin-\(1/2\) baryons transform as flavor octets, while spin-\(3/2\) baryons may transform in higher representations such as the decuplet \(\mathbf{10}\).
We build flavor singlets from trace monomials of these octet matrices and then impose three linear constraints: Cayley--Hamilton (CH) trace relations, decuplet or antidecuplet embedding conditions, and identical-particle symmetry.
Because these projectors do not commute, we extract the common subspace by the alternating-projection construction summarized in App.~\ref{app:flavor_projectors}.

The first reduction removes trace identities.
For any traceless \(3\times3\) matrix \(X\), the CH identity gives
\begin{equation}
	\label{eq:CH_traceless_matrix}
	X^3-\frac{1}{2}\Tr(X^2)\,X -\frac{1}{3}\Tr(X^3)\,\mathds{1}=0 .
\end{equation}
Multiplying this identity by additional matrices and polarizing the result gives the trace relations needed in the \(SU(3)\) trace algebra.
We impose these relations as linear constraints on the naively enumerated trace structures, thereby reducing the overcomplete trace list to independent representatives.
Equivalently, this reduction is implemented by a projector \(\mathcal P_{\rm CH}\) onto the CH-reduced trace space.

When decuplet baryons are present, we implement their flavor representation by embedding a decuplet or antidecuplet index into an auxiliary pair of octets.
The relevant decomposition is
\begin{equation}
	\label{eq:8x8_decomp}
	\mathbf{8}\otimes\mathbf{8}
	=
	\mathbf{27}\oplus\mathbf{10}\oplus\overline{\mathbf{10}}
	\oplus\mathbf{8}_S\oplus\mathbf{8}_A\oplus\mathbf{1}.
\end{equation}
Thus we introduce projectors \(\mathcal P_{10}\) and \(\mathcal P_{\overline{10}}\) acting on the auxiliary octet pair \(T^a,T^b\).
In adjoint-index notation,
\begin{equation}
	\label{eq:P10P10bar_tensor_compact}
	\bigl(\mathcal P_{10/\overline{10}}\bigr)^{ab,cd}
	=
	\frac{5}{36}(\delta^{ac}\delta^{bd}-\delta^{ad}\delta^{bc})
	-\frac{1}{6}
	\bigl(d^{ace}d^{bde}-d^{ade}d^{bce}\bigr)
	\mp
	\frac{i}{4}
	\bigl(d^{bce}f^{ade}+d^{ade}f^{bce}\bigr),
\end{equation}
with the \(SU(3)\) antisymmetric and symmetric structure constants
\begin{equation}
	[T^a,T^b]=i f^{abc}
	T^c, \qquad \{T^a,T^b\}=\frac{1}{3}\delta^{ab}\mathds{1}+d^{abc}T^c .
\end{equation}
The first two terms select the antisymmetric non-adjoint part of \(\mathbf{8}\otimes\mathbf{8}\), while the final term distinguishes \(\mathbf{10}\) from \(\overline{\mathbf{10}}\).
After translating this tensor projector into the trace basis, it acts as another linear map on the same flavor space.

The Cayley--Hamilton reduction and the representation embedding are independent constraints and do not commute,
\begin{equation}
	\label{eq:noncommute}
	[\mathcal P_{\rm CH},\,\mathcal P_{10/\overline{10}}]\neq 0 .
\end{equation}
Hence the physical flavor space is not obtained by simply multiplying projectors once.
Let \(\mathcal P_{\rm rep}\) denote the product of all required decuplet and antidecuplet embedding projectors.
We form the alternating-projection map
\begin{equation}
	\label{eq:P_CH_rep_def}
	\mathcal T
	=
	\mathcal P_{\rm CH}\,\mathcal P_{\rm rep},
\end{equation}
and extract the physical flavor basis as the fixed-point subspace of \(\mathcal T\).
The projector-limit formulation is given in App.~\ref{app:flavor_projectors}.

Finally, identical-particle symmetry is imposed on the full Lorentz--flavor operator basis.
We first keep identical external fields artificially labeled, \(X_1,\ldots,X_n\).
For these labeled fields, the candidate operator space is the tensor product of the Lorentz and flavor spaces,
\begin{equation}
	\label{eq:labeled_product_basis}
	\operatorname{span}\{\mathcal O_{i a}\}
	=
	\operatorname{span}\{\mathcal L_i\}
	\otimes
	\operatorname{span}\{\mathcal F_a\},
	\qquad
	\mathcal O_{i a}\equiv \mathcal L_i\,\mathcal F_a .
\end{equation}
Thus every Lorentz representative is combined with every flavor representative.
The product space is then projected onto the required permutation symmetry.

The permutation group \(S_n\) is generated by the transposition \(s=(12)\) and the cycle \(r=(12\cdots n)\).
Their actions on the Lorentz and flavor bases define matrices \(R_{\mathcal L}(s),R_{\mathcal L}(r)\) and \(R_{\mathcal F}(s),R_{\mathcal F}(r)\).
On the full operator space the induced actions are
\begin{equation}
	\label{eq:perm_generators_product_space}
	R(s)=R_{\mathcal L}(s)\otimes R_{\mathcal F}(s), \qquad R(r)=R_{\mathcal L}(r)\otimes R_{\mathcal F}(r).
\end{equation}
The bosonic and fermionic identical-particle projectors are then obtained by evaluating the corresponding Young operators on these matrices,
\begin{equation}
	\label{eq:identical_particle_projector}
	\mathcal P_{\rm B/F}
	=
	\mathcal Y_{\rm B/F}\left[R(s),R(r)\right]\,.
\end{equation}
Here the subscript \({\rm B/F}\) denotes identical bosons or fermions.
We keep the nonzero linearly independent vectors selected by the appropriate projector and only then remove the auxiliary labels \(X_1,\ldots,X_n\to X\).
The resulting operators are flavor-reduced and obey the required identical-particle exchange symmetry.
\subsection{\texorpdfstring{Factorized Sewing Construction for $\bar{B}
B u f_+$}{Factorized Sewing Construction for Bbar B u f+}} \label{sec:NNpigamma}

We use the positive-helicity \(\bar B B u f_+\) sector as a compact example of the factorized construction in Sec.~\ref{sec:onshell-heavy-dealing}.
The all-incoming ordering is \((\bar B,B,u,f_+)=(1,2,3,4)\).
The \(u_\mu\) leg is the massive spin-one object on leg \(3\), and the \(f_+\) source is the positive-helicity massless leg \(4\).
Flavor labels are restored only after the Lorentz sewing step.
We first construct the factorized heavy-pair left block, then construct the residual right block containing the light-sector local data, and finally sew the common auxiliary-current polarizations to obtain the local contact amplitude.

At fixed canonical amplitude dimension \(d_{\rm amp}\), the working logic is:
\begin{equation}
	\mathcal A_{\bar B B u f_+}^{(d_{\rm amp})}
	=
	\sum_{J_{12}}
	\mathcal A_{L}^{(J_{12})}\odot_{J_{12}}\mathcal A_{R}^{(J_{12})},
	\qquad
	\mathcal A_R \text{ retains the local }(u,f_+)\text{ data}.
\end{equation}
The auxiliary current carries the hard relative momentum
\begin{equation}
	Q=P_1-P_2,
	\qquad
	P_1+P_2\sim\mathcal O(p),
	\qquad
	P_1-P_2\sim\mathcal O(M),
\end{equation}
so \(Q\) contributes to the hard \(M\)-order counted in Eqs.~\eqref{eq:momentum_decompose} and \eqref{eq:hard_degree_rule}.

For a spin-\(\frac12\) heavy pair, the only closed current channel is \(J_{12}=0\).
The corresponding left factors are
\begin{equation}
	A_L^{J_{12}=0}= \{x,y\},
\end{equation}
while the open current channel \(J_{12}=1\) is represented by
\begin{equation}
	A_L^{J_{12}=1}= \left\{ \langle\mathbf{2}\mathbf{J}\rangle[\mathbf{1}\mathbf{J}], \langle\mathbf{1}\mathbf{J}\rangle[\mathbf{2}\mathbf{J}], [\mathbf{2}\mathbf{J}][\mathbf{1}\mathbf{J}], \langle\mathbf{1}\mathbf{J}\rangle\langle\mathbf{2}\mathbf{J}\rangle \right\}.
	\label{eq:BBuf_left_J1}
\end{equation}
The \(J_{12}=2\) channel is obtained by multiplying the same open basis by the hard-raising factor \(\langle \mathbf{Q}\mathbf{J}\rangle[\mathbf{Q}\mathbf{J}]\).
Higher channels repeat that factor.

The right residual sector is described by the complete SSYT family in the formal alphabet \(J<3<4<3'\).
We display one representative only.
The tableau
\[
	\begin{tikzpicture}[baseline=(yt.base)]
		\node[inner sep=0pt,outer sep=0pt] (yt) {$
				\begin{ytableau}
					\textcolor{blue}{J} & J & 4\\
					\textcolor{blue}{4} & 4 & 3'
				\end{ytableau}
			$};
		\draw[blue,line width=0.08em]
		([xshift=0.02em,yshift=-0.03em]yt.north west) rectangle
		([xshift=1.55em,yshift=-3.08em]yt.north west);
		\draw[blue,line width=0.08em]
		([xshift=0.02em,yshift=-1.56em]yt.north west) --
		([xshift=1.55em,yshift=-1.56em]yt.north west);
	\end{tikzpicture}
\]
corresponds to the right factor
\begin{equation}
	\mathcal A_R=\langle\mathbf{3}\mathbf{J}\rangle[4\mathbf{3}][4\mathbf{J}].
	\label{eq:BBuf_right_YT_example}
\end{equation}
The same right-channel construction can also be realized by two temporary massless auxiliary particles of spin \(1/2\); the temporary labels are used only to realize the formal \(\mathbf{J}\) dependence and are removed after sewing.

At \(d_{\rm amp}=3\), only the open \(J_{12}=1\) channel contributes.
A simple sewing example is
\begin{equation}
	\langle\mathbf{2}\mathbf{J}\rangle[\mathbf{1}\mathbf{J}]\odot
	\langle\mathbf{3}\mathbf{J}\rangle[\mathbf{3}4][4\mathbf{J}]
	=
	-\langle\mathbf{2}\mathbf{3}\rangle[\mathbf{3}4][\mathbf{1}4].
	\label{eq:BBuf_lowest_sewing_example}
\end{equation}
The two independent basis elements at \(d_{\rm amp}=3\) are
\begin{center}
	\small
	\setlength{\tabcolsep}{3pt}
	\renewcommand{\arraystretch}{1.35}
	\begin{tabularx}{\textwidth}{@{}c c >{\raggedright\arraybackslash}p{0.2\textwidth} >{\raggedright\arraybackslash}p{0.31\textwidth} >{\raggedright\arraybackslash}X@{}}
		\toprule
		index                                                            & \(J_{12}\) & \(A_L\) & \(A_R\) & \(A_{\rm sew}\) \\
		\midrule
		\(L^{(3)}_1\)                                                    & \(1\)      &
		\(\langle\mathbf{2}\mathbf{J}\rangle[\mathbf{1}\mathbf{J}]\)     &
		\(\langle\mathbf{3}\mathbf{J}\rangle[\mathbf{3}4][4\mathbf{J}]\) &
		\(\langle\mathbf{2}\mathbf{3}\rangle[\mathbf{3}4][\mathbf{1}4]\)                                                    \\
		\(L^{(3)}_2\)                                                    & \(1\)      &
		\(\langle\mathbf{1}\mathbf{J}\rangle[\mathbf{2}\mathbf{J}]\)     &
		\(\langle\mathbf{3}\mathbf{J}\rangle[\mathbf{3}4][4\mathbf{J}]\) &
		\(\langle\mathbf{1}\mathbf{3}\rangle[\mathbf{3}4][\mathbf{2}4]\)                                                    \\
		\bottomrule
	\end{tabularx}
\end{center}

At \(d_{\rm amp}=4\), the independent sewn structures come from two \(J_{12}=1\) residual blocks and one raised \(J_{12}=2\) block.
The relevant sewn expressions are
\begin{equation}
	\begin{aligned}
		[\mathbf{2}\mathbf{J}][\mathbf{1}\mathbf{J}]
		\odot
		\langle\mathbf{3}4\rangle[\mathbf{3}4][4\mathbf{J}][4\mathbf{J}]
		 & =
		\langle\mathbf{3}4\rangle[\mathbf{1}4][\mathbf{2}4][\mathbf{3}4],
		\\
		\langle\mathbf{1}\mathbf{J}\rangle\langle\mathbf{2}\mathbf{J}\rangle
		\odot
		\langle\mathbf{3}\mathbf{J}\rangle\langle\mathbf{3}\mathbf{J}\rangle
		[\mathbf{3}4][\mathbf{3}4]
		 & =
		\langle\mathbf{1}\mathbf{3}\rangle\langle\mathbf{2}\mathbf{3}\rangle
		[\mathbf{3}4]^2,
		\\
		[\mathbf{2}\mathbf{J}][\mathbf{1}\mathbf{J}]
		\langle\mathbf{Q}\mathbf{J}\rangle[\mathbf{Q}\mathbf{J}]
		\odot
		\langle\mathbf{3}\mathbf{J}\rangle[4\mathbf{J}][4\mathbf{J}]
		[\mathbf{3}\mathbf{J}]
		 & =
		\langle\mathbf{3}\mathbf{Q}\rangle[\mathbf{1}4][\mathbf{2}4]
		[\mathbf{3}\mathbf{Q}].
	\end{aligned}
\end{equation}
The corresponding basis elements are
\begin{center}
	\small
	\setlength{\tabcolsep}{3pt}
	\renewcommand{\arraystretch}{1.35}
	\begin{tabularx}{\textwidth}{@{}c c >{\raggedright\arraybackslash}p{0.2\textwidth} >{\raggedright\arraybackslash}p{0.31\textwidth} >{\raggedright\arraybackslash}X@{}}
		\toprule
		index                                                                                   & \(J_{12}\) & \(A_L\) & \(A_R\) & \(A_{\rm sew}\) \\
		\midrule
		\(L^{(4)}_1\)                                                                           & \(1\)      &
		\([\mathbf{2}\mathbf{J}][\mathbf{1}\mathbf{J}]\)                                        &
		\(\langle\mathbf{3}4\rangle[\mathbf{3}4][4\mathbf{J}][4\mathbf{J}]\)                    &
		\(\langle\mathbf{3}4\rangle[\mathbf{1}4][\mathbf{2}4][\mathbf{3}4]\)                                                                       \\
		\(L^{(4)}_2\)                                                                           & \(1\)      &
		\(\langle\mathbf{1}\mathbf{J}\rangle\langle\mathbf{2}\mathbf{J}\rangle\)                &
		\(\langle\mathbf{3}\mathbf{J}\rangle\langle\mathbf{3}\mathbf{J}\rangle[\mathbf{3}4]^2\) &
		\(\langle\mathbf{1}\mathbf{3}\rangle\langle\mathbf{2}\mathbf{3}\rangle[\mathbf{3}4]^2\)                                                    \\
		\(L^{(4)}_3\)                                                                           & \(2\)      &
		\([\mathbf{2}\mathbf{J}][\mathbf{1}\mathbf{J}]
		\langle\mathbf{Q}\mathbf{J}\rangle[\mathbf{Q}\mathbf{J}]\)                              &
		\(\langle\mathbf{3}\mathbf{J}\rangle[4\mathbf{J}][4\mathbf{J}]
		[\mathbf{3}\mathbf{J}]\)                                                                &
		\(\langle\mathbf{3}\mathbf{Q}\rangle[\mathbf{1}4][\mathbf{2}4]
		[\mathbf{3}\mathbf{Q}]\)                                                                                                                   \\
		\bottomrule
	\end{tabularx}
\end{center}

After sewing, the resulting overcomplete candidates are reduced in the common on-shell quotient, and only linearly independent representatives are retained.

The chiral order is assigned only after the sewing step.
For the corrected \(d_{\rm amp}=4\) structures, the relevant relative grading is
\begin{equation}
	d_{\rm rel}=d_{\rm amp}-J_{12}-n_x,
\end{equation}
with \(n_x=0\) for the displayed \(d_{\rm amp}=4\) rows.
In the \(\bar B B u f_+\) sector, \(\Delta_{\rm field}=1\) because the field strength has dimension two, so \(d_\chi=d_{\rm rel}+1\).
Thus \(L^{(3)}_1\), \(L^{(3)}_2\), and the raised \(J_{12}=2\) structure \(L^{(4)}_3\) are assigned to \(\mathcal O(p^3)\), while the two \(J_{12}=1\) structures \(L^{(4)}_1\) and \(L^{(4)}_2\) are assigned to \(\mathcal O(p^4)\).
The \(Q=P_1-P_2\) factors in the open-current structures are part of the hard current degree in Eq.~\eqref{eq:hard_degree_rule}, not additional soft powers.
For concision, the complete construction is displayed here through \(\mathcal O(p^3)\).
At \(\mathcal O(p^4)\), the table shows representative structures, while the complete result, including contributions descending from higher amplitude dimensions, is collected in App.~\ref{app:format-operator-bases}.

The resulting chiral assignments and operator representatives are summarized in Table~\ref{tab:BBuf-amplitude-operator}.
In this correspondence, \(\overleftrightarrow D_\mu\) is the baryon relative derivative representing \(Q_\mu=P_{1\mu}-P_{2\mu}\).
Its hard momentum weight is already included in \(J_{12}\), which is why the \(L^{(4)}_3\) representative remains at \(\mathcal O(p^3)\).

\begin{table}[htbp]
	\centering
	\renewcommand{\arraystretch}{1.25}
	\begin{tabularx}{\textwidth}{@{}c c X@{}}
		\toprule
		chiral/operator order & amplitudes    & amplitude--operator correspondence      \\
		\midrule
		\(\mathcal O(p^3)\)   & \(L^{(3)}_1\) &
		\(\begin{aligned}
			   & \bigl(\bar B^{L}_{a_1}\bar\sigma^{\nu\rho}\bar\sigma^\mu B^{L}_{a_2}\bigr)
			  u_{\mu a_3}f_{+,\nu\rho a_4}
		  \end{aligned}\) \\
		\(\mathcal O(p^3)\)   & \(L^{(3)}_2\) &
		\(\begin{aligned}
			   & \bigl(\bar B^{R}_{a_1}\sigma^\mu\bar\sigma^{\nu\rho}B^{R}_{a_2}\bigr)
			  u_{\mu a_3}f_{+,\nu\rho a_4}
		  \end{aligned}\)      \\
		\(\mathcal O(p^3)\)   & \(L^{(4)}_3\) &
		\(\begin{aligned}
			   & \bigl(\bar B^{L}_{a_1}\bar\sigma^{\nu\rho}
			  \overleftrightarrow D_{\mu}
			  B^{R}_{a_2}\bigr) \tr_\sigma\!
			  \left(\sigma^\sigma\bar\sigma^\mu\right)
			  u_{\sigma a_3}f_{+,\nu\rho a_4}
		  \end{aligned}\)                                 \\
		\(\mathcal O(p^4)\)   & \(L^{(4)}_1\) &
		\(\begin{aligned}
			   & \bigl(D_\mu f_{+,\nu\rho a_4}\bigr)
			  \bigl(\bar B^{L}_{a_1}\bar\sigma^\mu\sigma^\sigma
			  \bar\sigma^{\nu\rho}B^{R}_{a_2}\bigr)
			  u_{\sigma a_3}
		  \end{aligned}\)                              \\
		\(\mathcal O(p^4)\)   & \(L^{(4)}_2\) &
		\(\begin{aligned}
			   & \bigl(D_\mu u_{\sigma a_3}\bigr)
			  \bigl(\bar B^{R}_{a_1}\sigma^\mu\bar\sigma^{\nu\rho}
			  \bar\sigma^\sigma B^{L}_{a_2}\bigr)
			  f_{+,\nu\rho a_4}
		  \end{aligned}\)                           \\
		\(\mathcal O(p^4)\)   & \(\cdots\)    &
		Additional independent structures are collected in
		App.~\ref{app:format-operator-bases}.                                           \\
		\bottomrule
	\end{tabularx}
	\caption{Chiral-order assignment and amplitude--operator correspondence for the displayed \(\bar B B u f_+\) structures through \(\mathcal O(p^4)\).}
	\label{tab:BBuf-amplitude-operator}
\end{table}
\FloatBarrier

Thus the \(\mathcal O(p^3)\) basis contains three Lorentz structures in this sector, with the third supplied by the raised \(J_{12}=2\) sewn structure, matching the corresponding Dirac/heavy-baryon representative count in Refs.~\cite{Song:2024fae,Fettes:2000gb}.
Further sector-by-sector comparisons with existing representatives are collected in App.~\ref{app:comparison_chi_fchi}.
\section{Phenomenological Implications and Prospects}
\label{sec:pheno}

The on-shell static--recoil construction developed in this work yields a complete, linearly independent, and chiral-order-resolved basis for the local contact sector of baryonic chiral EFT.
Starting from a Lorentz-covariant relativistic on-shell amplitude basis, the static--recoil map separates the universal heavy-mass dependence from recoil structures and yields symmetry-adapted short-distance blocks that enter low-energy baryon amplitudes alongside pole and loop contributions.

A direct application is \(SU(3)\)-flavor breaking in octet--decuplet and pure decuplet observables.
The standard spurions \(\chi_\pm\) encode explicit chiral-symmetry breaking in the local terms contributing to baryon mass splittings, sigma terms, magnetic moments, and symmetry-suppressed transition amplitudes.
The minimal relativistic next-to-leading-order octet--decuplet Lagrangian contains structures for mass splittings, anomalous magnetic moments, decuplet--octet magnetic transitions, and an axial transition moment \cite{Holmberg:2018dtv}.
Static--recoil factorization organizes these structures into a basis with manifest heavy-baryon chiral order, facilitating combined analyses of local terms and chiral-loop contributions in decuplet mass and sigma-term studies \cite{Ren:2013oaa,Lutz:2023pba,Lutz:2024hbf}, as well as in radiative and weak hyperon transitions.
The decay \(\Xi(1530)^-\to\Xi^-\gamma\), which is suppressed in the exact U-spin limit, provides a sensitive probe of the corresponding \(SU(3)\)-breaking contact terms \cite{BESIII:2019jot}.
The \(\bar T T\chi_+ f_+\), \(\bar T T\chi_+\chi_+\), and \(\bar B B\chi_+ u f_+\) sectors in App.~\ref{app:format-operator-bases} provide explicit examples of flavor-breaking electromagnetic decuplet couplings, higher-order mass- and sigma-term structures, and source-dependent meson--baryon amplitudes, respectively.

A second application concerns the local contact sector accompanying explicit decuplet propagation in the \(\Delta(1232)\) region.
The small \(N\)--\(\Delta\) mass gap and the strong \(\pi N\Delta\) coupling make the \(\Delta\) an important explicit degree of freedom in low-energy \(\pi N\) scattering, pion photo- and electroproduction, and Compton scattering \cite{Hemmert:1996xg,Yao:2016vbz,Long:2009wq,OuYang:2024ykb,Pascalutsa:2006up}.
Static--recoil factorization provides a complete and linearly independent basis for the corresponding local contact sector, with manifest heavy-baryon power counting and the appropriate physical spin-\(\frac32\) constraints, while avoiding off-shell spin-\(\frac12\) artifacts.
The local operators can be matched onto the contact contributions to the magnetic-dipole (M1), electric-quadrupole (E2), and Coulomb-quadrupole (C2) components of electromagnetic \(N\to\Delta\) transition amplitudes \cite{Pascalutsa:2005ts,Pascalutsa:2006up}, as well as to local contributions to axial \(N\Delta\) responses in weak pion production and neutrino-induced reactions \cite{Geng:2008mf}.
The same construction applies to mixed octet--decuplet sectors and provides the local transition structures relevant to these amplitudes.

The same organization provides the analytic local sector used in matching chiral EFT to lattice-QCD results.
Chiral extrapolations of baryon masses, sigma terms, electromagnetic form factors, and decuplet--octet transition form factors combine loop-generated nonanalytic quark-mass dependence with analytic contributions from local operators \cite{Alexandrou:2003ea,Arndt:2003ww,Li:2017cfz,Lutz:2023pba,Lutz:2024hbf}.
The static--recoil basis organizes the local operators, and hence the associated low-energy constants, according to chiral order.
It therefore provides a convenient framework for combined analyses of spin-\(\frac12\) and spin-\(\frac32\) baryons in the presence of flavor breaking and external sources, including analyses in which finite-volume corrections are incorporated.

Beyond baryon \(\chi\)PT, the same static--recoil logic offers a natural organizing principle for theories in which hard particle masses coexist with soft residual kinematics.
In nuclear EFT and potential-based descriptions of nuclear forces, it can be used to organize symmetry-adapted short-range contact kernels that are combined with long-range pion exchange in few-body amplitudes \cite{Machleidt:1987hj,Gross:1991pm,Machleidt:2000ge}.
The Lorentz-covariant precursor basis may likewise provide a useful starting point for classifying local interpolating currents in exotic-hadron QCD sum rules, where canonical mass dimension organizes the currents and their operator-product expansion \cite{Chen:2016qju,Liu:2019zoy,Chen:2017rhl}.
These examples illustrate the broader applicability of the amplitude-based construction across chiral and nuclear effective theories and local-current classifications.

Taken together, these applications demonstrate how the static--recoil construction provides complete and linearly independent local contact bases for use in process-specific analyses.
Process-dependent pole and loop contributions, finite-volume effects, and other dynamical ingredients can then be combined with these bases while preserving a transparent separation between local operator structures and process-dependent dynamics, without reintroducing redundancies in the local contact sector.
\section{Conclusion}
\label{sec:conclusion}

We establish an on-shell static--recoil factorization for local operator bases in baryonic chiral perturbation theory.
The construction implements the separation between the static mass scale \(\mathcal O(M)\) and residual dynamics \(\mathcal O(p)\) directly in amplitude space.
Built entirely from physical on-shell states, the method yields complete and independent bases for higher-spin sectors without field-space inversion or a separate removal of unphysical lower-spin modes.

The central result is a new on-shell basis construction built from eigenstates of total angular momentum in the heavy-pair channel.
For each \(J_{12}\) eigenstate, the three-point heavy-pair current carries the hard mass degree, whereas the residual block contains the recoil and light-sector dependence.
After the hard factors are absorbed into low-energy constants, each sewn structure is reclassified from canonical dimension to its heavy-baryon chiral order.
The result is a complete, non-redundant chiral basis obtained without off-shell Lagrangian reduction.

The method applies uniformly at any heavy spin because it is formulated in physical massive little-group representations.
It reproduces the standard heavy-baryon organization for spin-\(\frac12\) and generates spin-\(\frac32\) and higher-spin bases without projected Rarita--Schwinger kinetic inversions.
Little-group covariance is preserved throughout, and the use of physical on-shell states eliminates the need for any separate lower-spin projection.

The construction first fixes the kinematic basis and its chiral grading.
Flavor tensors are then reduced independently by Cayley--Hamilton identities and the octet and decuplet representation constraints.
The direct product of the kinematic and flavor bases is finally projected onto the required identical-particle symmetry.
This yields complete, non-redundant flavor--Lorentz operator bases for octet and decuplet baryons, Goldstone modes, resonances, and external sources.

The Mathematica implementation carries out the complete construction.
Starting from the relativistic on-shell contact space and the flavor constraints, it performs the recoil-ordered reduction and outputs the explicit flavor--Lorentz operator blocks in App.~\ref{app:format-operator-bases}.
These bases establish the universal local layer on which pole, loop, finite-volume, and potential dynamics can be assembled with manifest chiral order.
They enable direct analyses of decuplet-resonance processes, \(SU(3)\)-breaking observables, lattice-QCD matching, and low-energy experimental amplitudes.
More broadly, because the same hard-mass separation applies whenever heavy particles carry soft residual momenta, the construction extends the on-shell basis program to a broad class of nonrelativistic effective field theories.

\begin{acknowledgments}
Z.D. is supported by the European Research Council, grant agreement n.
101039756.
T.M. is partly supported by Chinese Academy of Sciences Pioneer Initiative "Talent Introduction Plan" (grant No.\,E4ER6601A2), the Fundamental Research Funds for the Central Universities (grant No.\,E4EQ6602X2), and the National Natural Science Foundation of China (grant No.\,E514660101).
\end{acknowledgments}
\addtocontents{toc}{\protect\setcounter{tocdepth}{1}}
\appendix
\section{\texorpdfstring{The Projected Spin-\(3/2\) Field-Space Complexity}
{The Projected Spin-3/2 Field-Space Complexity}}
\label{app:inverse_C}

As discussed in Sec.~\ref{sec:HBP_highspin}, the field-space heavy-baryon projection requires the inverse of the heavy block \(\mathcal C_{\mu\nu}\).
For spin-\(1/2\), this block is essentially a scalar-like operator in spinor space.
For a Rarita--Schwinger field, however, \(\mathcal C_{\mu\nu}\) carries both Lorentz and Dirac indices and must be inverted only on the projected heavy tensor-spinor subspace.
This appendix summarizes the corresponding finite-dimensional operator algebra and explains why the direct symbolic inversion is not scalable beyond low orders.

For the parity-even spin-\(3/2\) operator space, we use the tensor product of four Dirac structures,
\begin{equation}
	\label{eq:RS-Dirac-basis}
	\Gamma^{(1)} = \mathbf{1},
	\qquad
	\Gamma^{(2)} = \slashed v,
	\qquad
	\Gamma^{(3)} = \frac{\slashed \partial}{M},
	\qquad
	\Gamma^{(4)} = \frac{\slashed \partial\,\slashed v}{M},
\end{equation}
with ten two-index Lorentz structures,
\begin{equation}
	\label{eq:RS-Lorentz-basis}
	\begin{aligned}
		T^{(1)}_{\mu\nu} & = g_{\mu\nu}, & T^{(2)}_{\mu\nu} & = \gamma_\mu \gamma_\nu, & T^{(3)}_{\mu\nu} & = \gamma_\mu v_\nu, & T^{(4)}_{\mu\nu} & = v_\mu \gamma_\nu, & T^{(5)}_{\mu\nu} & = v_\mu v_\nu, \\ T^{(6)}_{\mu\nu} &= \frac{\gamma_\mu \partial_\nu}{M}, & T^{(7)}_{\mu\nu} &= \frac{\partial_\mu \gamma_\nu}{M}, & T^{(8)}_{\mu\nu} &= \frac{v_\mu \partial_\nu}{M}, & T^{(9)}_{\mu\nu} &= \frac{\partial_\mu v_\nu}{M}, & T^{(10)}_{\mu\nu} &= \frac{\partial_\mu \partial_\nu}{M^2}.
	\end{aligned}
\end{equation}
Thus
\begin{equation}
	\label{eq:RS-basis-expansion}
	\Big\{\mathcal T_{\mu\nu}^{n}\Big\}_{n=1}^{40}
	\equiv
	\Big\{\Gamma^{(a)}\Big\}_{a=1}^{4}
	\times
	\Big\{T_{\mu\nu}^{(b)}\Big\}_{b=1}^{10}
\end{equation}
forms a \(40\)-dimensional parity-even tensor basis.
If parity-odd structures are included, this starting space doubles to dimension \(80\).

The counting also shows why the same field-space strategy becomes rapidly more complicated for higher spin.
A spin-\(J=m+1/2\) tensor carries \(m\) Lorentz indices on each side of the quadratic term.
At the ordered-index level, each uncontracted Lorentz index can be filled by one of \(\gamma_\mu\), \(v_\mu\), or \(\partial_\mu/M\), while pairs of indices can be contracted by metrics.
The number of parity-even Lorentz structures is therefore
\begin{equation}
	\label{eq:higher-spin-lorentz-count}
	N_L(m) = \sum_{\ell=0}^{m} \frac{(2m)!
	}{2^\ell \ell! (2m-2\ell)!}\,
	3^{\,2m-2\ell},
\end{equation}
where \(\ell\) counts the number of metric contractions.
Multiplying by the four Dirac structures in Eq.~\eqref{eq:RS-Dirac-basis} gives
\begin{equation}
	\label{eq:higher-spin-tensor-count}
	N_{\rm tens}(m)=4N_L(m).
\end{equation}
For spin-\(3/2\), \(m=1\), so \(N_L(1)=3^2+1=10\) and \(N_{\rm tens}(1)=40\).
For spin-\(5/2\), \(m=2\), and
\begin{equation}
	\label{eq:spin52-count}
	N_L(2)=3^4+\binom42 3^2+3=138, \qquad N_{\rm tens}(2)=4\times 138=552 .
\end{equation}
This \(552\) is the ordered parity-even starting space before imposing the spin-\(5/2\) heavy projector.
The projected dimension is a separate question and requires the explicit higher-spin projector.

We now return to spin-\(3/2\).
The full \(40\)-dimensional basis is useful for organizing \(\Lambda_{\mu\nu}\), but \(\mathcal C_{\mu\nu}\) is not invertible on this full space.
With the convention used in Sec.~\ref{sec:HBP_highspin},
\begin{equation}
	\label{eq:C-heavy-projected}
	\mathcal C_{\mu\nu}
	=
	- P^{\rm H}_{\mu\rho}\Lambda^{\rho\sigma}
	P^{\rm H}_{\sigma\nu}.
\end{equation}
Therefore \(\mathcal C\) acts only inside the double-sided heavy subspace \(P^{\rm H}\cdots P^{\rm H}\) and annihilates components outside it.
The inverse must be defined on this projected subspace, not on the complete tensor space.

For the parity-even spin-\(3/2\) basis above, the double-sided projection reduces the relevant dimension from \(40\) to
\begin{equation}
	\label{eq:projected-dimension}
	d_H=27 .
\end{equation}
Let \(\{E^\alpha_{\mu\nu}\}_{\alpha=1}^{d_H}\) be a basis of this projected subspace,
\begin{equation}
	P^{\rm H} E^\alpha P^{\rm H}=E^\alpha .
\end{equation}
The projected product closes,
\begin{equation}
	\label{eq:projected-algebra}
	(E^\alpha)_{\mu\rho}(E^\beta)^{\rho}{}_{\nu}
	=
	\tilde f^{\alpha\beta}{}_{\gamma}\,
	(E^\gamma)_{\mu\nu},
\end{equation}
with numerical structure constants \(\tilde f^{\alpha\beta}{}_{\gamma}\).
Writing
\begin{equation}
	\label{eq:C0-projected-expansion}
	\mathcal C^{(0)} = c^{(0)}_\alpha E^\alpha,
	\qquad
	P^{\rm H}=u_\alpha E^\alpha,
	\qquad
	(\mathcal C^{(0)})^{-1}=x^{(0)}_\alpha E^\alpha,
\end{equation}
the leading inverse is fixed by the projected identity condition
\begin{equation}
	\mathcal C^{(0)}(\mathcal C^{(0)})^{-1}=P^{\rm H}.
\end{equation}
Equivalently,
\begin{equation}
	\label{eq:subspace_linear_system}
	c^{(0)}_\alpha x^{(0)}_\beta
	\tilde f^{\alpha\beta}{}_{\gamma}
	=
	u_\gamma .
\end{equation}
Solving this \(27\)-dimensional linear problem gives the leading inverse.
This is the finite-dimensional version of the usual heavy-\(\Delta\) leading \(O(1/M)\) inverse.

The higher-order inverse is obtained by expanding the heavy block as
\begin{equation}
	\label{eq:C-order-expansion}
	\mathcal C
	=
	\mathcal C^{(0)}
	+
	\mathcal C^{(1)}
	+
	\mathcal C^{(2)}
	+\cdots ,
	\qquad
	\mathcal C^{(n)}\sim M^{1-n},
\end{equation}
so that
\begin{equation}
	\label{eq:C-inverse-order-expansion}
	\mathcal C^{-1}
	=
	X^{(0)}+X^{(1)}+X^{(2)}+\cdots ,
	\qquad
	X^{(n)}\sim M^{-n-1},
\end{equation}
with \(X^{(0)}=(\mathcal C^{(0)})^{-1}\).
The condition \(\mathcal C\,\mathcal C^{-1}=P^{\rm H}\) gives the recursion
\begin{equation}
	\label{eq:C-inverse-recursion}
	X^{(n)} = - X^{(0)} \sum_{k=1}^{n} \mathcal C^{(k)}X^{(n-k)} \qquad (n\geq 1).
\end{equation}
In particular,
\begin{equation}
	\begin{aligned}
		X^{(1)} & = - X^{(0)}\mathcal C^{(1)}X^{(0)}, \\ X^{(2)} &= X^{(0)}\mathcal C^{(1)}X^{(0)} \mathcal C^{(1)}X^{(0)} - X^{(0)}\mathcal C^{(2)}X^{(0)} .
	\end{aligned}
\end{equation}
Higher orders are obtained by the same recursion.
Thus an expansion through \(O(1/M^{\chi+1})\) requires all \(X^{(n)}\) with \(0\leq n\leq \chi\).

If the coefficients of the projected basis were numerical, this would be only a finite linear-algebra problem.
A dense multiplication in the \(d_H\)-dimensional projected algebra costs schematically \(O(d_H^3)\), and the recursive construction through order \(\chi\) gives a workload of order
\begin{equation}
	\label{eq:numerical-cost}
	O(d_H^3\chi^2) = O(27^3\chi^2).
\end{equation}
This numerical problem is not the main obstruction.

The EFT problem is harder because the entries of \(\Lambda\), and hence of \(\mathcal C^{(n)}\), are symbolic operators with undetermined Wilson coefficients.
To obtain a heavy-baryon Lagrangian through chiral order \(\chi\), one must include all Lorentz-covariant operators whose heavy-baryon expansion can contribute up to that order.
This includes not only operators appearing directly at order \(\chi\), but also lower-order covariant operators whose \(1/M\) recoil expansion feeds higher chiral orders.
Let the number of such covariant operators be \(N_\chi\).
Schematically,
\begin{equation}
	\label{eq:deltaC-symbolic}
	\delta\mathcal C_\chi
	=
	\sum_{a=1}^{N_\chi} c_a\,\Delta_a ,
\end{equation}
where the \(\Delta_a\) are noncommuting operator insertions and the \(c_a\) are Wilson coefficients.

An \(m\)-insertion contribution to \(\mathcal C^{-1}\) then contains ordered words of the form
\begin{equation}
	\label{eq:ordered-words}
	X^{(0)} \Delta_{a_1} X^{(0)} \Delta_{a_2} \cdots \Delta_{a_m} X^{(0)}, \qquad a_i=1,\ldots,N_\chi .
\end{equation}
Before using EoM, IBP, and spin-projector identities, the number of such words scales as \(N_\chi^m\).
Therefore the symbolic cost through order \(\chi\) is schematically
\begin{equation}
	\label{eq:symbolic-cost}
	O(27^3\chi^2 (N_\chi)^\chi),
\end{equation}
up to reductions and sparsity.
The essential point is not the precise numerical prefactor, but the symbolic proliferation of ordered noncommuting operator insertions.

After substituting the result into
\begin{equation}
	\label{eq:appendix-effective-block}
	\mathcal A
	+
	\widetilde{\mathcal B}\,
	\mathcal C^{-1}
	\mathcal B ,
\end{equation}
one obtains long polynomial combinations of Wilson coefficients.
These expressions are not yet a minimal heavy-baryon chiral basis; they still have to be reduced by the heavy-baryon EoM, IBP, spin-projector identities, and related algebraic identities.
Thus the spin-\(3/2\) inverse is well defined and useful at leading order, but the direct field-space block-elimination method is not a scalable symbolic route to a general higher-order effective Lagrangian.
For higher spin the starting operator space is already much larger, as illustrated by the \(552\) ordered spin-\(5/2\) structures above.
This is the field-space obstruction avoided by the on-shell construction used in the main text.

\section{Flavor Projectors and Alternating-Projection Reduction}
\label{app:flavor_projectors}

This appendix gives the linear-algebraic details used in the flavor and identical-particle reduction of Sec.~\ref{sec:flavor_symmetry}.
We work with a finite-dimensional space of trace monomials built from octet matrices.
The reduction has three ingredients: trace identities, decuplet or antidecuplet embedding constraints, and identical-particle exchange symmetry.
Each ingredient is implemented as a projector, and the physical basis is obtained from their common subspace.

\subsection*{Cayley--Hamilton Trace Identities}
The trace identities used in the flavor reduction are generated from the CH theorem.
For an \(N\times N\) matrix, the CH polynomial has the form
\begin{equation}
	G(X)\equiv X^N+c_1X^{N-1}+\cdots+c_N\mathds{1}=0 .
	\label{eq:CH_general_app}
\end{equation}
Multiplying this matrix identity by an auxiliary word \(Y\), taking the trace, and then polarizing the result gives multilinear trace identities,
\begin{equation}
	G(X)Y \longrightarrow \Tr(G(X)Y) \longrightarrow F(X_1,\ldots,X_N,Y)=0 .
	\label{eq:polarization_scheme_app}
\end{equation}
Here polarization means replacing the repeated copies of \(X\) by independent matrices and extracting the multilinear part.
Repeating the same construction with longer auxiliary words generates trace identities at higher trace length.

For the present \(SU(3)\) application, the octet matrices are traceless \(3\times3\) matrices.
In the traceless \(N=3\) case, the first nontrivial multilinear identity is the length-four relation
\begin{align}
	 & \Tr(ABCD)+\Tr(ACBD)+\Tr(ABDC)
	+\Tr(ACDB)+\Tr(ADBC)+\Tr(ADCB)
	\nonumber                        \\
	 & \hspace{2em}
	-\Tr(AB)\Tr(CD)
	-\Tr(AC)\Tr(BD)
	-\Tr(AD)\Tr(BC)
	=0 ,
	\label{eq:FTI_SU3_app}
\end{align}
for traceless matrices \(A,B,C,D\).
This is the traceless specialization of the standard fundamental trace identity: terms involving one-cycles vanish because all matrices are traceless.
Higher-length trace relations are obtained by the same CH multiplication and polarization procedure.

In implementation, the naively enumerated trace monomials span a vector space \(\mathcal V_{\rm tr}\).
The CH-generated trace identities are collected as linear constraints,
\begin{equation}
	C_{\rm CH}\,v=0 .
	\label{eq:CCH_app}
\end{equation}
Thus the CH-reduced trace space is \(\ker C_{\rm CH}\).
After choosing an inner product on \(\mathcal V_{\rm tr}\), let the columns of \(Q_{\rm CH}\) form an orthonormal basis of this kernel.
The corresponding orthogonal projector is
\begin{equation}
	\mathcal P_{\rm CH}=Q_{\rm CH}
	Q_{\rm CH}^{\dagger}.
	\label{eq:PCH_construct_app}
\end{equation}

\subsection*{Decuplet Embedding}

The decuplet and antidecuplet projectors in Sec.~\ref{sec:flavor_symmetry} act on a designated auxiliary octet pair.
To convert this adjoint-index projector into a matrix acting on trace monomials, we choose two auxiliary octet matrices \(A=A^aT^a\) and \(B=B^bT^b\), apply the projector to their adjoint indices, and rewrite the result back into the chosen trace basis.

The induced action depends only on whether the two auxiliary matrices appear in the same trace or in different traces.
Let \(X\) and \(Y\) denote possibly empty words made from the remaining octet matrices.
For a same-trace structure,
\begin{align}
	\mathcal P^{(A,B)}_{10}\Tr(A X B Y)
	={} &
	\frac14\Big[
		\Tr(A X B Y)-\Tr(B X A Y)
		\Big]
	\nonumber                            \\
	    & +\frac{1}{12}\Tr(X)\Tr([A,B]Y)
	+\frac{1}{12}\Tr(Y)\Tr([B,A]X)
	\nonumber                            \\
	    & -\frac{1}{12}\Tr([B,A][X,Y])
	\nonumber                            \\
	    & +\frac14\Big[
		\Tr(BY)\Tr(AX)-\Tr(AY)\Tr(BX)
		\Big].
	\label{eq:P10_singletrace_app}
\end{align}
For a different-trace structure,
\begin{align}
	\mathcal P^{(A,B)}_{10}\!\left[\Tr(AX)\Tr(BY)\right]
	={} &
	\frac{5}{36}\Big[
		\Tr(AX)\Tr(BY)-\Tr(BX)\Tr(AY)
		\Big]
	\nonumber           \\
	    & +\frac14\Big[
		\Tr(A X B Y)-\Tr(A Y X B)
		\Big]
	\nonumber           \\
	    & -\frac13\Big[
		\Tr([B,A]X)\Tr(Y)+\Tr([B,A]Y)\Tr(X)
		\Big].
	\label{eq:P10_doubletrace_app}
\end{align}
The antidecuplet case is obtained by replacing \(\mathcal P_{10}^{(A,B)}\) with \(\mathcal P_{\overline{10}}^{(A,B)}\) in the same construction.

Equations~\eqref{eq:P10_singletrace_app} and \eqref{eq:P10_doubletrace_app} give the matrix representation of the representation projector on the trace basis.
They also show that the projector can mix single- and double-trace topologies.
This is why the representation projection need not commute with the CH trace reduction,
\begin{equation}
	[\mathcal P_{\rm CH},\mathcal P_{10/\overline{10}}]\neq 0 .
	\label{eq:PCH_Prep_noncomm_app}
\end{equation}
For sectors with several decuplet or antidecuplet legs, the corresponding embedding projectors are applied to all designated auxiliary pairs; their product is denoted by \(\mathcal P_{\rm rep}\).

\subsection*{Identical-Particle Projectors}

Identical-particle symmetry is imposed after Lorentz and flavor structures have been combined.
For a set of labeled identical fields \(X_1,\ldots,X_n\), each permutation \(\sigma\in S_n\) acts on the operator basis by permuting the labels and rewriting the result in the chosen basis.
We denote the resulting matrix by \(R(\sigma)\).

For identical bosons and fermions, the projectors are
\begin{equation}
	\mathcal P_{\rm B}
	=
	\frac{1}{|S_n|}
	\sum_{\sigma\in S_n}
	R(\sigma), \qquad \mathcal P_{\rm F} = \frac{1}{|S_n|} \sum_{\sigma\in S_n} \operatorname{sgn}(\sigma)R(\sigma).
	\label{eq:PB_PF_app}
\end{equation}
Equivalently, one may construct the same projectors from the generators \(s=(12)\) and \(r=(12\cdots n)\), as in the main text.
In practice, we compute the image of the appropriate projector, keep a linearly independent basis, and only then remove the auxiliary labels \(X_1,\ldots,X_n\to X\).
The resulting operators have the required exchange symmetry.

\subsection*{Common Subspace}

The physical operator space is the common subspace satisfying all imposed constraints.
After choosing an inner product, each constraint is represented by an orthogonal projector.
If two such projectors commute, their product is the projector onto the intersection.
In the present problem the CH reduction, representation embedding, and identical-particle projection are independent constraints and need not commute, so a single product of projectors should not be treated as the final projector.

We therefore use the standard alternating-projection construction.
After lifting the flavor projectors to the full Lorentz--flavor operator space, define the cyclic product
\begin{equation}
	\mathcal T
	=
	\mathcal P_{\rm CH}\mathcal P_{\rm rep}\mathcal P_{\rm B/F},
	\label{eq:T_all_app}
\end{equation}
where \(\mathcal P_{\rm B/F}\) denotes the bosonic or fermionic identical-particle projector appropriate to the sector under consideration.
The projector onto the common subspace is obtained as the repeated-product limit,
\begin{equation}
	\mathcal P_{\cap}
	=
	\lim_{n\to\infty}\mathcal T^n,
	\qquad
	\operatorname{Im}(\mathcal P_{\cap})
	=
	\operatorname{Im}(\mathcal P_{\rm CH})
	\cap
	\operatorname{Im}(\mathcal P_{\rm rep})
	\cap
	\operatorname{Im}(\mathcal P_{\rm B/F}) .
	\label{eq:Pcap_app}
\end{equation}
In finite-dimensional implementation, we do not need to iterate this limit explicitly.
The same common subspace is extracted as the fixed subspace of \(\mathcal T\),
\begin{equation}
	\mathcal V_{\rm phys}
	=
	\operatorname{Fix}(\mathcal T)
	=
	\ker(\mathds{1}-\mathcal T).
	\label{eq:Vphys_fixed_app}
\end{equation}
A basis of this kernel gives the independent operator structures after flavor reduction and identical-particle symmetrization.
\newcommand{\Ohh}{\mathcal O_{\scriptscriptstyle[\frac12,\frac12]}}
\newcommand{\Oth}{\mathcal O_{\scriptscriptstyle[\frac32,\frac12]}}
\newcommand{\Ott}{\mathcal O_{\scriptscriptstyle[\frac32,\frac32]}}
\providecommand{\TrF}{\operatorname{Tr}_F}
\providecommand{\trsigma}{\operatorname{tr}_\sigma}

\section{Conventions and Heavy-Pair Amplitude--Operator Correspondence}
\label{app:conventions}

This appendix collects the conventions used to relate the on-shell spinor-helicity basis to local operator structures.
The standard spacetime and spinor conventions follow those of our previous work.
The only new ingredient needed in the present paper is the amplitude--operator correspondence for the heavy-pair factors appearing in the static--recoil factorization discussed in Sec.~\ref{sec:onshell-heavy-dealing}.

\subsection{Spacetime, Spinor, and Flavor Conventions}

We use
\begin{equation}
	g_{\mu\nu}=\operatorname{diag}(+1,-1,-1,-1),
	\qquad
	\epsilon^{0123}=-\epsilon_{0123}=+1,
	\qquad
	\epsilon^{12}=\epsilon_{21}=+1.
\end{equation}
For the two-component sigma matrices, we adopt
\begin{equation}
	\sigma^\mu_{\alpha\dot\alpha}=(1,\vec{\sigma}),
	\qquad
	\bar{\sigma}^{\mu\,\dot\alpha\alpha}
	=
	\epsilon^{\alpha\beta}\epsilon^{\dot\alpha\dot\beta}
	\sigma^\mu_{\beta\dot\beta}
	=
	(1,-\vec{\sigma}),
\end{equation}
together with
\begin{equation}
	\sigma^\mu_{\alpha\dot\alpha}\bar{\sigma}^{\nu\,\dot\alpha\beta}
	+
	\sigma^\nu_{\alpha\dot\alpha}\bar{\sigma}^{\mu\,\dot\alpha\beta}
	=
	2g^{\mu\nu}\delta_\alpha{}^\beta,
	\qquad
	\trsigma\!\left(\sigma^\mu\bar{\sigma}^\nu\right)=2g^{\mu\nu}.
\end{equation}
The antisymmetric sigma matrices are defined by
\begin{equation}
	(\sigma^{\mu\nu})_\alpha{}^\beta
	=
	\frac{i}{2}
	\left(
	\sigma^\mu_{\alpha\dot\alpha}\bar{\sigma}^{\nu\,\dot\alpha\beta}
	-
	\sigma^\nu_{\alpha\dot\alpha}\bar{\sigma}^{\mu\,\dot\alpha\beta}
	\right),
\end{equation}
\begin{equation}
	(\bar{\sigma}^{\mu\nu})^{\dot\alpha}{}_{\dot\beta}
	=
	\frac{i}{2}
	\left(
	\bar{\sigma}^{\mu\,\dot\alpha\alpha}\sigma^\nu_{\alpha\dot\beta}
	-
	\bar{\sigma}^{\nu\,\dot\alpha\alpha}\sigma^\mu_{\alpha\dot\beta}
	\right).
\end{equation}

For a massive momentum \(P^\mu\), we use the standard massive spinor-helicity decomposition
\begin{equation}
	P_\mu \sigma^\mu_{\alpha\dot\alpha}
	=
	|i_I\rangle_\alpha [i^I|_{\dot\alpha},
		\qquad
		P_\mu \bar{\sigma}^{\mu\,\dot\alpha\alpha}
		=
		|i^I]^{\dot\alpha}\langle i_I|^\alpha,
\end{equation}
with symmetrized little-group indices for higher-spin states.
In particular,
\begin{equation}
	\psi_{L,\alpha}\;\leftrightarrow\;|i_I\rangle_\alpha,
	\qquad
	\psi_R^{\dot\alpha}\;\leftrightarrow\;|i^I]^{\dot\alpha},
\end{equation}
\begin{equation}
	F^+_{\mu\nu}\,\bar{\sigma}^{\mu\nu\,\dot\alpha\dot\beta}
	\;\leftrightarrow\;
	|i_{\{I_1}]^{\dot\alpha}|i_{I_2\}}]^{\dot\beta},
	\qquad
	F^-_{\mu\nu}\,\sigma^{\mu\nu}_{\alpha\beta}
	\;\leftrightarrow\;
	|i_{\{I_1}\rangle_\alpha |i_{I_2\}}\rangle_\beta,
\end{equation}
\begin{equation}
	mA_\mu \sigma^\mu_{\alpha\dot\alpha}
	\;\leftrightarrow\;
	|i_{\{I_1}\rangle_\alpha |i_{I_2\}}]_{\dot\alpha},
	\qquad
	mA_\mu \bar{\sigma}^{\mu\,\dot\alpha\alpha}
	\;\leftrightarrow\;
	|i_{\{I_1}]^{\dot\alpha}\langle i_{I_2\}}|^\alpha.
\end{equation}
For massless particles, the little-group indices are omitted.

We write Dirac spinors in the Weyl basis,
\begin{equation}
	\Psi=
	\begin{pmatrix}
		\psi_L \\[2pt]
		\psi_R
	\end{pmatrix},
	\qquad
	\bar{\Psi}
	=
	\bigl(
	\psi_R^\dagger,\,
	\psi_L^\dagger
	\bigr),
\end{equation}
with
\begin{equation}
	\gamma^\mu=
	\begin{pmatrix}
		0                                   & \sigma^\mu_{\alpha\dot\beta} \\
		\bar{\sigma}^{\mu\,\dot\alpha\beta} & 0
	\end{pmatrix},
	\qquad
	\gamma_5=
	\begin{pmatrix}
		-1 & 0 \\
		0  & 1
	\end{pmatrix},
	\qquad
	C=i\gamma^0\gamma^2=
	\begin{pmatrix}
		\epsilon_{\alpha\beta} & 0                              \\
		0                      & \epsilon^{\dot\alpha\dot\beta}
	\end{pmatrix}.
\end{equation}

For a massive spin-\(\tfrac32\) state, we use a vector-spinor field \(\Psi_\mu\).
Its totally symmetric rank-three massive little-group representation contains only the physical spin-\(\tfrac32\) external state.
Under the amplitude--operator correspondence it maps to a vector-spinor satisfying \(P^\mu\Psi_\mu=0\) and \(\gamma^\mu\Psi_\mu=0\), so transversality and gamma-tracelessness are built into the external-state representation.
The explicit little-group indices are suppressed below and only the Lorentz vector index is kept manifest.
The tensors \(\mathcal P_{xx}\), \(\mathcal P_{xy}\), and \(\mathcal P_{yy}\) encode the static--recoil operator translation on these physical states and introduce no lower-spin component.

Flavor traces are denoted by \(\TrF\).
This notation is used to distinguish flavor traces from traces over spinor or Lorentz matrices, such as \(\trsigma(\sigma^\mu\bar\sigma^\nu)\).
Thus, for flavor structures we write
\begin{equation}
	\TrF(\cdots),
\end{equation}
and, for example,
\begin{equation}
	\TrF\!\left(T^{a_1}T^{a_2}\cdots T^{a_n}\right),
\end{equation}
where \(T^a\) denotes the chosen fundamental \(SU(3)\) generator.
If \(T^a=\lambda^a/2\) is used, the corresponding normalization should be kept uniform throughout the flavor-basis construction.

\subsection{Flavor Group of External Sources}
Since the scalar and pseudoscalar spurions are general flavor matrices, their homogeneous transformation under the residual \(SU(3)_V\) symmetry contains both singlet and adjoint components.
We decompose
\[
\chi_\pm=\widehat{\chi}_\pm+\frac{1}{3}\langle\chi_\pm\rangle\,\mathbf{1}, \qquad \langle \widehat{\chi}_\pm\rangle=0 ,
\]
with \(\langle\cdots\rangle\equiv {\rm Tr}_F(\cdots)\).
The traceless component transforms as
\[
\widehat{\chi}_\pm\to h\,\widehat{\chi}_\pm\,h^\dagger ,
\]
and therefore belongs to the adjoint representation, while the trace component is invariant,
\[
\langle\chi_\pm\rangle\to \langle\chi_\pm\rangle .
\]
Equivalently,
\[
	\chi_\pm\in \mathbf{1}\oplus \mathbf{8}.
\]
Throughout the flavor construction, adjoint flavor labels carried by \(\chi_\pm\) denote the traceless component \(\widehat{\chi}_\pm\), whereas \(\langle\chi_\pm\rangle\) is included separately as an \(SU(3)_V\)-singlet insertion of chiral order \(O(p^2)\).

\subsection{Heavy-Pair Amplitude--Operator Correspondence}
\label{app:heavy_pair_translation}

We follow the notation of Secs.~\ref{sec:heavy_spinor_projection} and \ref{HBP-amp}.
The contact amplitude is organized by the \(12\)-channel angular momentum \(J_{12}\), displayed through the auxiliary current \(\mathcal J_{J_{12}}\).
This appendix specifies the amplitude--operator correspondence for the left three-point factor \(\mathcal V_{12\to\mathcal J_{J_{12}}}\).
In low angular-momentum channels this factor contains explicit \(x,y\) factors; the remaining current-attached spinors are matched to local fields by the standard correspondence in Table~\ref{tab:dm_buildingblocks}.

For the heavy baryon--antibaryon pair carried by legs \(1,2\), the elementary heavy-pair scalar factors are
\begin{equation}
	x \equiv [\mathbf{1}\mathbf{2}]-\langle \mathbf{1}\mathbf{2}\rangle,
	\qquad
	y \equiv [\mathbf{1}\mathbf{2}]+\langle \mathbf{1}\mathbf{2}\rangle,
	\qquad
	x\sim\mathcal O(M),
	\qquad
	y\sim\mathcal O(p).
\end{equation}
Bold external labels denote massive external legs; the symmetrized massive little-group indices carried by their spinors and fields are suppressed.
The \(x\) structure isolates the static heavy-pair scalar factor, while \(y\) contains one recoil insertion.

For a spin-\(\frac12\) heavy pair, \(J_{12}=0\) is the only channel with an explicit \(x,y\) factor.
The corresponding amplitude--operator correspondence is
\begin{equation}
	x
	\;\longleftrightarrow\;
	\bar{\Psi}_{\mathbf{1}}\,\Psi_{\mathbf{2}},
	\qquad
	y
	\;\longleftrightarrow\;
	\bar{\Psi}_{\mathbf{1}}\,\gamma_5\,\Psi_{\mathbf{2}}.
\end{equation}
With the Weyl-basis convention above, the scalar and pseudoscalar bilinears correspond to the two independent relative signs between left- and right-chiral spinor contractions.
In the heavy-pair convention used here, the static combination is identified with the scalar bilinear, while the recoil combination is identified with the \(\gamma_5\) bilinear.

For \(J_{12}\ge1\), all spin-\(\frac12\) heavy indices are attached to the current, so no additional \(x,y\) factor is needed.

For the spin-\(\frac32\) heavy pair, it is useful to display the corresponding left three-point factors before writing their operator form.
We write the auxiliary-current spinor label as \(\mathsf J\).
Up to the reflected square labels used by the implementation, the left factors have the schematic form
\begin{equation}
	\mathcal V^{(\frac32)}_{12\to\mathcal J_{J_{12}}}
	\in
	\begin{cases}
		\{x^3,x^2y,xy^2,y^3\},
		 & J_{12}=0,   \\[2pt]
		\{x^2,xy,y^2\}\,
		\langle1\mathsf J\rangle^a[1\mathsf J]^{1-a}
		\langle2\mathsf J\rangle^b[2\mathsf J]^{1-b},
		 & J_{12}=1,   \\[2pt]
		\{x,y\}\,
		\langle1\mathsf J\rangle^a[1\mathsf J]^{2-a}
		\langle2\mathsf J\rangle^b[2\mathsf J]^{2-b},
		 & J_{12}=2,   \\[2pt]
		\langle1\mathsf J\rangle^a[1\mathsf J]^{3-a}
		\langle2\mathsf J\rangle^b[2\mathsf J]^{3-b}
		\bigl(\langle Q\mathsf J\rangle[Q\mathsf J]\bigr)^{J_{12}-3},
		 & J_{12}\ge3.
	\end{cases}
	\label{eq:spin32-left-three-point-structure}
\end{equation}
In the \(J_{12}=1\) line \(a,b=0,1\), in the \(J_{12}=2\) line \(a,b=0,1,2\), and in the \(J_{12}\ge3\) line \(a,b=0,1,2,3\).
Equation~\eqref{eq:spin32-left-three-point-structure} gives the universal \(x,y\) degree carried by every angular momentum channel.
The angle- and square-spinor choices complete the corresponding left-current tensors; they are sewn to the right residual structures and reduced as in Sec.~\ref{HBP-amp}.

Thus the low angular-momentum channels \(J_{12}=0,1,2\) require an amplitude--operator correspondence for the displayed \(x,y\) factors, while \(J_{12}\ge3\) uses the standard spin-\(\frac32\) correspondence in Table~\ref{tab:dm_buildingblocks}.
The projection tensors below are not a second definition of \(J_{12}\); they specify how the \(x,y\) factors in Eq.~\eqref{eq:spin32-left-three-point-structure} are represented by Rarita-Schwinger fields.
We use the all-incoming convention of Sec.~\ref{HBP-amp}, with \(\widehat P_1\equiv P_1\) and \(\widehat P_2\equiv -P_2\):
\begin{equation}
	\mathcal P_x\equiv \mathbf{1},\qquad
	\mathcal P_y\equiv \gamma_5,\qquad
	Q_i^\pm\equiv \frac12\left(1\pm\frac{\slashed{\widehat P}_i}{M}\right).
\end{equation}
Leg \(\mathbf{2}\) is still the all-incoming antibaryon leg; \(\widehat P_2\) is only the positive-energy momentum entering \(Q_i^\pm\).
We then define the projection tensors
\begin{equation}
	\begin{aligned}
		\mathcal P_{xx}^{\mu\nu}
		 & \equiv
		\operatorname{tr}\!\left(Q_1^+\gamma^\mu Q_2^-\gamma^\nu\right),
		\\
		\mathcal P_{xy}^{\mu\nu}
		 & \equiv
		\operatorname{tr}\!\left(Q_1^+\gamma^\mu Q_2^-\gamma^\nu\gamma_5\right),
		\\
		\mathcal P_{yy}^{\mu\nu}
		 & \equiv
		\operatorname{tr}\!\left(Q_1^+\gamma^\mu\gamma_5 Q_2^-\gamma^\nu\gamma_5\right).
	\end{aligned}
	\label{eq:spin32-closed-projectors}
\end{equation}
Here \(\operatorname{tr}\) denotes the Dirac trace.
These tensors encode the degree-two \(x,y\) part of the spin-\(\frac32\) left factor.
Since \(v\!
\cdot\!\Psi_{\mathbf{i}}\sim0\), only the transverse part of each vector index is physical; in the static limit the definitions above give \(\mathcal P_{xx}^{\mu\nu}=2g_\perp^{\mu\nu}\), \(\mathcal P_{xy}^{\mu\nu}=0+\mathcal O(p/M)\), and \(\mathcal P_{yy}^{\mu\nu}=0+\mathcal O(p^2/M^2)\), with \(g_\perp^{\mu\nu}=g^{\mu\nu}-v^\mu v^\nu\).
Thus these projection tensors carry the same recoil hierarchy as the stripped factors \(x^2,xy,y^2\).
The displayed assignments fix the monomial convention used below; no separate sum over placements is implied.

For \(J_{12}=0\), the left factor is cubic in \(x,y\).
No spinor label is attached to the auxiliary current, so the object multiplying the heavy pair is a scalar right operator.
The four amplitude--operator assignments are
\begin{equation}
	\begin{aligned}
		x^3\,\mathcal O
		 & \longleftrightarrow
		\mathcal P_{xx}^{\mu\nu}\,
		\bar\Psi_{\mathbf{1}\mu}\mathcal P_x\Psi_{\mathbf{2}\nu}\,\mathcal O,
		\\
		x^2y\,\mathcal O
		 & \longleftrightarrow
		\mathcal P_{xy}^{\mu\nu}\,
		\bar\Psi_{\mathbf{1}\mu}\mathcal P_x\Psi_{\mathbf{2}\nu}\,\mathcal O,
		\\
		xy^2\,\mathcal O
		 & \longleftrightarrow
		\mathcal P_{xy}^{\mu\nu}\,
		\bar\Psi_{\mathbf{1}\mu}\mathcal P_y\Psi_{\mathbf{2}\nu}\,\mathcal O,
		\\
		y^3\,\mathcal O
		 & \longleftrightarrow
		\mathcal P_{yy}^{\mu\nu}\,
		\bar\Psi_{\mathbf{1}\mu}\mathcal P_y\Psi_{\mathbf{2}\nu}\,\mathcal O .
	\end{aligned}
	\label{eq:spin32-N3-translation}
\end{equation}
Equivalently, the four cases are represented by the projector tensors assignments \(xx\,x\), \(xy\,x\), \(xy\,y\), and \(yy\,y\), respectively.
Thus the two mixed monomials use the same \(xy\) projection tensor and differ by the remaining spinor-chain insertion, \(\mathcal P_x\) or \(\mathcal P_y\).

For \(J_{12}=1\), the left factor contains a degree-two \(x,y\) factor and one current spinor from each heavy leg.
For the representative with left-handed remaining spinors,
\begin{equation}
	x_ax_b\,\lambda_{\mathbf{1}}^{\alpha}\lambda_{\mathbf{2}}^{\beta}\,\mathcal O_{\alpha\beta}
	\quad\longleftrightarrow\quad
	\mathcal P_{x_ax_b}^{\mu\nu}\,
	(\bar\Psi^L_{\mathbf{1}\mu})_\alpha\mathcal O^{\alpha\beta}
	(\Psi^L_{\mathbf{2}\nu})_\beta,
	\qquad
	x_ax_b=xx,xy,yy .
	\label{eq:spin32-N2-translation}
\end{equation}
Here the shorthand \(xx,xy,yy\) on the left-hand side denotes the three prefactors \(x^2,xy,y^2\), and the same labels select the corresponding projection tensors \(\mathcal P_{xx}^{\mu\nu},\mathcal P_{xy}^{\mu\nu},\mathcal P_{yy}^{\mu\nu}\).
Moreover, \(\Psi^L_{\mathbf{i}\mu}\equiv(\Psi_{\mathbf{i}\mu})_L\) is the left-handed component of the same Rarita-Schwinger field, not a separate spin-\(\frac12\) field.
The two residual spinors on legs \(\mathbf{1},\mathbf{2}\) can independently be replaced by \(\tilde\lambda_{\mathbf{i}}\), with the corresponding \(\Psi^L_{\mathbf{i}\mu}\to\Psi^R_{\mathbf{i}\mu}\) replacement on that leg.

For \(J_{12}=2\), the left factor contains one \(x\) or \(y\) factor and two current spinors from each heavy leg.
We denote the corresponding residual vector structures by \(R_{\mathbf{i}}^\mu\).
Then
\begin{equation}
	x_a\,R_{\mathbf{1}}^\mu R_{\mathbf{2}}^\nu\,\mathcal O_{\mu\nu}
	\quad\longleftrightarrow\quad
	\bar\Psi_{\mathbf{1}\mu}\,\mathcal P_{x_a}\,\mathcal O^{\mu\nu}\,\Psi_{\mathbf{2}\nu},
	\qquad
	x_a=x,y .
	\label{eq:spin32-N1-translation}
\end{equation}
Here \(R_{\mathbf{i}}^\mu\) may be represented by \(\lambda_{\mathbf{i}}\lambda_{\mathbf{i}}\), \(\lambda_{\mathbf{i}}\tilde\lambda_{\mathbf{i}}\), or \(\tilde\lambda_{\mathbf{i}}\tilde\lambda_{\mathbf{i}}\), following the ordinary spin-1 correspondence above.
For \(J_{12}\ge3\), no explicit \(x,y\) factor remains; all three spinors on each heavy leg are attached to the current, and one simply uses the standard spin-\(\frac32\) entry of Table~\ref{tab:dm_buildingblocks}.
\section{Comparison of the \texorpdfstring{$\bar B B\chi_+\chi_-$}{BB chi chi} and
\texorpdfstring{$\bar B B f_+\chi_+$}{BB f chi} sectors}
\label{app:comparison_chi_fchi}

This appendix verifies the agreement between the two-component representatives obtained in the present construction and the corresponding Dirac/heavy-baryon representatives in Ref.~\cite{Fettes:2000gb,Oller:2006yh,Frink:2006hx,Oller:2007qd,Jiang:2016tff,Song:2024fae,Sun:2025zuk}.
We compare the representative spaces, including their chiral-order and parity decompositions, in the two sectors for which a direct comparison can be made.
The comparison is organized separately for $\bar B B\chi_+\chi_-$ and $\bar B B f_+\chi_+$, and at each order we display the linear map from our representatives to the reference representatives.

We identify the scalar sources according to
\begin{equation}
	\Sigma_\pm \leftrightarrow \chi_\pm.
\end{equation}
For the $f_+\chi_+\bar B B$ sector we use the dual-field convention of Ref.~\cite{Sun:2025zuk},
\begin{equation}
	\widetilde f_+^{\mu\nu}=\epsilon^{\mu\nu\rho\sigma}f_{+,\rho\sigma},
	\label{eq:tilde_convention}
\end{equation}
with no additional factor of $1/2$.
In this convention
\begin{equation}
	\widetilde{\widetilde f}_{+,\mu\nu}=-4f_{+,\mu\nu}.
\end{equation}
According to the static--recoil expansion rules of Sec.~\ref{sec:heavy_spinor_projection}, the projected heavy-baryon representatives generated by \(\gamma^\mu\) and by the baryon relative derivative \(\LRD^\mu\) are identical.
Throughout this appendix, \(\LRD^\mu\) represents the hard relative momentum \(p_-^\mu=P_1^\mu-P_2^\mu\) and consequently carries zero additional chiral weight after the static--recoil factorization.

\subsection{Type I: \texorpdfstring{$\bar B B\chi_+\chi_-$}{BB chi chi}}

\subsubsection{LO: \texorpdfstring{$p^4$}{p4}}

At leading order the present construction gives a single representative,
\begin{equation}
	O_{\chi\chi}^{(4)}=(\bar B B)\chi_-\chi_+ .
	\label{eq:our_chichi_lo}
\end{equation}
The corresponding representative in Ref.~\cite{Sun:2025zuk} is
\begin{equation}
	R_{\chi\chi}^{(4)}=(\bar N N)\Sigma_-\Sigma_+ .
	\label{eq:jhy_chichi_lo}
\end{equation}
With the source identification above, the matching is one-to-one:
\begin{center}
	\renewcommand{\arraystretch}{1.35}
	\begin{tabularx}{\textwidth}{L{0.20\textwidth}|Y|Y|c}
		\hline
		order & reference representative & Combination of our structures & $P$ \\
		\hline
		$p^4$ & $R_{\chi\chi}^{(4)}$     & $O_{\chi\chi}^{(4)}$          & $-$ \\
		\hline
	\end{tabularx}
\end{center}
Thus
\begin{equation}
	N_{\rm ours}^{(4)}=N_{\rm ref}^{(4)}=1, \qquad N_{\rm ours}^{(4),P-}=N_{\rm ref}^{(4),P-}=1 .
\end{equation}
The parity assignment follows from
\begin{equation}
	P(\bar B B)=+,
	\qquad P(\chi_+)=+,
	\qquad P(\chi_-)=-,
\end{equation}
and hence $P(O_{\chi\chi}^{(4)})=-$.

\subsubsection{NLO: \texorpdfstring{$p^5$}{p5}}

At next-to-leading order there are three independent representatives in this sector: one pseudoscalar bilinear and two Weyl-current structures,
\begin{equation}
	O_P^{(5)}=(\bar B\gamma_5 B)\chi_-\chi_+,
	\label{eq:our_chichi_pseudo}
\end{equation}
\begin{equation}
	O_L^{(5)}=-(D_\mu\chi_+)(\bar B_L\bar\sigma^\mu B_L)\chi_-, \qquad O_R^{(5)}=-(D_\mu\chi_+)(\bar B_R\sigma^\mu B_R)\chi_- .
	\label{eq:our_chichi_currents}
\end{equation}
The corresponding reference representatives may be written as
\begin{align}
	R_P^{(5)} & =(\bar N\gamma_5 N)\Sigma_-\Sigma_+, \\ R_V^{(5)}&=(D_\mu\Sigma_+)(\bar N\gamma^\mu N)\Sigma_-, \\ R_A^{(5)}&=(D_\mu\Sigma_+)(\bar N\gamma_5\gamma^\mu N)\Sigma_- .
\end{align}
The two sets are related by the following linear combinations:
\begin{center}
	\renewcommand{\arraystretch}{1.35}
	\begin{tabularx}{\textwidth}{L{0.18\textwidth}|Y|Y|c}
		\hline
		order & reference representative & Combination of our structures & $P$ \\
		\hline
		$p^5$ & $R_P^{(5)}$              & $O_P^{(5)}$                   & $+$ \\
		$p^5$ & $R_V^{(5)}$              & $-(O_L^{(5)}+O_R^{(5)})$      & $-$ \\
		$p^5$ & $R_A^{(5)}$              & $O_L^{(5)}-O_R^{(5)}$         & $+$ \\
		\hline
	\end{tabularx}
\end{center}
It follows that
\begin{equation}
	N_{\rm ours}^{(5)}=N_{\rm ref}^{(5)}=3,
\end{equation}
with
\begin{equation}
	N_{\rm ours}^{(5),P+}=N_{\rm ref}^{(5),P+}=2, \qquad N_{\rm ours}^{(5),P-}=N_{\rm ref}^{(5),P-}=1 .
\end{equation}

The current rows follow directly from the standard two-component Dirac dictionary,
\begin{equation}
	J_V^\mu=\bar B_L\bar\sigma^\mu B_L+\bar B_R\sigma^\mu B_R
	=\bar B\gamma^\mu B,
\end{equation}
\begin{equation}
	J_A^\mu=-\bar B_L\bar\sigma^\mu B_L+\bar B_R\sigma^\mu B_R =\bar B\gamma_5\gamma^\mu B.
\end{equation}
Indeed,
\begin{equation}
	-(O_L^{(5)}+O_R^{(5)})=(D_\mu\chi_+)(\bar B\gamma^\mu B)\chi_-,
\end{equation}
whereas
\begin{equation}
	O_L^{(5)}-O_R^{(5)}=(D_\mu\chi_+)(\bar B\gamma_5\gamma^\mu B)\chi_- .
\end{equation}
Combining the leading and next-to-leading orders, the full $\bar B B\chi_+\chi_-$ type therefore gives
\begin{equation}
	N_{\rm ours}=N_{\rm ref}=4, \qquad N_{\rm ours}^{P+}=N_{\rm ref}^{P+}=2, \qquad N_{\rm ours}^{P-}=N_{\rm ref}^{P-}=2 .
\end{equation}

\subsection{Type II: \texorpdfstring{$\bar B B f_+\chi_+$}{BB f chi}}

\subsubsection{LO: \texorpdfstring{$p^4$}{p4}}

At leading order the two-component construction gives two Weyl tensor representatives,
\begin{equation}
	W_1^{(4)}=-(\bar B_L\bar\sigma^{\mu\nu}B_R)f_{+,\mu\nu}\chi_+,
	\label{eq:W14}
\end{equation}
\begin{equation}
	W_2^{(4)}=-(\bar B_R\sigma^{\mu\nu}B_L)\widetilde f_{+,\mu\nu}\chi_+ .
	\label{eq:W24}
\end{equation}
The corresponding Dirac representatives in Ref.~\cite{Sun:2025zuk} are the ordinary and dual tensor insertions,
\begin{equation}
	R_f^{(4)}=(\bar N\sigma^{\mu\nu}N)f_{+,\mu\nu}\Sigma_+,
\end{equation}
\begin{equation}
	R_{\widetilde f}^{(4)}=(\bar N\sigma^{\mu\nu}N)\widetilde f_{+,\mu\nu}\Sigma_+ .
\end{equation}
In the convention of Eq.~\eqref{eq:tilde_convention}, the two descriptions are related as follows:
\begin{center}
	\renewcommand{\arraystretch}{1.45}
	\begin{tabularx}{\textwidth}{L{0.18\textwidth}|Y|Y|c}
		\hline
		order & reference representative & Combination of our Weyl structures   & $P$ \\
		\hline
		$p^4$ & $R_f^{(4)}$              & $-W_1^{(4)}+\dfrac{\ii}{2}W_2^{(4)}$ & $+$ \\
		$p^4$ & $R_{\widetilde f}^{(4)}$ & $2\ii W_1^{(4)}-W_2^{(4)}$           & $-$ \\
		\hline
	\end{tabularx}
\end{center}
Consequently,
\begin{equation}
	N_{\rm ours}^{(4)}=N_{\rm ref}^{(4)}=2, \qquad N_{\rm ours}^{(4),P+}=N_{\rm ref}^{(4),P+}=1, \qquad N_{\rm ours}^{(4),P-}=N_{\rm ref}^{(4),P-}=1 .
\end{equation}

The conversion used in the table follows from
\begin{equation}
	\sigma^{\mu\nu}\widetilde f_{+,\mu\nu}
	=2\ii\sigma^{\mu\nu}f_{+,\mu\nu},
	\qquad
	\bar\sigma^{\mu\nu}\widetilde f_{+,\mu\nu}
	=-2\ii\bar\sigma^{\mu\nu}f_{+,\mu\nu}.
	\label{eq:dual_sigma_id}
\end{equation}
Thus the two Weyl tensor structures in Eqs.~\eqref{eq:W14} and \eqref{eq:W24} generate the ordinary and dual tensor reference representatives.

\subsubsection{NLO: \texorpdfstring{$p^5$}{p5}}

At next-to-leading order we use, for this comparison, the canonical representative convention in which the two-component basis is written as four current-type chains and two tensor-current chains.
Within this subsection we denote these six representatives by $W_1,\ldots,W_6$.
The current-type chains are
\begin{align}
	W_1 & =(D_\mu\chi_+)(\bar B_L\bar\sigma_{\nu\rho}\bar\sigma^\mu B_L) f_+^{\nu\rho}, \\ W_2&=-(D_\mu\chi_+)(\bar B_R\sigma^\mu\bar\sigma_{\nu\rho}B_R) f_+^{\nu\rho}, \\ W_3&=-(D_\mu\chi_+)(\bar B_L\bar\sigma^\mu\sigma_{\nu\rho}B_L) \widetilde f_+^{\nu\rho}, \\ W_4&=(D_\mu\chi_+)(\bar B_R\sigma_{\nu\rho}\sigma^\mu B_R) \widetilde f_+^{\nu\rho} .
\end{align}
The tensor-current chains are
\begin{equation}
	W_5=-(D_\nu\chi_+)
	(\bar B_L\bar\sigma^\nu\sigma^\mu\bar\sigma_{\rho\sigma}
	\LRD_\mu B_R)f_+^{\rho\sigma},
	\label{eq:W5tensor}
\end{equation}
\begin{equation}
	W_6=(D_\nu\chi_+) (\bar B_R\sigma_{\rho\sigma}\LRD_\mu B_L) \trsig(\sigma^\mu\bar\sigma^\nu)\widetilde f_+^{\rho\sigma} .
	\label{eq:W6tensor}
\end{equation}
The six Dirac/heavy-baryon representatives used for the comparison are
\begin{align}
	R_{V_f}^{(5)} & =(D^\nu\Sigma_+)(\bar N\gamma^\mu N)f_{+,\mu\nu}, \\ R_{V_{\widetilde f}}^{(5)}&=(D^\nu\Sigma_+)(\bar N\gamma^\mu N) \widetilde f_{+,\mu\nu}, \\ R_{A_f}^{(5)}&=(D^\nu\Sigma_+)(\bar N\gamma_5\gamma^\mu N)f_{+,\mu\nu}, \\ R_{A_{\widetilde f}}^{(5)}&=(D^\nu\Sigma_+)(\bar N\gamma_5\gamma^\mu N) \widetilde f_{+,\mu\nu}, \\ R_{T_f}^{(5)}&=(D_\alpha\Sigma_+)(\bar N\sigma^{\mu\nu}\LRD^\alpha N)f_{+,\mu\nu}, \\ R_{T_{\widetilde f}}^{(5)}&=(D_\alpha\Sigma_+)(\bar N\sigma^{\mu\nu}\LRD^\alpha N) \widetilde f_{+,\mu\nu} .
\end{align}
The linear map from the two-component representatives to these Dirac/heavy-baryon representatives is
\begin{center}
	\renewcommand{\arraystretch}{1.45}
	\begin{tabularx}{\textwidth}{L{0.20\textwidth}|Y|Y|c}
		\hline
		reference representative                                                        & Dirac/HB reference structure & Combination of our structures & $P$ \\
		\hline
		$R_{V_f}^{(5)}$                                                                 &
		$(D^\nu\Sigma_+)(\bar N\gamma^\mu N)f_{+,\mu\nu}$                               &
		$\dfrac{\ii}{4}(W_1+W_2)+\dfrac18(W_3+W_4)$                                     & $+$                                                                \\

		$R_{V_{\widetilde f}}^{(5)}$                                                    &
		$(D^\nu\Sigma_+)(\bar N\gamma^\mu N)\widetilde f_{+,\mu\nu}$                    &
		$\dfrac12(W_1+W_2)+\dfrac{\ii}{4}(W_3+W_4)$                                     & $-$                                                                \\

		$R_{A_f}^{(5)}$                                                                 &
		$(D^\nu\Sigma_+)(\bar N\gamma_5\gamma^\mu N)f_{+,\mu\nu}$                       &
		$\dfrac{\ii}{4}(-W_1+W_2)+\dfrac18(-W_3+W_4)$                                   & $-$                                                                \\

		$R_{A_{\widetilde f}}^{(5)}$                                                    &
		$(D^\nu\Sigma_+)(\bar N\gamma_5\gamma^\mu N)\widetilde f_{+,\mu\nu}$            &
		$\dfrac12(-W_1+W_2)+\dfrac{\ii}{4}(-W_3+W_4)$                                   & $+$                                                                \\

		$R_{T_f}^{(5)}$                                                                 &
		$(D_\alpha\Sigma_+)(\bar N\sigma^{\mu\nu}\LRD^\alpha N)f_{+,\mu\nu}$            &
		$W_5-\dfrac{\ii}{4}W_6$                                                         & $+$                                                                \\

		$R_{T_{\widetilde f}}^{(5)}$                                                    &
		$(D_\alpha\Sigma_+)(\bar N\sigma^{\mu\nu}\LRD^\alpha N)\widetilde f_{+,\mu\nu}$ &
		$-2\ii W_5+\dfrac12 W_6$                                                        & $-$                                                                \\
		\hline
	\end{tabularx}
\end{center}
This gives
\begin{equation}
	N_{\rm ours}^{(5)}=N_{\rm ref}^{(5)}=6,
\end{equation}
with
\begin{equation}
	N_{\rm ours}^{(5),P+}=N_{\rm ref}^{(5),P+}=3, \qquad N_{\rm ours}^{(5),P-}=N_{\rm ref}^{(5),P-}=3 .
\end{equation}

The entries of the table are obtained as follows.
For the four current rows, introduce
\begin{equation}
	L_f=(D_\mu\chi_+)f_+^{\mu\lambda}
	(\bar B_L\bar\sigma_\lambda B_L),
	\qquad
	R_f=(D_\mu\chi_+)f_+^{\mu\lambda}
	(\bar B_R\sigma_\lambda B_R),
\end{equation}
and define $L_{\widetilde f},R_{\widetilde f}$ by the replacement $f_+\to\widetilde f_+$.
Reducing the sigma chains gives
\begin{equation}
	L_f=\frac{\ii}{4}W_1+\frac18 W_3,
	\qquad
	R_f=\frac{\ii}{4}W_2+\frac18 W_4,
	\label{eq:LfRf}
\end{equation}
\begin{equation}
	L_{\widetilde f}=\frac12 W_1+\frac{\ii}{4}W_3, \qquad R_{\widetilde f}=\frac12 W_2+\frac{\ii}{4}W_4.
	\label{eq:LftRfT}
\end{equation}
The vector and axial-vector representatives are then obtained by combining left- and right-handed currents as
\begin{equation}
	J_V^\lambda=L^\lambda+R^\lambda,
	\qquad
	J_A^\lambda=-L^\lambda+R^\lambda,
\end{equation}
which gives the first four rows of the table.

It remains to relate the two tensor-current rows to $W_5$ and $W_6$.
The left-handed branch follows from the reduction of the four-sigma chain in $W_5$,
\begin{equation}
	\bar\sigma^\nu\sigma^\mu\bar\sigma_{\rho\sigma}f_+^{\rho\sigma}
	=2\ii f_+^{\mu\nu}-\widetilde f_+^{\mu\nu}
	+(2f_+^{\mu\lambda}+\ii\widetilde f_+^{\mu\lambda})
	\bar\sigma^\nu{}_{\lambda}.
	\label{eq:W5_reduction}
\end{equation}
This separates $W_5$ into
\begin{equation}
	W_5=C_{5,D}+W_{5,{\rm mis}},
\end{equation}
where
\begin{equation}
	C_{5,D}
	=-(D_\nu\chi_+)(2\ii f_+^{\mu\nu}-\widetilde f_+^{\mu\nu})
	(\bar B_L\LRD_\mu B_R),
	\label{eq:C5D}
\end{equation}
and
\begin{equation}
	W_{5,{\rm mis}} =-(D_\nu\chi_+)S^{\mu\lambda} (\bar B_L\bar\sigma^\nu{}_{\lambda}\LRD_\mu B_R), \qquad S^{\mu\lambda}=2f_+^{\mu\lambda}+\ii\widetilde f_+^{\mu\lambda}.
	\label{eq:W5mis}
\end{equation}

Using the projected heavy-baryon representative relation described above, the first term is mapped to a left-handed current representative,
\begin{equation}
	C_{5,D}=2\ii L_f-L_{\widetilde f}.
\end{equation}
Equations~\eqref{eq:LfRf} and \eqref{eq:LftRfT} then give
\begin{equation}
	C_{5,D} =2\ii\left(\frac{\ii}{4}W_1+\frac18W_3\right) -\left(\frac12W_1+\frac{\ii}{4}W_3\right) =-W_1 .
	\label{eq:C5D_minus_W1}
\end{equation}
For the misaligned term we use
\begin{equation}
	\bar\sigma^\nu{}_{\lambda}
	=\frac{\ii}{2}
	\left(
	\bar\sigma^\nu\sigma_\lambda
	-\bar\sigma_\lambda\sigma^\nu
	\right).
\end{equation}
The antisymmetric part gives the aligned left-handed tensor-current representative.
In the same projected representative space,
\begin{equation}
	S^{\mu\lambda}
	\left(
	\bar\sigma^\nu{}_{\lambda}\LRD_\mu
	-\bar\sigma^\nu{}_{\mu}\LRD_\lambda
	\right)
	=S^{\mu\lambda}\bar\sigma_{\lambda\mu},
	\label{eq:W5_to_tensor}
\end{equation}
which corresponds to
\begin{equation}
	(D_\alpha\chi_+)
	(\bar B_L\bar\sigma_{\mu\nu}\LRD^\alpha B_R)f_+^{\mu\nu}.
\end{equation}
The remaining three-sigma chain reduces as
\begin{equation}
	S^{\mu\lambda}\bar\sigma_\lambda\sigma^\nu\bar\sigma_\mu =-2S^{\nu\alpha}\bar\sigma_\alpha .
	\label{eq:W5_leftover_current}
\end{equation}
It therefore gives the second current-type remainder
\begin{equation}
	C_{5,\sigma} =(D_\nu\chi_+) (\widetilde f_+^{\nu\alpha}-2\ii f_+^{\nu\alpha}) (\bar B_L\bar\sigma_\alpha B_L) =L_{\widetilde f}-2\ii L_f.
\end{equation}
Using the same current-block relations,
\begin{equation}
	C_{5,\sigma} =\left(\frac12W_1+\frac{\ii}{4}W_3\right) -2\ii\left(\frac{\ii}{4}W_1+\frac18W_3\right) =W_1 .
	\label{eq:C5sigma_plus_W1}
\end{equation}
The two current remainders cancel,
\begin{equation}
	C_{5,D}+C_{5,\sigma}=-W_1+W_1=0.
\end{equation}
Thus the complete $W_5$ chain represents the left-handed tensor branch in the heavy-baryon convention of Ref.~\cite{Sun:2025zuk}.
No separate projection of $W_5$ is needed in the tensor rows of the matching table.

The right-handed tensor branch is obtained from $W_6$.
Since
\begin{equation}
	\trsig(\sigma^\mu\bar\sigma^\nu)=2g^{\mu\nu},
\end{equation}
Eq.~\eqref{eq:W6tensor} becomes
\begin{equation}
	W_6=2(D_\lambda\chi_+) (\bar B_R\sigma_{\rho\sigma}\LRD^\lambda B_L) \widetilde f_+^{\rho\sigma}.
\end{equation}
Together with
\begin{equation}
	\sigma_{\rho\sigma}\widetilde f_+^{\rho\sigma}
	=2\ii\sigma_{\rho\sigma}f_+^{\rho\sigma},
\end{equation}
this gives
\begin{equation}
	T_R^f
	=(D_\lambda\chi_+)(\bar B_R\sigma_{\rho\sigma}\LRD^\lambda B_L)
	f_+^{\rho\sigma}
	=-\frac{\ii}{4}W_6,
\end{equation}
and
\begin{equation}
	T_R^{\widetilde f} =(D_\lambda\chi_+)(\bar B_R\sigma_{\rho\sigma}\LRD^\lambda B_L) \widetilde f_+^{\rho\sigma} =\frac12 W_6.
\end{equation}
Together with the left-handed tensor representative supplied by $W_5$, these relations give the two tensor rows in the matching table.

After combining the $p^4$ and $p^5$ contributions, the $\bar B B f_+\chi_+$ type gives
\begin{equation}
	N_{\rm ours}=N_{\rm ref}=8,
\end{equation}
with
\begin{equation}
	N_{\rm ours}^{P+}=N_{\rm ref}^{P+}=4, \qquad N_{\rm ours}^{P-}=N_{\rm ref}^{P-}=4 .
\end{equation}

\section{Explicit Operator Bases}
\label{app:format-operator-bases}
This appendix demonstrates the output of the construction through explicit flavor--Lorentz direct-product bases for octet- and decuplet-baryon operator sectors.
The same construction applies directly to additional field contents and chiral orders.
The bases are first organized by the chiral dimension \(\dchi\).
The bases listed below are Lorentz- and \(SU(3)_V\)-covariant contact representatives before imposing parity, charge-conjugation, and Hermitian-conjugation constraints.
A physical symmetry sector is obtained by the corresponding linear projection and h.c. pairing.
Within a fixed \(\dchi\) and basis set \(r\), we use
\begin{equation}
	\mathcal O^{X,\dchi,r}_{ik}
	=
	C^{X,r}_i L^{X,\dchi,r}_k ,
	\qquad
	i=1,\ldots,n_C^r,\quad
	k=1,\ldots,n_L^{X,\dchi,r} .
\end{equation}
The table at the beginning of each sector should be read as follows.
``Insertion'' is the displayed source/spurion representative.
The column ``set'' is the value of \(r\).
If the same insertion appears with more than one set number, it is not a duplicate: each set has its own flavor basis \(C_i^{X,r}\) and Lorentz basis \(L_k^{X,\dchi,r}\), and the full operators in that set are all products \(C_i^{X,r}L_k^{X,\dchi,r}\).
For each displayed set, \(N=n_Cn_L\).
Within each set, the Lorentz structures are ordered only for readability; this ordering does not define additional operator classes.
The power-counting assignment is summarized in table~\ref{tab:operator-power-counting}.

\begin{table}[tbp]
	\centering
	\small
	\begin{tabular}{c|ccccccc}
		\toprule
		\(X\)                         &
		\(u_\mu\)                     &
		\(\hat\alpha_{\parallel\mu}\) &
		\(\widehat{\chi}_\pm\)        &
		\(f_\pm^{\mu\nu}\)            &
		\(\widetilde f_\pm^{\mu\nu}\) &
		\(V_{\mu\nu}\)                &
		\(D_\mu^{\rm mes}\)                                   \\
		\midrule
		\(d_\chi(X)\)                 &
		1                             & 1 & 2 & 2 & 2 & 2 & 1 \\
		\bottomrule
	\end{tabular}
	\qquad
	\begin{tabular}{c|ccccc}
		\toprule
		\(X\)                         &
		\(\overleftrightarrow D_\mu\) &
		\(\gamma_5\)                  &
		\(\mathcal P_{xx}\)           &
		\(\mathcal P_{xy}\)           &
		\(\mathcal P_{yy}\)                           \\
		\midrule
		\(d_\chi(X)\)                 &
		0                             & 1 & 0 & 1 & 2 \\
		\bottomrule
	\end{tabular}

	\caption{Power-counting assignment used in this appendix.
	\(D_\mu^{\rm mes}\) denotes a derivative acting on mesonic or source building blocks.}
	\label{tab:operator-power-counting}
\end{table}
\FloatBarrier

Following the baryon relative-derivative convention of Table~\ref{tab:building_blocks}, \(\overleftrightarrow D_\mu\) carries \(d_\chi=0\) after its hard mass factor is absorbed into the Wilson coefficient.

\subsection{\texorpdfstring{$\bar B B u f_+$}{BbarB-u-Fp}}

\label{app:hbp-BbarB-u-Fp}

\par\smallskip
\begingroup
\centering
\scriptsize
\setlength{\tabcolsep}{5pt}
\begin{tabular}{@{}lccrrr@{}}
	\toprule
	Insertion & $d_\chi$          & set & $n_C$ & $n_L$ & $N$ \\
	\midrule
	$u f_+$   & $\mathcal O(p^3)$ & 1   & 8     & 3     & 24  \\
	$u f_+$   & $\mathcal O(p^4)$ & 1   & 8     & 5     & 40  \\
	\bottomrule
\end{tabular}
\par
\endgroup
\smallskip

\noindent The following sets give explicit representatives.

\subsubsection*{$\dchi=3$, $\mathcal O(p^3)$}

\paragraph{$u f_+$.}

\emph{Set 1, $\dchi=3$, $\mathcal O(p^3)$.}

Flavor basis $C_i^{1}$:
\begin{align}
	C^{1}_{1} & = \TrF(a_{1}, a_{2})\,\TrF(a_{3}, a_{4}), \\
	C^{1}_{2} & = \TrF(a_{1}, a_{2}, a_{3}, a_{4}),       \\
	C^{1}_{3} & = \TrF(a_{1}, a_{2}, a_{4}, a_{3}),       \\
	C^{1}_{4} & = \TrF(a_{1}, a_{3}, a_{4}, a_{2}),       \\
	C^{1}_{5} & = \TrF(a_{1}, a_{3})\,\TrF(a_{2}, a_{4}), \\
	C^{1}_{6} & = \TrF(a_{1}, a_{3}, a_{2}, a_{4}),       \\
	C^{1}_{7} & = \TrF(a_{1}, a_{4}, a_{3}, a_{2}),       \\
	C^{1}_{8} & = \TrF(a_{1}, a_{4}, a_{2}, a_{3})        \\
\end{align}

Lorentz/operator basis $L_k^{1}$:
\begin{align}
	L^{1}_{1} & = (\Bar{B}^{L}_{a_{1}} \bar{\sigma}^{\nu \rho} \overleftrightarrow{D}_{\mu} B^{R}_{a_{2}}) \text{Tr}(\sigma^{\sigma} \bar{\sigma}^{\mu}) u_{\sigma a_{3}} f_{+, \nu \rho a_{4}}, \\
	L^{1}_{2} & = (\Bar{B}^{R}_{a_{1}} \sigma^{\mu} \bar{\sigma}^{\nu \rho} B^{R}_{a_{2}}) u_{\mu a_{3}} f_{+, \nu \rho a_{4}},                                                                  \\
	L^{1}_{3} & = (\Bar{B}^{L}_{a_{1}} \bar{\sigma}^{\nu \rho} \bar{\sigma}^{\mu} B^{L}_{a_{2}}) u_{\mu a_{3}} f_{+, \nu \rho a_{4}}                                                             \\
\end{align}

\subsubsection*{$\dchi=4$, $\mathcal O(p^4)$}

\paragraph{$u f_+$.}

\emph{Set 1, $\dchi=4$, $\mathcal O(p^4)$.}

Flavor basis $C_i^{1}$:
\begin{align}
	C^{1}_{1} & = \TrF(a_{1}, a_{2})\,\TrF(a_{3}, a_{4}), \\
	C^{1}_{2} & = \TrF(a_{1}, a_{2}, a_{3}, a_{4}),       \\
	C^{1}_{3} & = \TrF(a_{1}, a_{2}, a_{4}, a_{3}),       \\
	C^{1}_{4} & = \TrF(a_{1}, a_{3}, a_{4}, a_{2}),       \\
	C^{1}_{5} & = \TrF(a_{1}, a_{3})\,\TrF(a_{2}, a_{4}), \\
	C^{1}_{6} & = \TrF(a_{1}, a_{3}, a_{2}, a_{4}),       \\
	C^{1}_{7} & = \TrF(a_{1}, a_{4}, a_{3}, a_{2}),       \\
	C^{1}_{8} & = \TrF(a_{1}, a_{4}, a_{2}, a_{3})        \\
\end{align}

Lorentz/operator basis $L_k^{1}$:
\begin{align}
	L^{1}_{1} & = (D_{\rho} f_{+, \sigma \alpha a_{4}}) (\Bar{B}^{L}_{a_{1}} \bar{\sigma}^{\rho} \sigma^{\mu} \bar{\sigma}^{\sigma \alpha} \overleftrightarrow{D}_{\nu} \overleftrightarrow{D}_{\mu} B^{R}_{a_{2}}) \text{Tr}(\sigma^{\beta} \bar{\sigma}^{\nu}) u_{\beta a_{3}}, \\
	L^{1}_{2} & = (D_{\nu} f_{+, \sigma \alpha a_{4}}) (\Bar{B}^{R}_{a_{1}} \sigma^{\rho} \bar{\sigma}^{\sigma \alpha} \bar{\sigma}^{\mu} \sigma^{\nu} \overleftrightarrow{D}_{\mu} B^{R}_{a_{2}}) u_{\rho a_{3}},                                                                \\
	L^{1}_{3} & = (D_{\nu} f_{+, \sigma \alpha a_{4}}) (\Bar{B}^{L}_{a_{1}} \bar{\sigma}^{\nu} \sigma^{\mu} \bar{\sigma}^{\sigma \alpha} \bar{\sigma}^{\rho} \overleftrightarrow{D}_{\mu} B^{L}_{a_{2}}) u_{\rho a_{3}},                                                          \\
	L^{1}_{4} & = (D_{\mu} u_{\sigma a_{3}}) (\Bar{B}^{R}_{a_{1}} \sigma^{\mu} \bar{\sigma}^{\nu \rho} \bar{\sigma}^{\sigma} B^{L}_{a_{2}}) f_{+, \nu \rho a_{4}},                                                                                                                \\
	L^{1}_{5} & = (D_{\mu} f_{+, \nu \rho a_{4}}) (\Bar{B}^{L}_{a_{1}} \bar{\sigma}^{\mu} \sigma^{\sigma} \bar{\sigma}^{\nu \rho} B^{R}_{a_{2}}) u_{\sigma a_{3}}                                                                                                                 \\
\end{align}

\subsection{\texorpdfstring{$\bar B B f_+ f_-$}{BbarB-Fp-Fm}}

\label{app:hbp-BbarB-Fp-Fm}

\par\smallskip
\begingroup
\centering
\scriptsize
\setlength{\tabcolsep}{5pt}
\begin{tabular}{@{}lccrrr@{}}
	\toprule
	Insertion & $d_\chi$          & set & $n_C$ & $n_L$ & $N$ \\
	\midrule
	$f_+ f_-$ & $\mathcal O(p^4)$ & 1   & 8     & 2     & 16  \\
	$f_+ f_-$ & $\mathcal O(p^5)$ & 1   & 8     & 4     & 32  \\
	\bottomrule
\end{tabular}
\par
\endgroup
\smallskip

\noindent The following sets give explicit representatives.

\subsubsection*{$\dchi=4$, $\mathcal O(p^4)$}

\paragraph{$f_+ f_-$.}

\emph{Set 1, $\dchi=4$, $\mathcal O(p^4)$.}

Flavor basis $C_i^{1}$:
\begin{align}
	C^{1}_{1} & = \TrF(a_{1}, a_{2})\,\TrF(a_{3}, a_{4}), \\
	C^{1}_{2} & = \TrF(a_{1}, a_{2}, a_{3}, a_{4}),       \\
	C^{1}_{3} & = \TrF(a_{1}, a_{2}, a_{4}, a_{3}),       \\
	C^{1}_{4} & = \TrF(a_{1}, a_{3}, a_{4}, a_{2}),       \\
	C^{1}_{5} & = \TrF(a_{1}, a_{3})\,\TrF(a_{2}, a_{4}), \\
	C^{1}_{6} & = \TrF(a_{1}, a_{3}, a_{2}, a_{4}),       \\
	C^{1}_{7} & = \TrF(a_{1}, a_{4}, a_{3}, a_{2}),       \\
	C^{1}_{8} & = \TrF(a_{1}, a_{4}, a_{2}, a_{3})        \\
\end{align}

Lorentz/operator basis $L_k^{1}$:
\begin{align}
	L^{1}_{1} & = (\Bar{B}_{a_{1}} B_{a_{2}}) \text{Tr}(\bar{\sigma}^{\mu \nu} \bar{\sigma}^{\rho \sigma}) f_{+, \rho \sigma a_{3}} f_{-, \mu \nu a_{4}}, \\
	L^{1}_{2} & = (\Bar{B}^{L}_{a_{1}} \bar{\sigma}^{\rho \sigma} \bar{\sigma}^{\mu \nu} B^{R}_{a_{2}}) f_{+, \rho \sigma a_{3}} f_{-, \mu \nu a_{4}}     \\
\end{align}

\subsubsection*{$\dchi=5$, $\mathcal O(p^5)$}

\paragraph{$f_+ f_-$.}

\emph{Set 1, $\dchi=5$, $\mathcal O(p^5)$.}

Flavor basis $C_i^{1}$:
\begin{align}
	C^{1}_{1} & = \TrF(a_{1}, a_{2})\,\TrF(a_{3}, a_{4}), \\
	C^{1}_{2} & = \TrF(a_{1}, a_{2}, a_{3}, a_{4}),       \\
	C^{1}_{3} & = \TrF(a_{1}, a_{2}, a_{4}, a_{3}),       \\
	C^{1}_{4} & = \TrF(a_{1}, a_{3}, a_{4}, a_{2}),       \\
	C^{1}_{5} & = \TrF(a_{1}, a_{3})\,\TrF(a_{2}, a_{4}), \\
	C^{1}_{6} & = \TrF(a_{1}, a_{3}, a_{2}, a_{4}),       \\
	C^{1}_{7} & = \TrF(a_{1}, a_{4}, a_{3}, a_{2}),       \\
	C^{1}_{8} & = \TrF(a_{1}, a_{4}, a_{2}, a_{3})        \\
\end{align}

Lorentz/operator basis $L_k^{1}$:
\begin{align}
	L^{1}_{1} & = (D_{\nu} f_{-, \alpha \beta a_{4}}) (\Bar{B}^{L}_{a_{1}} \bar{\sigma}^{\rho \sigma} \bar{\sigma}^{\alpha \beta} \bar{\sigma}^{\mu} \sigma^{\nu} \overleftrightarrow{D}_{\mu} B^{R}_{a_{2}}) f_{+, \rho \sigma a_{3}}, \\
	L^{1}_{2} & = (D_{\mu} f_{-, \nu \rho a_{4}}) (\Bar{B}^{L}_{a_{1}} \bar{\sigma}^{\mu} B^{L}_{a_{2}}) \text{Tr}(\bar{\sigma}^{\nu \rho} \bar{\sigma}^{\sigma \alpha}) f_{+, \sigma \alpha a_{3}},                                    \\
	L^{1}_{3} & = (D_{\mu} f_{-, \nu \rho a_{4}}) (\Bar{B}^{R}_{a_{1}} \sigma^{\mu} B^{R}_{a_{2}}) \text{Tr}(\bar{\sigma}^{\nu \rho} \bar{\sigma}^{\sigma \alpha}) f_{+, \sigma \alpha a_{3}},                                          \\
	L^{1}_{4} & = (\Bar{B}_{a_{1}} \gamma_5 B_{a_{2}}) \text{Tr}(\bar{\sigma}^{\mu \nu} \bar{\sigma}^{\rho \sigma}) f_{+, \rho \sigma a_{3}} f_{-, \mu \nu a_{4}}                                                                       \\
\end{align}

\subsection{\texorpdfstring{$\bar B B a_\perp f_+$}{BbarB-aperpL-Fp}}

\label{app:hbp-BbarB-aperpL-Fp}

\par\smallskip
\begingroup
\centering
\scriptsize
\setlength{\tabcolsep}{5pt}
\begin{tabular}{@{}lccrrr@{}}
	\toprule
	Insertion     & $d_\chi$          & set & $n_C$ & $n_L$ & $N$ \\
	\midrule
	$a_\perp f_+$ & $\mathcal O(p^3)$ & 1   & 8     & 3     & 24  \\
	$a_\perp f_+$ & $\mathcal O(p^4)$ & 1   & 8     & 5     & 40  \\
	\bottomrule
\end{tabular}
\par
\endgroup
\smallskip

\noindent The following sets give explicit representatives.

\subsubsection*{$\dchi=3$, $\mathcal O(p^3)$}

\paragraph{$a_\perp f_+$.}

\emph{Set 1, $\dchi=3$, $\mathcal O(p^3)$.}

Flavor basis $C_i^{1}$:
\begin{align}
	C^{1}_{1} & = \TrF(a_{1}, a_{2})\,\TrF(a_{3}, a_{4}), \\
	C^{1}_{2} & = \TrF(a_{1}, a_{2}, a_{3}, a_{4}),       \\
	C^{1}_{3} & = \TrF(a_{1}, a_{2}, a_{4}, a_{3}),       \\
	C^{1}_{4} & = \TrF(a_{1}, a_{3}, a_{4}, a_{2}),       \\
	C^{1}_{5} & = \TrF(a_{1}, a_{3})\,\TrF(a_{2}, a_{4}), \\
	C^{1}_{6} & = \TrF(a_{1}, a_{3}, a_{2}, a_{4}),       \\
	C^{1}_{7} & = \TrF(a_{1}, a_{4}, a_{3}, a_{2}),       \\
	C^{1}_{8} & = \TrF(a_{1}, a_{4}, a_{2}, a_{3})        \\
\end{align}

Lorentz/operator basis $L_k^{1}$:
\begin{align}
	L^{1}_{1} & = (\Bar{B}^{L}_{a_{1}} \bar{\sigma}^{\nu \rho} \overleftrightarrow{D}_{\mu} B^{R}_{a_{2}}) \text{Tr}(\sigma^{\sigma} \bar{\sigma}^{\mu}) \hat\alpha_{\perp, \sigma a_{3}} f_{+, \nu \rho a_{4}}, \\
	L^{1}_{2} & = (\Bar{B}^{R}_{a_{1}} \sigma^{\mu} \bar{\sigma}^{\nu \rho} B^{R}_{a_{2}}) \hat\alpha_{\perp, \mu a_{3}} f_{+, \nu \rho a_{4}},                                                                  \\
	L^{1}_{3} & = (\Bar{B}^{L}_{a_{1}} \bar{\sigma}^{\nu \rho} \bar{\sigma}^{\mu} B^{L}_{a_{2}}) \hat\alpha_{\perp, \mu a_{3}} f_{+, \nu \rho a_{4}}                                                             \\
\end{align}

\subsubsection*{$\dchi=4$, $\mathcal O(p^4)$}

\paragraph{$a_\perp f_+$.}

\emph{Set 1, $\dchi=4$, $\mathcal O(p^4)$.}

Flavor basis $C_i^{1}$:
\begin{align}
	C^{1}_{1} & = \TrF(a_{1}, a_{2})\,\TrF(a_{3}, a_{4}), \\
	C^{1}_{2} & = \TrF(a_{1}, a_{2}, a_{3}, a_{4}),       \\
	C^{1}_{3} & = \TrF(a_{1}, a_{2}, a_{4}, a_{3}),       \\
	C^{1}_{4} & = \TrF(a_{1}, a_{3}, a_{4}, a_{2}),       \\
	C^{1}_{5} & = \TrF(a_{1}, a_{3})\,\TrF(a_{2}, a_{4}), \\
	C^{1}_{6} & = \TrF(a_{1}, a_{3}, a_{2}, a_{4}),       \\
	C^{1}_{7} & = \TrF(a_{1}, a_{4}, a_{3}, a_{2}),       \\
	C^{1}_{8} & = \TrF(a_{1}, a_{4}, a_{2}, a_{3})        \\
\end{align}

Lorentz/operator basis $L_k^{1}$:
\begin{align}
	L^{1}_{1} & = (D_{\rho} f_{+, \sigma \alpha a_{4}}) (\Bar{B}^{L}_{a_{1}} \bar{\sigma}^{\rho} \sigma^{\mu} \bar{\sigma}^{\sigma \alpha} \overleftrightarrow{D}_{\nu} \overleftrightarrow{D}_{\mu} B^{R}_{a_{2}}) \text{Tr}(\sigma^{\beta} \bar{\sigma}^{\nu}) \hat\alpha_{\perp, \beta a_{3}}, \\
	L^{1}_{2} & = (D_{\nu} f_{+, \sigma \alpha a_{4}}) (\Bar{B}^{R}_{a_{1}} \sigma^{\rho} \bar{\sigma}^{\sigma \alpha} \bar{\sigma}^{\mu} \sigma^{\nu} \overleftrightarrow{D}_{\mu} B^{R}_{a_{2}}) \hat\alpha_{\perp, \rho a_{3}},                                                                \\
	L^{1}_{3} & = (D_{\nu} f_{+, \sigma \alpha a_{4}}) (\Bar{B}^{L}_{a_{1}} \bar{\sigma}^{\nu} \sigma^{\mu} \bar{\sigma}^{\sigma \alpha} \bar{\sigma}^{\rho} \overleftrightarrow{D}_{\mu} B^{L}_{a_{2}}) \hat\alpha_{\perp, \rho a_{3}},                                                          \\
	L^{1}_{4} & = (D_{\mu} \hat\alpha_{\perp, \sigma a_{3}}) (\Bar{B}^{R}_{a_{1}} \sigma^{\mu} \bar{\sigma}^{\nu \rho} \bar{\sigma}^{\sigma} B^{L}_{a_{2}}) f_{+, \nu \rho a_{4}},                                                                                                                \\
	L^{1}_{5} & = (D_{\mu} f_{+, \nu \rho a_{4}}) (\Bar{B}^{L}_{a_{1}} \bar{\sigma}^{\mu} \sigma^{\sigma} \bar{\sigma}^{\nu \rho} B^{R}_{a_{2}}) \hat\alpha_{\perp, \sigma a_{3}}                                                                                                                 \\
\end{align}

\subsection{\texorpdfstring{$\bar B B A f_+$}{BbarB-AperpT-Fp}}

\label{app:hbp-BbarB-AperpT-Fp}

\par\smallskip
\begingroup
\centering
\scriptsize
\setlength{\tabcolsep}{5pt}
\begin{tabular}{@{}lccrrr@{}}
	\toprule
	Insertion & $d_\chi$          & set & $n_C$ & $n_L$ & $N$ \\
	\midrule
	$A f_+$   & $\mathcal O(p^4)$ & 1   & 8     & 4     & 32  \\
	$A f_+$   & $\mathcal O(p^5)$ & 1   & 8     & 8     & 64  \\
	\bottomrule
\end{tabular}
\par
\endgroup
\smallskip

\noindent The following sets give explicit representatives.

\subsubsection*{$\dchi=4$, $\mathcal O(p^4)$}

\paragraph{$A f_+$.}

\emph{Set 1, $\dchi=4$, $\mathcal O(p^4)$.}

Flavor basis $C_i^{1}$:
\begin{align}
	C^{1}_{1} & = \TrF(a_{1}, a_{2})\,\TrF(a_{3}, a_{4}), \\
	C^{1}_{2} & = \TrF(a_{1}, a_{2}, a_{3}, a_{4}),       \\
	C^{1}_{3} & = \TrF(a_{1}, a_{2}, a_{4}, a_{3}),       \\
	C^{1}_{4} & = \TrF(a_{1}, a_{3}, a_{4}, a_{2}),       \\
	C^{1}_{5} & = \TrF(a_{1}, a_{3})\,\TrF(a_{2}, a_{4}), \\
	C^{1}_{6} & = \TrF(a_{1}, a_{3}, a_{2}, a_{4}),       \\
	C^{1}_{7} & = \TrF(a_{1}, a_{4}, a_{3}, a_{2}),       \\
	C^{1}_{8} & = \TrF(a_{1}, a_{4}, a_{2}, a_{3})        \\
\end{align}

Lorentz/operator basis $L_k^{1}$:
\begin{align}
	L^{1}_{1} & = (\Bar{B}^{R}_{a_{1}} \sigma^{\nu \rho} \sigma^{\mu} \bar{\sigma}^{\sigma \alpha} \overleftrightarrow{D}_{\mu} B^{R}_{a_{2}}) A_{\nu \rho a_{3}} f_{+, \sigma \alpha a_{4}},       \\
	L^{1}_{2} & = (\Bar{B}^{L}_{a_{1}} \bar{\sigma}^{\sigma \alpha} \bar{\sigma}^{\mu} \sigma^{\nu \rho} \overleftrightarrow{D}_{\mu} B^{L}_{a_{2}}) A_{\nu \rho a_{3}} f_{+, \sigma \alpha a_{4}}, \\
	L^{1}_{3} & = (\Bar{B}_{a_{1}} B_{a_{2}}) \text{Tr}(\bar{\sigma}^{\mu \nu} \bar{\sigma}^{\rho \sigma}) A_{\rho \sigma a_{3}} f_{+, \mu \nu a_{4}},                                              \\
	L^{1}_{4} & = (\Bar{B}^{L}_{a_{1}} \bar{\sigma}^{\mu \nu} \bar{\sigma}^{\rho \sigma} B^{R}_{a_{2}}) A_{\rho \sigma a_{3}} f_{+, \mu \nu a_{4}}                                                  \\
\end{align}

\subsubsection*{$\dchi=5$, $\mathcal O(p^5)$}

\paragraph{$A f_+$.}

\emph{Set 1, $\dchi=5$, $\mathcal O(p^5)$.}

Flavor basis $C_i^{1}$:
\begin{align}
	C^{1}_{1} & = \TrF(a_{1}, a_{2})\,\TrF(a_{3}, a_{4}), \\
	C^{1}_{2} & = \TrF(a_{1}, a_{2}, a_{3}, a_{4}),       \\
	C^{1}_{3} & = \TrF(a_{1}, a_{2}, a_{4}, a_{3}),       \\
	C^{1}_{4} & = \TrF(a_{1}, a_{3}, a_{4}, a_{2}),       \\
	C^{1}_{5} & = \TrF(a_{1}, a_{3})\,\TrF(a_{2}, a_{4}), \\
	C^{1}_{6} & = \TrF(a_{1}, a_{3}, a_{2}, a_{4}),       \\
	C^{1}_{7} & = \TrF(a_{1}, a_{4}, a_{3}, a_{2}),       \\
	C^{1}_{8} & = \TrF(a_{1}, a_{4}, a_{2}, a_{3})        \\
\end{align}

Lorentz/operator basis $L_k^{1}$:
\begin{align}
	L^{1}_{1} & = (D_{\rho} f_{+, \beta \gamma a_{4}}) (\Bar{B}^{R}_{a_{1}} \sigma^{\sigma \alpha} \sigma^{\mu} \bar{\sigma}^{\beta \gamma} \bar{\sigma}^{\nu} \sigma^{\rho} \overleftrightarrow{D}_{\nu} \overleftrightarrow{D}_{\mu} B^{R}_{a_{2}}) A_{\sigma \alpha a_{3}},       \\
	L^{1}_{2} & = (D_{\rho} f_{+, \beta \gamma a_{4}}) (\Bar{B}^{L}_{a_{1}} \bar{\sigma}^{\rho} \sigma^{\nu} \bar{\sigma}^{\beta \gamma} \bar{\sigma}^{\mu} \sigma^{\sigma \alpha} \overleftrightarrow{D}_{\nu} \overleftrightarrow{D}_{\mu} B^{L}_{a_{2}}) A_{\sigma \alpha a_{3}}, \\
	L^{1}_{3} & = (D_{\nu} A_{\rho \sigma a_{3}}) (\Bar{B}^{R}_{a_{1}} \sigma^{\rho \sigma} \overleftrightarrow{D}_{\mu} B^{L}_{a_{2}}) \text{Tr}(\bar{\sigma}^{\alpha \beta} \bar{\sigma}^{\mu} \sigma^{\nu}) f_{+, \alpha \beta a_{4}},                                            \\
	L^{1}_{4} & = (D_{\nu} f_{+, \rho \sigma a_{4}}) (\Bar{B}^{L}_{a_{1}} \bar{\sigma}^{\nu} \sigma^{\mu} \bar{\sigma}^{\alpha \beta} \bar{\sigma}^{\rho \sigma} \overleftrightarrow{D}_{\mu} B^{R}_{a_{2}}) A_{\alpha \beta a_{3}},                                                 \\
	L^{1}_{5} & = (D_{\nu} f_{+, \rho \sigma a_{4}}) (\Bar{B}^{L}_{a_{1}} \bar{\sigma}^{\nu} \sigma^{\alpha \beta} \sigma^{\mu} \bar{\sigma}^{\rho \sigma} \overleftrightarrow{D}_{\mu} B^{R}_{a_{2}}) A_{\alpha \beta a_{3}},                                                       \\
	L^{1}_{6} & = (D_{\mu} f_{+, \nu \rho a_{4}}) (\Bar{B}^{L}_{a_{1}} \bar{\sigma}^{\mu} B^{L}_{a_{2}}) \text{Tr}(\bar{\sigma}^{\nu \rho} \bar{\sigma}^{\sigma \alpha}) A_{\sigma \alpha a_{3}},                                                                                    \\
	L^{1}_{7} & = (D_{\mu} f_{+, \nu \rho a_{4}}) (\Bar{B}^{R}_{a_{1}} \sigma^{\mu} B^{R}_{a_{2}}) \text{Tr}(\bar{\sigma}^{\nu \rho} \bar{\sigma}^{\sigma \alpha}) A_{\sigma \alpha a_{3}},                                                                                          \\
	L^{1}_{8} & = (\Bar{B}_{a_{1}} \gamma_5 B_{a_{2}}) \text{Tr}(\bar{\sigma}^{\mu \nu} \bar{\sigma}^{\rho \sigma}) A_{\rho \sigma a_{3}} f_{+, \mu \nu a_{4}}                                                                                                                       \\
\end{align}

\subsection{\texorpdfstring{$\bar B B \widehat{\chi}_+ u f_+$}{BbarB-chip-u-Fp}}

\label{app:hbp-BbarB-chip-u-Fp}

\par\smallskip
\begingroup
\centering
\scriptsize
\setlength{\tabcolsep}{5pt}
\begin{tabular}{@{}lccrrr@{}}
	\toprule
	Insertion                & $d_\chi$          & set & $n_C$ & $n_L$ & $N$ \\
	\midrule
	$\widehat{\chi}_+ u f_+$ & $\mathcal O(p^5)$ & 1   & 32    & 3     & 96  \\
	$\widehat{\chi}_+ u f_+$ & $\mathcal O(p^6)$ & 1   & 32    & 6     & 192 \\
	\bottomrule
\end{tabular}
\par
\endgroup
\smallskip

\noindent The following sets give explicit representatives.

\subsubsection*{$\dchi=5$, $\mathcal O(p^5)$}

\paragraph{$\widehat{\chi}_+ u f_+$.}

\emph{Set 1, $\dchi=5$, $\mathcal O(p^5)$.}

Flavor basis $C_i^{1}$:
\begin{align}
	C^{1}_{1}  & = \TrF(a_{1}, a_{2})\,\TrF(a_{3}, a_{4}, a_{5}), \\
	C^{1}_{2}  & = \TrF(a_{1}, a_{2})\,\TrF(a_{3}, a_{5}, a_{4}), \\
	C^{1}_{3}  & = \TrF(a_{4}, a_{5})\,\TrF(a_{1}, a_{2}, a_{3}), \\
	C^{1}_{4}  & = \TrF(a_{1}, a_{2}, a_{3}, a_{4}, a_{5}),       \\
	C^{1}_{5}  & = \TrF(a_{1}, a_{2}, a_{3}, a_{5}, a_{4}),       \\
	C^{1}_{6}  & = \TrF(a_{1}, a_{2}, a_{4}, a_{5}, a_{3}),       \\
	C^{1}_{7}  & = \TrF(a_{3}, a_{5})\,\TrF(a_{1}, a_{2}, a_{4}), \\
	C^{1}_{8}  & = \TrF(a_{1}, a_{2}, a_{4}, a_{3}, a_{5}),       \\
	C^{1}_{9}  & = \TrF(a_{1}, a_{2}, a_{5}, a_{4}, a_{3}),       \\
	C^{1}_{10} & = \TrF(a_{1}, a_{2}, a_{5}, a_{3}, a_{4}),       \\
	C^{1}_{11} & = \TrF(a_{4}, a_{5})\,\TrF(a_{1}, a_{3}, a_{2}), \\
	C^{1}_{12} & = \TrF(a_{1}, a_{3}, a_{4}, a_{5}, a_{2}),       \\
	C^{1}_{13} & = \TrF(a_{1}, a_{3}, a_{5}, a_{4}, a_{2}),       \\
	C^{1}_{14} & = \TrF(a_{1}, a_{3})\,\TrF(a_{2}, a_{4}, a_{5}), \\
	C^{1}_{15} & = \TrF(a_{1}, a_{3}, a_{2}, a_{4}, a_{5}),       \\
	C^{1}_{16} & = \TrF(a_{1}, a_{3}, a_{5}, a_{2}, a_{4}),       \\
	C^{1}_{17} & = \TrF(a_{2}, a_{4})\,\TrF(a_{1}, a_{3}, a_{5}), \\
	C^{1}_{18} & = \TrF(a_{1}, a_{3})\,\TrF(a_{2}, a_{5}, a_{4}), \\
	C^{1}_{19} & = \TrF(a_{1}, a_{3}, a_{2}, a_{5}, a_{4}),       \\
	C^{1}_{20} & = \TrF(a_{2}, a_{5})\,\TrF(a_{1}, a_{3}, a_{4}), \\
	C^{1}_{21} & = \TrF(a_{1}, a_{4}, a_{5}, a_{3}, a_{2}),       \\
	C^{1}_{22} & = \TrF(a_{3}, a_{5})\,\TrF(a_{1}, a_{4}, a_{2}), \\
	C^{1}_{23} & = \TrF(a_{1}, a_{4}, a_{3}, a_{5}, a_{2}),       \\
	C^{1}_{24} & = \TrF(a_{1}, a_{4}, a_{5}, a_{2}, a_{3}),       \\
	C^{1}_{25} & = \TrF(a_{1}, a_{4})\,\TrF(a_{2}, a_{3}, a_{5}), \\
	C^{1}_{26} & = \TrF(a_{1}, a_{4}, a_{2}, a_{5}, a_{3}),       \\
	C^{1}_{27} & = \TrF(a_{2}, a_{5})\,\TrF(a_{1}, a_{4}, a_{3}), \\
	C^{1}_{28} & = \TrF(a_{1}, a_{4})\,\TrF(a_{2}, a_{5}, a_{3}), \\
	C^{1}_{29} & = \TrF(a_{1}, a_{5}, a_{4}, a_{3}, a_{2}),       \\
	C^{1}_{30} & = \TrF(a_{1}, a_{5}, a_{3}, a_{4}, a_{2}),       \\
	C^{1}_{31} & = \TrF(a_{1}, a_{5}, a_{4}, a_{2}, a_{3}),       \\
	C^{1}_{32} & = \TrF(a_{1}, a_{5}, a_{2}, a_{3}, a_{4})        \\
\end{align}

Lorentz/operator basis $L_k^{1}$:
\begin{align}
	L^{1}_{1} & = (\Bar{B}^{L}_{a_{1}} \bar{\sigma}^{\nu \rho} \overleftrightarrow{D}_{\mu} B^{R}_{a_{2}}) \text{Tr}(\sigma^{\sigma} \bar{\sigma}^{\mu}) u_{\sigma a_{4}} f_{+, \nu \rho a_{5}} \widehat{\chi}_{+, a_{3}}, \\
	L^{1}_{2} & = (\Bar{B}^{R}_{a_{1}} \sigma^{\mu} \bar{\sigma}^{\nu \rho} B^{R}_{a_{2}}) u_{\mu a_{4}} f_{+, \nu \rho a_{5}} \widehat{\chi}_{+, a_{3}},                                                                  \\
	L^{1}_{3} & = (\Bar{B}^{L}_{a_{1}} \bar{\sigma}^{\nu \rho} \bar{\sigma}^{\mu} B^{L}_{a_{2}}) u_{\mu a_{4}} f_{+, \nu \rho a_{5}} \widehat{\chi}_{+, a_{3}}                                                             \\
\end{align}

\subsubsection*{$\dchi=6$, $\mathcal O(p^6)$}

\paragraph{$\widehat{\chi}_+ u f_+$.}

\emph{Set 1, $\dchi=6$, $\mathcal O(p^6)$.}

Flavor basis $C_i^{1}$:
\begin{align}
	C^{1}_{1}  & = \TrF(a_{1}, a_{2})\,\TrF(a_{3}, a_{4}, a_{5}), \\
	C^{1}_{2}  & = \TrF(a_{1}, a_{2})\,\TrF(a_{3}, a_{5}, a_{4}), \\
	C^{1}_{3}  & = \TrF(a_{4}, a_{5})\,\TrF(a_{1}, a_{2}, a_{3}), \\
	C^{1}_{4}  & = \TrF(a_{1}, a_{2}, a_{3}, a_{4}, a_{5}),       \\
	C^{1}_{5}  & = \TrF(a_{1}, a_{2}, a_{3}, a_{5}, a_{4}),       \\
	C^{1}_{6}  & = \TrF(a_{1}, a_{2}, a_{4}, a_{5}, a_{3}),       \\
	C^{1}_{7}  & = \TrF(a_{3}, a_{5})\,\TrF(a_{1}, a_{2}, a_{4}), \\
	C^{1}_{8}  & = \TrF(a_{1}, a_{2}, a_{4}, a_{3}, a_{5}),       \\
	C^{1}_{9}  & = \TrF(a_{1}, a_{2}, a_{5}, a_{4}, a_{3}),       \\
	C^{1}_{10} & = \TrF(a_{1}, a_{2}, a_{5}, a_{3}, a_{4}),       \\
	C^{1}_{11} & = \TrF(a_{4}, a_{5})\,\TrF(a_{1}, a_{3}, a_{2}), \\
	C^{1}_{12} & = \TrF(a_{1}, a_{3}, a_{4}, a_{5}, a_{2}),       \\
	C^{1}_{13} & = \TrF(a_{1}, a_{3}, a_{5}, a_{4}, a_{2}),       \\
	C^{1}_{14} & = \TrF(a_{1}, a_{3})\,\TrF(a_{2}, a_{4}, a_{5}), \\
	C^{1}_{15} & = \TrF(a_{1}, a_{3}, a_{2}, a_{4}, a_{5}),       \\
	C^{1}_{16} & = \TrF(a_{1}, a_{3}, a_{5}, a_{2}, a_{4}),       \\
	C^{1}_{17} & = \TrF(a_{2}, a_{4})\,\TrF(a_{1}, a_{3}, a_{5}), \\
	C^{1}_{18} & = \TrF(a_{1}, a_{3})\,\TrF(a_{2}, a_{5}, a_{4}), \\
	C^{1}_{19} & = \TrF(a_{1}, a_{3}, a_{2}, a_{5}, a_{4}),       \\
	C^{1}_{20} & = \TrF(a_{2}, a_{5})\,\TrF(a_{1}, a_{3}, a_{4}), \\
	C^{1}_{21} & = \TrF(a_{1}, a_{4}, a_{5}, a_{3}, a_{2}),       \\
	C^{1}_{22} & = \TrF(a_{3}, a_{5})\,\TrF(a_{1}, a_{4}, a_{2}), \\
	C^{1}_{23} & = \TrF(a_{1}, a_{4}, a_{3}, a_{5}, a_{2}),       \\
	C^{1}_{24} & = \TrF(a_{1}, a_{4}, a_{5}, a_{2}, a_{3}),       \\
	C^{1}_{25} & = \TrF(a_{1}, a_{4})\,\TrF(a_{2}, a_{3}, a_{5}), \\
	C^{1}_{26} & = \TrF(a_{1}, a_{4}, a_{2}, a_{5}, a_{3}),       \\
	C^{1}_{27} & = \TrF(a_{2}, a_{5})\,\TrF(a_{1}, a_{4}, a_{3}), \\
	C^{1}_{28} & = \TrF(a_{1}, a_{4})\,\TrF(a_{2}, a_{5}, a_{3}), \\
	C^{1}_{29} & = \TrF(a_{1}, a_{5}, a_{4}, a_{3}, a_{2}),       \\
	C^{1}_{30} & = \TrF(a_{1}, a_{5}, a_{3}, a_{4}, a_{2}),       \\
	C^{1}_{31} & = \TrF(a_{1}, a_{5}, a_{4}, a_{2}, a_{3}),       \\
	C^{1}_{32} & = \TrF(a_{1}, a_{5}, a_{2}, a_{3}, a_{4})        \\
\end{align}

Lorentz/operator basis $L_k^{1}$:
\begin{align}
	L^{1}_{1} & = (D_{\mu} \widehat{\chi}_{+, a_{3}}) (\Bar{B}_{a_{1}} B_{a_{2}}) \text{Tr}(\bar{\sigma}^{\nu \rho} \bar{\sigma}^{\mu} \sigma^{\sigma}) u_{\sigma a_{4}} f_{+, \nu \rho a_{5}},         \\
	L^{1}_{2} & = (D_{\mu} \widehat{\chi}_{+, a_{3}}) (\Bar{B}^{R}_{a_{1}} \sigma^{\mu} \bar{\sigma}^{\nu \rho} \bar{\sigma}^{\sigma} B^{L}_{a_{2}}) u_{\sigma a_{4}} f_{+, \nu \rho a_{5}},            \\
	L^{1}_{3} & = (D_{\mu} \widehat{\chi}_{+, a_{3}}) (\Bar{B}^{L}_{a_{1}} \bar{\sigma}^{\nu \rho} B^{R}_{a_{2}}) \text{Tr}(\sigma^{\sigma} \bar{\sigma}^{\mu}) u_{\sigma a_{4}} f_{+, \nu \rho a_{5}}, \\
	L^{1}_{4} & = (D_{\mu} \widehat{\chi}_{+, a_{3}}) (\Bar{B}^{L}_{a_{1}} \bar{\sigma}^{\mu} \sigma^{\sigma} \bar{\sigma}^{\nu \rho} B^{R}_{a_{2}}) u_{\sigma a_{4}} f_{+, \nu \rho a_{5}},            \\
	L^{1}_{5} & = (D_{\mu} u_{\sigma a_{4}}) (\Bar{B}^{R}_{a_{1}} \sigma^{\mu} \bar{\sigma}^{\nu \rho} \bar{\sigma}^{\sigma} B^{L}_{a_{2}}) f_{+, \nu \rho a_{5}} \widehat{\chi}_{+, a_{3}},            \\
	L^{1}_{6} & = (D_{\mu} f_{+, \nu \rho a_{5}}) (\Bar{B}^{L}_{a_{1}} \bar{\sigma}^{\mu} \sigma^{\sigma} \bar{\sigma}^{\nu \rho} B^{R}_{a_{2}}) u_{\sigma a_{4}} \widehat{\chi}_{+, a_{3}}             \\
\end{align}

\subsection{\texorpdfstring{$\bar T T \widehat{\chi}_+ f_+$}{TbarT-chip-Fp}}

\label{app:hbp-TbarT-chip-Fp}

\par\smallskip
\begingroup
\centering
\scriptsize
\setlength{\tabcolsep}{5pt}
\begin{tabular}{@{}lccrrr@{}}
	\toprule
	Insertion              & $d_\chi$          & set & $n_C$ & $n_L$ & $N$ \\
	\midrule
	$\widehat{\chi}_+ f_+$ & $\mathcal O(p^4)$ & 1   & 4     & 1     & 4   \\
	$\widehat{\chi}_+ f_+$ & $\mathcal O(p^5)$ & 1   & 4     & 1     & 4   \\
	\bottomrule
\end{tabular}
\par
\endgroup
\smallskip

\noindent The following sets give explicit representatives.

\subsubsection*{$\dchi=4$, $\mathcal O(p^4)$}

\paragraph{$\widehat{\chi}_+ f_+$.}

\emph{Set 1, $\dchi=4$, $\mathcal O(p^4)$.}

Flavor basis $C_i^{1}$:
\begin{align}
	C^{1}_{1} & = \TrF(a_{1}, a_{3})\,\TrF(a_{2}, a_{4})\,\TrF(a_{5}, a_{6}), \\
	C^{1}_{2} & = \TrF(a_{1}, a_{3})\,\TrF(a_{2}, a_{4}, a_{5}, a_{6}),       \\
	C^{1}_{3} & = \TrF(a_{1}, a_{3})\,\TrF(a_{2}, a_{4}, a_{6}, a_{5}),       \\
	C^{1}_{4} & = \TrF(a_{1}, a_{3}, a_{5})\,\TrF(a_{2}, a_{4}, a_{6})        \\
\end{align}

Lorentz/operator basis $L_k^{1}$:
\begin{align}
	L^{1}_{1} & = (\mathcal{P}_{xx}^{\mu \nu} \bar{T}^{L}_{\mu a_{1} a_{2}} \bar{\sigma}^{\rho \sigma} T^{R}_{\nu a_{3} a_{4}}) f_{+, \rho \sigma a_{6}} \widehat{\chi}_{+, a_{5}} \\
\end{align}

\subsubsection*{$\dchi=5$, $\mathcal O(p^5)$}

\paragraph{$\widehat{\chi}_+ f_+$.}

\emph{Set 1, $\dchi=5$, $\mathcal O(p^5)$.}

Flavor basis $C_i^{1}$:
\begin{align}
	C^{1}_{1} & = \TrF(a_{1}, a_{3})\,\TrF(a_{2}, a_{4})\,\TrF(a_{5}, a_{6}), \\
	C^{1}_{2} & = \TrF(a_{1}, a_{3})\,\TrF(a_{2}, a_{4}, a_{5}, a_{6}),       \\
	C^{1}_{3} & = \TrF(a_{1}, a_{3})\,\TrF(a_{2}, a_{4}, a_{6}, a_{5}),       \\
	C^{1}_{4} & = \TrF(a_{1}, a_{3}, a_{5})\,\TrF(a_{2}, a_{4}, a_{6})        \\
\end{align}

Lorentz/operator basis $L_k^{1}$:
\begin{align}
	L^{1}_{1} & = (\mathcal{P}_{xy}^{\mu \nu} \bar{T}^{L}_{\mu a_{1} a_{2}} \bar{\sigma}^{\rho \sigma} T^{R}_{\nu a_{3} a_{4}}) f_{+, \rho \sigma a_{6}} \widehat{\chi}_{+, a_{5}} \\
\end{align}

\subsection{\texorpdfstring{$\bar T T f_- f_+$}{TbarT-Fm-Fp}}

\label{app:hbp-TbarT-Fm-Fp}

\par\smallskip
\begingroup
\centering
\scriptsize
\setlength{\tabcolsep}{5pt}
\begin{tabular}{@{}lccrrr@{}}
	\toprule
	Insertion & $d_\chi$          & set & $n_C$ & $n_L$ & $N$ \\
	\midrule
	$f_- f_+$ & $\mathcal O(p^4)$ & 1   & 4     & 3     & 12  \\
	$f_- f_+$ & $\mathcal O(p^5)$ & 1   & 4     & 3     & 12  \\
	\bottomrule
\end{tabular}
\par
\endgroup
\smallskip

\noindent The following sets give explicit representatives.

\subsubsection*{$\dchi=4$, $\mathcal O(p^4)$}

\paragraph{$f_- f_+$.}

\emph{Set 1, $\dchi=4$, $\mathcal O(p^4)$.}

Flavor basis $C_i^{1}$:
\begin{align}
	C^{1}_{1} & = \TrF(a_{1}, a_{3})\,\TrF(a_{2}, a_{4})\,\TrF(a_{5}, a_{6}), \\
	C^{1}_{2} & = \TrF(a_{1}, a_{3})\,\TrF(a_{2}, a_{4}, a_{5}, a_{6}),       \\
	C^{1}_{3} & = \TrF(a_{1}, a_{3})\,\TrF(a_{2}, a_{4}, a_{6}, a_{5}),       \\
	C^{1}_{4} & = \TrF(a_{1}, a_{3}, a_{5})\,\TrF(a_{2}, a_{4}, a_{6})        \\
\end{align}

Lorentz/operator basis $L_k^{1}$:
\begin{align}
	L^{1}_{1} & = (\bar{T}_{\rho \sigma a_{1} a_{2}} T_{\gamma \delta a_{3} a_{4}}) \text{Tr}(\bar{\sigma}^{\mu \nu} \bar{\sigma}^{\rho \sigma}) \text{Tr}(\bar{\sigma}^{\alpha \beta} \bar{\sigma}^{\gamma \delta}) f_{-, \mu \nu a_{5}} f_{+, \alpha \beta a_{6}}, \\
	L^{1}_{2} & = (\mathcal{P}_{xx}^{\mu \nu} \bar{T}_{\mu a_{1} a_{2}} T_{\nu a_{3} a_{4}}) \text{Tr}(\bar{\sigma}^{\rho \sigma} \bar{\sigma}^{\alpha \beta}) f_{-, \alpha \beta a_{5}} f_{+, \rho \sigma a_{6}},                                                   \\
	L^{1}_{3} & = (\mathcal{P}_{xx}^{\mu \nu} \bar{T}^{L}_{\mu a_{1} a_{2}} \bar{\sigma}^{\alpha \beta} \bar{\sigma}^{\rho \sigma} T^{R}_{\nu a_{3} a_{4}}) f_{-, \alpha \beta a_{5}} f_{+, \rho \sigma a_{6}}                                                       \\
\end{align}

\subsubsection*{$\dchi=5$, $\mathcal O(p^5)$}

\paragraph{$f_- f_+$.}

\emph{Set 1, $\dchi=5$, $\mathcal O(p^5)$.}

Flavor basis $C_i^{1}$:
\begin{align}
	C^{1}_{1} & = \TrF(a_{1}, a_{3})\,\TrF(a_{2}, a_{4})\,\TrF(a_{5}, a_{6}), \\
	C^{1}_{2} & = \TrF(a_{1}, a_{3})\,\TrF(a_{2}, a_{4}, a_{5}, a_{6}),       \\
	C^{1}_{3} & = \TrF(a_{1}, a_{3})\,\TrF(a_{2}, a_{4}, a_{6}, a_{5}),       \\
	C^{1}_{4} & = \TrF(a_{1}, a_{3}, a_{5})\,\TrF(a_{2}, a_{4}, a_{6})        \\
\end{align}

Lorentz/operator basis $L_k^{1}$:
\begin{align}
	L^{1}_{1} & = (\bar{T}_{\rho \sigma a_{1} a_{2}} \gamma_5 T_{\gamma \delta a_{3} a_{4}}) \text{Tr}(\bar{\sigma}^{\mu \nu} \bar{\sigma}^{\rho \sigma}) \text{Tr}(\bar{\sigma}^{\alpha \beta} \bar{\sigma}^{\gamma \delta}) f_{-, \mu \nu a_{5}} f_{+, \alpha \beta a_{6}}, \\
	L^{1}_{2} & = (\mathcal{P}_{xy}^{\mu \nu} \bar{T}_{\mu a_{1} a_{2}} T_{\nu a_{3} a_{4}}) \text{Tr}(\bar{\sigma}^{\rho \sigma} \bar{\sigma}^{\alpha \beta}) f_{-, \alpha \beta a_{5}} f_{+, \rho \sigma a_{6}},                                                            \\
	L^{1}_{3} & = (\mathcal{P}_{xy}^{\mu \nu} \bar{T}^{L}_{\mu a_{1} a_{2}} \bar{\sigma}^{\alpha \beta} \bar{\sigma}^{\rho \sigma} T^{R}_{\nu a_{3} a_{4}}) f_{-, \alpha \beta a_{5}} f_{+, \rho \sigma a_{6}}                                                                \\
\end{align}

\subsection{\texorpdfstring{$\bar T T \hat\alpha_\parallel f_+$}{TbarT-V-Fp}}

\label{app:hbp-TbarT-V-Fp}

\par\smallskip
\begingroup
\centering
\scriptsize
\setlength{\tabcolsep}{5pt}
\begin{tabular}{@{}lccrrr@{}}
	\toprule
	Insertion                  & $d_\chi$          & set & $n_C$ & $n_L$ & $N$ \\
	\midrule
	$\hat\alpha_\parallel f_+$ & $\mathcal O(p^3)$ & 1   & 4     & 4     & 16  \\
	$\hat\alpha_\parallel f_+$ & $\mathcal O(p^4)$ & 1   & 4     & 4     & 16  \\
	\bottomrule
\end{tabular}
\par
\endgroup
\smallskip

\noindent The following sets give explicit representatives.

\subsubsection*{$\dchi=3$, $\mathcal O(p^3)$}

\paragraph{$\hat\alpha_\parallel f_+$.}

\emph{Set 1, $\dchi=3$, $\mathcal O(p^3)$.}

Flavor basis $C_i^{1}$:
\begin{align}
	C^{1}_{1} & = \TrF(a_{1}, a_{3})\,\TrF(a_{2}, a_{4})\,\TrF(a_{5}, a_{6}), \\
	C^{1}_{2} & = \TrF(a_{1}, a_{3})\,\TrF(a_{2}, a_{4}, a_{5}, a_{6}),       \\
	C^{1}_{3} & = \TrF(a_{1}, a_{3})\,\TrF(a_{2}, a_{4}, a_{6}, a_{5}),       \\
	C^{1}_{4} & = \TrF(a_{1}, a_{3}, a_{5})\,\TrF(a_{2}, a_{4}, a_{6})        \\
\end{align}

Lorentz/operator basis $L_k^{1}$:
\begin{align}
	L^{1}_{1} & = (\bar{T}_{\alpha \beta a_{1} a_{2}} T_{\nu a_{3} a_{4}}) \text{Tr}(\bar{\sigma}^{\mu} \sigma^{\nu}) \text{Tr}(\bar{\sigma}^{\rho \sigma} \bar{\sigma}^{\alpha \beta}) \hat\alpha_{\parallel, \mu a_{5}} f_{+, \rho \sigma a_{6}}, \\
	L^{1}_{2} & = (\bar{T}_{\rho a_{1} a_{2}} T_{\alpha \beta a_{3} a_{4}}) \text{Tr}(\bar{\sigma}^{\mu \nu} \bar{\sigma}^{\rho} \sigma^{\sigma} \bar{\sigma}^{\alpha \beta}) \hat\alpha_{\parallel, \sigma a_{5}} f_{+, \mu \nu a_{6}},            \\
	L^{1}_{3} & = (\mathcal{P}_{xx}^{\mu \nu} \bar{T}^{R}_{\mu a_{1} a_{2}} \sigma^{\rho} \bar{\sigma}^{\sigma \alpha} T^{R}_{\nu a_{3} a_{4}}) \hat\alpha_{\parallel, \rho a_{5}} f_{+, \sigma \alpha a_{6}},                                      \\
	L^{1}_{4} & = (\mathcal{P}_{xx}^{\mu \nu} \bar{T}^{L}_{\mu a_{1} a_{2}} \bar{\sigma}^{\sigma \alpha} \bar{\sigma}^{\rho} T^{L}_{\nu a_{3} a_{4}}) \hat\alpha_{\parallel, \rho a_{5}} f_{+, \sigma \alpha a_{6}}                                 \\
\end{align}

\subsubsection*{$\dchi=4$, $\mathcal O(p^4)$}

\paragraph{$\hat\alpha_\parallel f_+$.}

\emph{Set 1, $\dchi=4$, $\mathcal O(p^4)$.}

Flavor basis $C_i^{1}$:
\begin{align}
	C^{1}_{1} & = \TrF(a_{1}, a_{3})\,\TrF(a_{2}, a_{4})\,\TrF(a_{5}, a_{6}), \\
	C^{1}_{2} & = \TrF(a_{1}, a_{3})\,\TrF(a_{2}, a_{4}, a_{5}, a_{6}),       \\
	C^{1}_{3} & = \TrF(a_{1}, a_{3})\,\TrF(a_{2}, a_{4}, a_{6}, a_{5}),       \\
	C^{1}_{4} & = \TrF(a_{1}, a_{3}, a_{5})\,\TrF(a_{2}, a_{4}, a_{6})        \\
\end{align}

Lorentz/operator basis $L_k^{1}$:
\begin{align}
	L^{1}_{1} & = (\bar{T}_{\rho a_{1} a_{2}} \gamma_5 T_{\alpha \beta a_{3} a_{4}}) \text{Tr}(\bar{\sigma}^{\mu \nu} \bar{\sigma}^{\rho} \sigma^{\sigma} \bar{\sigma}^{\alpha \beta}) \hat\alpha_{\parallel, \sigma a_{5}} f_{+, \mu \nu a_{6}},            \\
	L^{1}_{2} & = (\bar{T}_{\alpha \beta a_{1} a_{2}} \gamma_5 T_{\nu a_{3} a_{4}}) \text{Tr}(\bar{\sigma}^{\mu} \sigma^{\nu}) \text{Tr}(\bar{\sigma}^{\rho \sigma} \bar{\sigma}^{\alpha \beta}) \hat\alpha_{\parallel, \mu a_{5}} f_{+, \rho \sigma a_{6}}, \\
	L^{1}_{3} & = (\mathcal{P}_{xy}^{\mu \nu} \bar{T}^{R}_{\mu a_{1} a_{2}} \sigma^{\rho} \bar{\sigma}^{\sigma \alpha} T^{R}_{\nu a_{3} a_{4}}) \hat\alpha_{\parallel, \rho a_{5}} f_{+, \sigma \alpha a_{6}},                                               \\
	L^{1}_{4} & = (\mathcal{P}_{xy}^{\mu \nu} \bar{T}^{L}_{\mu a_{1} a_{2}} \bar{\sigma}^{\sigma \alpha} \bar{\sigma}^{\rho} T^{L}_{\nu a_{3} a_{4}}) \hat\alpha_{\parallel, \rho a_{5}} f_{+, \sigma \alpha a_{6}}                                          \\
\end{align}

\subsection{\texorpdfstring{$\bar T T u u$}{TbarT-uu}}

\label{app:hbp-TbarT-uu}

\par\smallskip
\begingroup
\centering
\scriptsize
\setlength{\tabcolsep}{5pt}
\begin{tabular}{@{}lccrrr@{}}
	\toprule
	Insertion & $d_\chi$          & set & $n_C$ & $n_L$ & $N$ \\
	\midrule
	$u u$     & $\mathcal O(p^2)$ & 1   & 3     & 4     & 12  \\
	$u u$     & $\mathcal O(p^2)$ & 2   & 1     & 1     & 1   \\
	$u u$     & $\mathcal O(p^3)$ & 1   & 2     & 6     & 12  \\
	$u u$     & $\mathcal O(p^3)$ & 2   & 1     & 10    & 10  \\
	$u u$     & $\mathcal O(p^3)$ & 3   & 1     & 3     & 3   \\
	\bottomrule
\end{tabular}
\par
\endgroup
\smallskip

\noindent The following sets give explicit representatives.

\subsubsection*{$\dchi=2$, $\mathcal O(p^2)$}

\paragraph{$u u$.}

\emph{Set 1, $\dchi=2$, $\mathcal O(p^2)$.}

Flavor basis $C_i^{1}$:
\begin{align}
	C^{1}_{1} & = \TrF(a_{1}, a_{3})\,\TrF(a_{2}, a_{4})\,\TrF(a_{5}, a_{6}), \\
	C^{1}_{2} & = \TrF(a_{1}, a_{3})\,\TrF(a_{2}, a_{4}, a_{5}, a_{6}),       \\
	C^{1}_{3} & = \TrF(a_{1}, a_{3}, a_{5})\,\TrF(a_{2}, a_{4}, a_{6})        \\
\end{align}

Lorentz/operator basis $L_k^{1}$:
\begin{align}
	L^{1}_{1} & = (\bar{T}_{\nu a_{1} a_{2}} T_{\sigma a_{3} a_{4}}) \text{Tr}(\sigma^{\mu} \bar{\sigma}^{\nu} \sigma^{\rho} \bar{\sigma}^{\sigma}) u_{\rho a_{5}} u_{\mu a_{6}},                           \\
	L^{1}_{2} & = (\bar{T}_{\nu \rho a_{1} a_{2}} T_{\alpha \beta a_{3} a_{4}}) \text{Tr}(\bar{\sigma}^{\mu} \sigma^{\nu \rho} \sigma^{\sigma} \bar{\sigma}^{\alpha \beta}) u_{\sigma a_{5}} u_{\mu a_{6}}, \\
	L^{1}_{3} & = (\bar{T}_{\alpha \beta a_{1} a_{2}} T_{\nu \rho a_{3} a_{4}}) \text{Tr}(\bar{\sigma}^{\mu} \sigma^{\nu \rho} \sigma^{\sigma} \bar{\sigma}^{\alpha \beta}) u_{\sigma a_{5}} u_{\mu a_{6}}, \\
	L^{1}_{4} & = (\mathcal{P}_{xx}^{\mu \nu} \bar{T}_{\mu a_{1} a_{2}} T_{\nu a_{3} a_{4}}) \text{Tr}(\bar{\sigma}^{\rho} \sigma^{\sigma}) u_{\sigma a_{5}} u_{\rho a_{6}}                                 \\
\end{align}

\emph{Set 2, $\dchi=2$, $\mathcal O(p^2)$.}

Flavor basis $C_i^{2}$:
\begin{align}
	C^{2}_{1} & = \TrF(a_{1}, a_{3})\,\TrF(a_{2}, a_{4}, a_{6}, a_{5}) \\
\end{align}

Lorentz/operator basis $L_k^{2}$:
\begin{align}
	L^{2}_{1} & = (\mathcal{P}_{xx}^{\mu \nu} \bar{T}_{\mu a_{1} a_{2}} T_{\nu a_{3} a_{4}}) \text{Tr}(\bar{\sigma}^{\rho} \sigma^{\sigma}) u_{\sigma a_{5}} u_{\rho a_{6}} \\
\end{align}

\subsubsection*{$\dchi=3$, $\mathcal O(p^3)$}

\paragraph{$u u$.}

\emph{Set 1, $\dchi=3$, $\mathcal O(p^3)$.}

Flavor basis $C_i^{1}$:
\begin{align}
	C^{1}_{1} & = \TrF(a_{1}, a_{3})\,\TrF(a_{2}, a_{4})\,\TrF(a_{5}, a_{6}), \\
	C^{1}_{2} & = \TrF(a_{1}, a_{3}, a_{5})\,\TrF(a_{2}, a_{4}, a_{6})        \\
\end{align}

Lorentz/operator basis $L_k^{1}$:
\begin{align}
	L^{1}_{1} & = (\bar{T}_{\nu a_{1} a_{2}} \gamma_5 T_{\sigma a_{3} a_{4}}) \text{Tr}(\sigma^{\mu} \bar{\sigma}^{\nu} \sigma^{\rho} \bar{\sigma}^{\sigma}) u_{\rho a_{5}} u_{\mu a_{6}},                           \\
	L^{1}_{2} & = (\bar{T}_{\nu \rho a_{1} a_{2}} \gamma_5 T_{\alpha \beta a_{3} a_{4}}) \text{Tr}(\bar{\sigma}^{\mu} \sigma^{\nu \rho} \sigma^{\sigma} \bar{\sigma}^{\alpha \beta}) u_{\sigma a_{5}} u_{\mu a_{6}}, \\
	L^{1}_{3} & = (\bar{T}_{\alpha \beta a_{1} a_{2}} \gamma_5 T_{\nu \rho a_{3} a_{4}}) \text{Tr}(\bar{\sigma}^{\mu} \sigma^{\nu \rho} \sigma^{\sigma} \bar{\sigma}^{\alpha \beta}) u_{\sigma a_{5}} u_{\mu a_{6}}, \\
	L^{1}_{4} & = (\mathcal{P}_{xx}^{\mu \nu} \bar{T}^{R}_{\mu a_{1} a_{2}} \sigma^{\rho} \bar{\sigma}^{\sigma} \sigma^{\alpha} T^{R}_{\nu a_{3} a_{4}}) (D_{\rho} u_{\alpha a_{6}}) u_{\sigma a_{5}},               \\
	L^{1}_{5} & = (\mathcal{P}_{xx}^{\mu \nu} \bar{T}^{L}_{\mu a_{1} a_{2}} \bar{\sigma}^{\alpha} \sigma^{\sigma} \bar{\sigma}^{\rho} T^{L}_{\nu a_{3} a_{4}}) (D_{\rho} u_{\alpha a_{6}}) u_{\sigma a_{5}},         \\
	L^{1}_{6} & = (\mathcal{P}_{xy}^{\mu \nu} \bar{T}_{\mu a_{1} a_{2}} T_{\nu a_{3} a_{4}}) \text{Tr}(\bar{\sigma}^{\rho} \sigma^{\sigma}) u_{\sigma a_{5}} u_{\rho a_{6}}                                          \\
\end{align}

\emph{Set 2, $\dchi=3$, $\mathcal O(p^3)$.}

Flavor basis $C_i^{2}$:
\begin{align}
	C^{2}_{1} & = \TrF(a_{1}, a_{3})\,\TrF(a_{2}, a_{4}, a_{5}, a_{6}) \\
\end{align}

Lorentz/operator basis $L_k^{2}$:
\begin{align}
	L^{2}_{1}  & = (D_{\mu} u_{\sigma a_{6}}) (\bar{T}^{R}_{\alpha a_{1} a_{2}} \sigma^{\nu} \bar{\sigma}^{\rho} \sigma^{\sigma} T^{R}_{\rho a_{3} a_{4}}) \text{Tr}(\sigma^{\alpha} \bar{\sigma}^{\mu}) u_{\nu a_{5}},                      \\
	L^{2}_{2}  & = (D_{\mu} u_{\alpha a_{6}}) (\bar{T}^{L}_{\nu a_{1} a_{2}} \bar{\sigma}^{\alpha} \sigma^{\sigma} \bar{\sigma}^{\rho} \sigma^{\nu} \bar{\sigma}^{\mu} T^{L}_{\sigma a_{3} a_{4}}) u_{\rho a_{5}},                           \\
	L^{2}_{3}  & = (\bar{T}_{\nu a_{1} a_{2}} \gamma_5 T_{\sigma a_{3} a_{4}}) \text{Tr}(\sigma^{\mu} \bar{\sigma}^{\nu} \sigma^{\rho} \bar{\sigma}^{\sigma}) u_{\rho a_{5}} u_{\mu a_{6}},                                                  \\
	L^{2}_{4}  & = (D_{\mu} u_{\gamma a_{6}}) (\bar{T}^{R}_{\alpha \beta a_{1} a_{2}} \sigma^{\mu} \bar{\sigma}^{\nu \rho} \bar{\sigma}^{\sigma} \sigma^{\alpha \beta} \sigma^{\gamma} T^{R}_{\nu \rho a_{3} a_{4}}) u_{\sigma a_{5}},       \\
	L^{2}_{5}  & = (D_{\mu} u_{\gamma a_{6}}) (\bar{T}^{L}_{\nu \rho a_{1} a_{2}} \bar{\sigma}^{\gamma} \sigma^{\alpha \beta} \sigma^{\sigma} \bar{\sigma}^{\nu \rho} \bar{\sigma}^{\mu} T^{L}_{\alpha \beta a_{3} a_{4}}) u_{\sigma a_{5}}, \\
	L^{2}_{6}  & = (\bar{T}_{\nu \rho a_{1} a_{2}} \gamma_5 T_{\alpha \beta a_{3} a_{4}}) \text{Tr}(\bar{\sigma}^{\mu} \sigma^{\nu \rho} \sigma^{\sigma} \bar{\sigma}^{\alpha \beta}) u_{\sigma a_{5}} u_{\mu a_{6}},                        \\
	L^{2}_{7}  & = (\bar{T}_{\alpha \beta a_{1} a_{2}} \gamma_5 T_{\nu \rho a_{3} a_{4}}) \text{Tr}(\bar{\sigma}^{\mu} \sigma^{\nu \rho} \sigma^{\sigma} \bar{\sigma}^{\alpha \beta}) u_{\sigma a_{5}} u_{\mu a_{6}},                        \\
	L^{2}_{8}  & = (\mathcal{P}_{xx}^{\mu \nu} \bar{T}^{R}_{\mu a_{1} a_{2}} \sigma^{\rho} \bar{\sigma}^{\sigma} \sigma^{\alpha} T^{R}_{\nu a_{3} a_{4}}) (D_{\rho} u_{\alpha a_{6}}) u_{\sigma a_{5}},                                      \\
	L^{2}_{9}  & = (\mathcal{P}_{xx}^{\mu \nu} \bar{T}^{L}_{\mu a_{1} a_{2}} \bar{\sigma}^{\alpha} \sigma^{\sigma} \bar{\sigma}^{\rho} T^{L}_{\nu a_{3} a_{4}}) (D_{\rho} u_{\alpha a_{6}}) u_{\sigma a_{5}},                                \\
	L^{2}_{10} & = (\mathcal{P}_{xy}^{\mu \nu} \bar{T}_{\mu a_{1} a_{2}} T_{\nu a_{3} a_{4}}) \text{Tr}(\bar{\sigma}^{\rho} \sigma^{\sigma}) u_{\sigma a_{5}} u_{\rho a_{6}}                                                                 \\
\end{align}

\emph{Set 3, $\dchi=3$, $\mathcal O(p^3)$.}

Flavor basis $C_i^{3}$:
\begin{align}
	C^{3}_{1} & = \TrF(a_{1}, a_{3})\,\TrF(a_{2}, a_{4}, a_{6}, a_{5}) \\
\end{align}

Lorentz/operator basis $L_k^{3}$:
\begin{align}
	L^{3}_{1} & = (\mathcal{P}_{xx}^{\mu \nu} \bar{T}^{R}_{\mu a_{1} a_{2}} \sigma^{\rho} \bar{\sigma}^{\sigma} \sigma^{\alpha} T^{R}_{\nu a_{3} a_{4}}) (D_{\rho} u_{\alpha a_{6}}) u_{\sigma a_{5}},       \\
	L^{3}_{2} & = (\mathcal{P}_{xx}^{\mu \nu} \bar{T}^{L}_{\mu a_{1} a_{2}} \bar{\sigma}^{\alpha} \sigma^{\sigma} \bar{\sigma}^{\rho} T^{L}_{\nu a_{3} a_{4}}) (D_{\rho} u_{\alpha a_{6}}) u_{\sigma a_{5}}, \\
	L^{3}_{3} & = (\mathcal{P}_{xy}^{\mu \nu} \bar{T}_{\mu a_{1} a_{2}} T_{\nu a_{3} a_{4}}) \text{Tr}(\bar{\sigma}^{\rho} \sigma^{\sigma}) u_{\sigma a_{5}} u_{\rho a_{6}}                                  \\
\end{align}

\subsection{\texorpdfstring{$\bar T T \widehat{\chi}_+ \widehat{\chi}_+$}{TbarT-chipp-chipp-adaptive}}

\label{app:hbp-TbarT-chipp-chipp-adaptive}

\par\smallskip
\begingroup
\centering
\scriptsize
\setlength{\tabcolsep}{5pt}
\begin{tabular}{@{}lccrrr@{}}
	\toprule
	Insertion                           & $d_\chi$          & set & $n_C$ & $n_L$ & $N$ \\
	\midrule
	$\widehat{\chi}_+ \widehat{\chi}_+$ & $\mathcal O(p^4)$ & 1   & 3     & 1     & 3   \\
	$\widehat{\chi}_+ \widehat{\chi}_+$ & $\mathcal O(p^5)$ & 1   & 2     & 1     & 2   \\
	$\widehat{\chi}_+ \widehat{\chi}_+$ & $\mathcal O(p^5)$ & 2   & 1     & 3     & 3   \\
	\bottomrule
\end{tabular}
\par
\endgroup
\smallskip

\noindent The following sets give explicit representatives.

\subsubsection*{$\dchi=4$, $\mathcal O(p^4)$}

\paragraph{$\widehat{\chi}_+ \widehat{\chi}_+$.}

\emph{Set 1, $\dchi=4$, $\mathcal O(p^4)$.}

Flavor basis $C_i^{1}$:
\begin{align}
	C^{1}_{1} & = \TrF(a_{1}, a_{3})\,\TrF(a_{2}, a_{4})\,\TrF(a_{5}, a_{6}), \\
	C^{1}_{2} & = \TrF(a_{1}, a_{3})\,\TrF(a_{2}, a_{4}, a_{5}, a_{6}),       \\
	C^{1}_{3} & = \TrF(a_{1}, a_{3}, a_{5})\,\TrF(a_{2}, a_{4}, a_{6})        \\
\end{align}

Lorentz/operator basis $L_k^{1}$:
\begin{align}
	L^{1}_{1} & = (\mathcal{P}_{xx}^{\mu \nu} \bar{T}_{\mu a_{1} a_{2}} T_{\nu a_{3} a_{4}}) \widehat{\chi}_{+, a_{5}} \widehat{\chi}_{+, a_{6}} \\
\end{align}

\subsubsection*{$\dchi=5$, $\mathcal O(p^5)$}

\paragraph{$\widehat{\chi}_+ \widehat{\chi}_+$.}

\emph{Set 1, $\dchi=5$, $\mathcal O(p^5)$.}

Flavor basis $C_i^{1}$:
\begin{align}
	C^{1}_{1} & = \TrF(a_{1}, a_{3})\,\TrF(a_{2}, a_{4})\,\TrF(a_{5}, a_{6}), \\
	C^{1}_{2} & = \TrF(a_{1}, a_{3}, a_{5})\,\TrF(a_{2}, a_{4}, a_{6})        \\
\end{align}

Lorentz/operator basis $L_k^{1}$:
\begin{align}
	L^{1}_{1} & = (\mathcal{P}_{xy}^{\mu \nu} \bar{T}_{\mu a_{1} a_{2}} T_{\nu a_{3} a_{4}}) \widehat{\chi}_{+, a_{5}} \widehat{\chi}_{+, a_{6}} \\
\end{align}

\emph{Set 2, $\dchi=5$, $\mathcal O(p^5)$.}

Flavor basis $C_i^{2}$:
\begin{align}
	C^{2}_{1} & = \TrF(a_{1}, a_{3})\,\TrF(a_{2}, a_{4}, a_{5}, a_{6}) \\
\end{align}

Lorentz/operator basis $L_k^{2}$:
\begin{align}
	L^{2}_{1} & = (\mathcal{P}_{xx}^{\mu \nu} \bar{T}^{L}_{\mu a_{1} a_{2}} \bar{\sigma}^{\rho} T^{L}_{\nu a_{3} a_{4}}) (D_{\rho} \widehat{\chi}_{+, a_{6}}) \widehat{\chi}_{+, a_{5}}, \\
	L^{2}_{2} & = (\mathcal{P}_{xx}^{\mu \nu} \bar{T}^{R}_{\mu a_{1} a_{2}} \sigma^{\rho} T^{R}_{\nu a_{3} a_{4}}) (D_{\rho} \widehat{\chi}_{+, a_{6}}) \widehat{\chi}_{+, a_{5}},       \\
	L^{2}_{3} & = (\mathcal{P}_{xy}^{\mu \nu} \bar{T}_{\mu a_{1} a_{2}} T_{\nu a_{3} a_{4}}) \widehat{\chi}_{+, a_{5}} \widehat{\chi}_{+, a_{6}}                                         \\
\end{align}

\bibliographystyle{JHEP}
\bibliography{references}
\end{document}